%% file: plos_opinion_dynamics_arxiv_v3.tex
\documentclass[10pt,letterpaper]{article}
\usepackage{geometry}
\usepackage{changepage}

\usepackage[utf8]{inputenc}

\usepackage{textcomp,marvosym}

\usepackage{fixltx2e}

\usepackage{amsmath,amssymb}

\usepackage{cite}

\usepackage{nameref,hyperref}

\usepackage[right]{lineno}

\usepackage{microtype}
\DisableLigatures[f]{encoding = *, family = * }

\usepackage{rotating}

\usepackage{subcaption}
\usepackage{mwe}

\usepackage[aboveskip=1pt,labelfont=bf,labelsep=period,justification=justified,singlelinecheck=off]{caption}

\makeatletter
\renewcommand{\@biblabel}[1]{\quad#1.}
\makeatother

\date{}

\input{new_commands_and_packages.tex}

\usepackage{lastpage,fancyhdr,graphicx}
\usepackage{epstopdf}
\begin{document}
\vspace*{0.35in}

\begin{flushleft}
{\Large
\textbf\newline{Modelling influence and opinion evolution in online collective behaviour}
}



Corentin Vande Kerckhove\textsuperscript{1,+},
Samuel Martin\textsuperscript{2,+,*},
Pascal Gend\textsuperscript{2},
Peter J. Rentfrow\textsuperscript{3},
Julien M. Hendrickx\textsuperscript{1},
Vincent D. Blondel\textsuperscript{1}
\\
\bigskip
\bf{1} Large graphs and networks group, Universit\'{e} catholique de Louvain, Avenue Georges Lemaitre, 4 B-1348 Louvain-la-Neuve, Belgium
\\
\bf{2} Universit\'e de Lorraine, CRAN, UMR 7039 and CNRS, CRAN, UMR 7039, 2 Avenue de la For\^et de Haye, Vandoeuvre-les-Nancy, France
\\
\bf{3} Department of Psychology, University of Cambridge
\\
\bigskip

%
%
+ These authors contributed equally to this work.





* samuel.martin@univ-lorraine.fr

\end{flushleft}
\section*{Abstract}
\input{abstract.tex}






\section*{Introduction}

Many individual judgments are mediated by observing others' judgments. This is true for buying products, voting for a political party or choosing to donate blood. This is particularly noticeable on the online world. The availability of online data has lead to a recent surge in trying to understand how online social influence impact human behaviour.
Some \textit{in vivo} large scale online experiments were devoted to understand how information and behaviours spread in online social networks~\cite{Centola2010}, others focused on determining which sociological attributes such as gender or age were involved in social influence processes~\cite{aral2012identifying}.

Although decision outcomes are often tied to an objective best choice, outcomes can hardly be fully inferred from this supposedly best choice. For instance, predicting the popularity of songs in a cultural market requires more than just knowing the actual song quality~\cite{salganik2006experimental}. The decision outcome is rather determined by the social influence process at work~\cite{lorenz2011social}. Hence, there is a need for opinion dynamics models with a predictive power.

Complementarily to the \textit{in vivo} experiments, other recent studies used online \textit{in vitro} experiment to identify the micro-level mechanisms susceptible to explain the way social influence impacts human decision making~\cite{moussaid2013social,chacoma2015opinion,mavrodiev2013quantifying}. These recent online \textit{in vitro} studies have lead to posit that the so-called linear consensus model may be appropriate to describe the way individuals revise their judgment when exposed to judgments of others. The predictive power of such a mechanism remains to be assessed. 

\samsc{
Trying to describe how individuals revise their judgment when subject to social influence has a long history in the psychological and social sciences. The consensus model used in this article draws from this line of work. These works were \cocoreview{originally developed} to better understand small group decision making. This occurs for instance when a jury in civil trials has to decide the amount of compensation awarded to plaintiffs~\cite{hastie1983inside, horowitz1996effects,hinsz1995assimilation}. Various types of tasks have been explored by researchers. These includes the forecasts of future events e.g., predicting market sales based on previous prices and other cues~\cite{fischer1999combining, harvey2000using}, the price of products~\cite{schrah2006no, sniezek2004improving}, the probability of event occurrence~\cite{budescu2000confidence,budescu2003effects}, such as the number of future cattle deaths~\cite{harvey1997taking}, or regional temperatures~\cite{harries2004combining}.
The central ingredient entering in models of judgment revision is the weight which individuals put on the judgments of others, termed \textit{influenceability} in the present article. This quantity is also known as the \textit{advice taking weight}~\cite{harvey1997taking,sniezek2004improving} or the \textit{weight of advice}~\cite{yaniv2004benefit,yaniv2004receiving,gino2008we}. It is represented by a number taking $0$ value when the individual is not influenced and $1$ when they entirely forget their own opinion to adopt the one from other individuals in the group. It has been observed that in a \cocoreview{vast} majority of the cases, the final judgment falls between the initial one and the ones from the rest of the group. Said otherwise, the influenceability lies between $0$ and $1$. This has been shown to sensibly improve the accuracy of decisions~\cite{yaniv2007using}. A $20\%$ improvement has been found in an experiment when individuals considered the opinion of another person only~\cite{yaniv2004benefit}. However, individuals do not weight themselves and others equally. They rather overweight their own opinions~\cite{bonaccio2006advice}. This has been coined egocentric discounting~\cite{yaniv2000advice}.
Many factors affect influenceability. These include the perceived \cocoreview{expertise} of the adviser~\cite{harvey1997taking,soll2009strategies} which may result from age, education, life experience~\cite{feng2006predicting}, the difficulty of the task~\cite{gino2008we}, whether the individual feels powerful~\cite{see2011detrimental} or angry~\cite{gino2008blinded}, the size of the group~\cite{mannes2009we}, among others. A sensitivity analysis has been carried out to determine which factors most affect advice taking~\cite{azen2003dominance}.}

This line of work has focused on determining the factors impacting influenceability. None has yet answered whether judgment revision models could be used to predict future decisions. Instead, the models were validated on the data which served to calibrate the models themselves. This pitfall tends to favor more complex models over more simple ones and may result in overfitting. The model would then be unable to predict judgment revision from a new dataset. One reason for this literature gap could be the lack of access to large judgment revision database at the time, now made more readily available via online \cocoreview{\textit{in vitro}} experiments.
The predictability assessment is a necessary step to grow confidence in our understanding and in turn use this mechanism as a building block to design efficient online social systems. Revising judgments after being exposed to others' judgments takes an important role in many online social systems such as recommendation system~\cite{pope2009reacting,dellarocas2003digitization} or viral marketing campaign~\cite{bessi2015science} among others.
Unlike previous research, the present work provides an assessment of the model predictive power through crossvalidation of the proposed judgment revision model.

The prediction accuracy of a model is limited to the extent the judgment revision process is a deterministic process. However, \cocoreview{there is} theoretical~\cite{steyvers2006probabilistic,kersten2003bayesian,ma2006bayesian} and empirical~\cite{vul2008measuring} evidence showing that the opinion individuals display is a sample of an internal probabilistic distribution. For instance, Vul and Pashler~\cite{vul2008measuring} showed that when participants were asked to provide their opinion twice with some delay in between, participants provided two different answers. Following these results, the present article details a new methodology to estimate the unpredictability level of the judgment revision mechanism. This quantifies the highest prediction accuracy one can expect.

The results presented in this article were derived using \cocoreview{\textit{in vitro}} online experiments, where each participant repeated several times estimation tasks in very similar conditions. These repeated experiments \cocoreview{yielded two} complementary sets of results. 
First, it is shown that, in presence of social influence, the way individuals revise their judgment can be modeled using a quantitative model. Unlike the previously discussed studies, the gathered data allow assessing the predictive power of the model. The model casts individuals' behaviours according to their \textit{influenceability}, the factor quantifying to what extent one takes external opinions into account. Secondly, a measure of intrinsic unpredictability in judgment revision is provided. Estimating the intrinsic unpredictability provides a limit beyond which no one can expect to improve predictions. This last result was made possible through a specific \cocoreview{\textit{in vitro}} control experiment. 
Although models of opinion dynamics have been widely studied for decades by sociologist~\cite{French1956} from a theoretical standpoint, to the best of our knowledge, it is the first time the predictive power of a quantitative model of opinion dynamics is tested against a real dataset.

\section*{\cocoreview{Results and Discussion}}\label{sec:results}

\corentin{To quantify opinion dynamics subject to social influence, we carried out online experiments in which participants had to estimate some quantities while receiving information regarding opinions from other participants.
\samgreen{In a first round, a participant expresses their opinion $x_i$ corresponding to their estimation related to the task. In the two subsequent rounds, the participant is exposed to a set of opinions $x_j$ of other participants who performed the same task independently, and gets to update their own opinion. The objective of the study is to model and predict how an individual revises their judgment when exposed to other opinions.}
 Two types of games were designed : the \gauginggame,}
 \corentinb{in which the participants evaluated color proportions}
 \corentin{and the~\countinggame,}
\corentinb{where the task required to guess amounts of items displayed in a picture \cocoreview{(see \textit{Experiment} section in \textit{Material and Methods})}.} 
Participants to 
\corentin{this}
online crowdsourced study were involved in $3$-round judgment revision games.
Judgment revision is modeled using
\corentin{a}
time-varying influenceability consensus model.
 \corentin{In mathematical terms,} 
\corentin{$x_i(r)$ denotes the opinion of individual $i$ at round $r$ and its evolution is described as}
\begin{equation}\label{eq:consensus-alpha-to-mean}
x_i(r+1) \;=\; x_i(r) \;+\; \alpha_i(r) \cdot \left(\bar{x}(r) - x_i(r)\right) 
\end{equation}
where
\corentin{$r=1,2$ }
\corentin{and where $\bar{x}(r)$ is the mean opinion of the group at round $r$}
(see \textit{Opinion revision model} section in \textit{Material and Methods}
for details \samreview{and supplementary section~\ref{SI-sec:linearity-test} for a test of the validity of the linearity assumption}).
This model is based on the \textit{influenceability} $\alpha_i(r)$ of participants, a factor representing to what extent a participant incorporates external judgments. 

\subsection*{Influenceability of participants}

\corentin{The influenceability is described }
\corentin{for each participant}
by two parameters~: 
 \corentin{$\alpha_i(1)$}, 
 the influenceability after first social influence and 
 \corentin{$\alpha_i(2)$}, 
 the influenceability after second social influence. 

\begin{figure}[!ht]
\centering
\textbf{(A) Gauging}\\
\vspace{0.1cm}
\includegraphics[clip=true,scale=0.3]{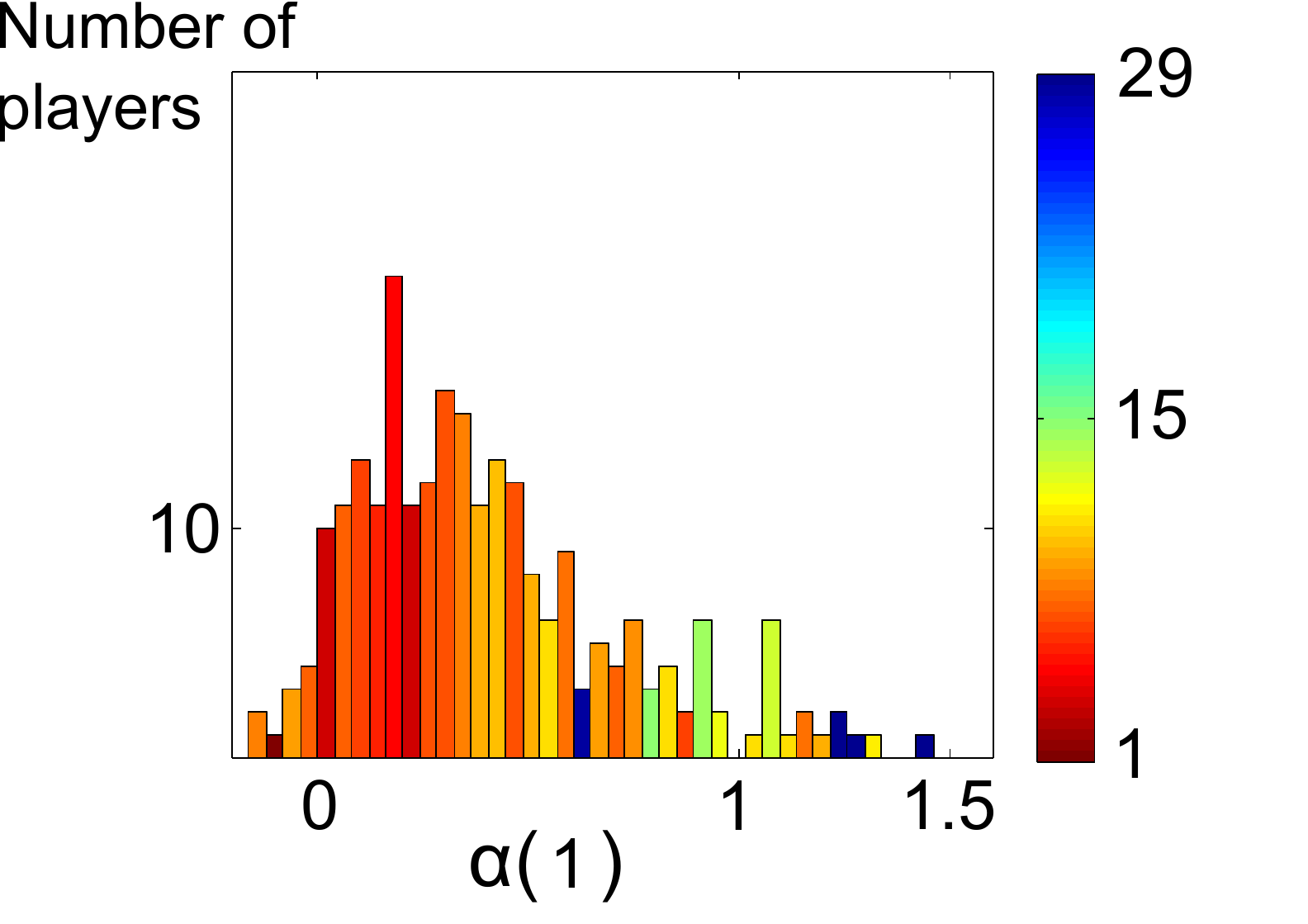}
\includegraphics[clip=true,scale=0.3]{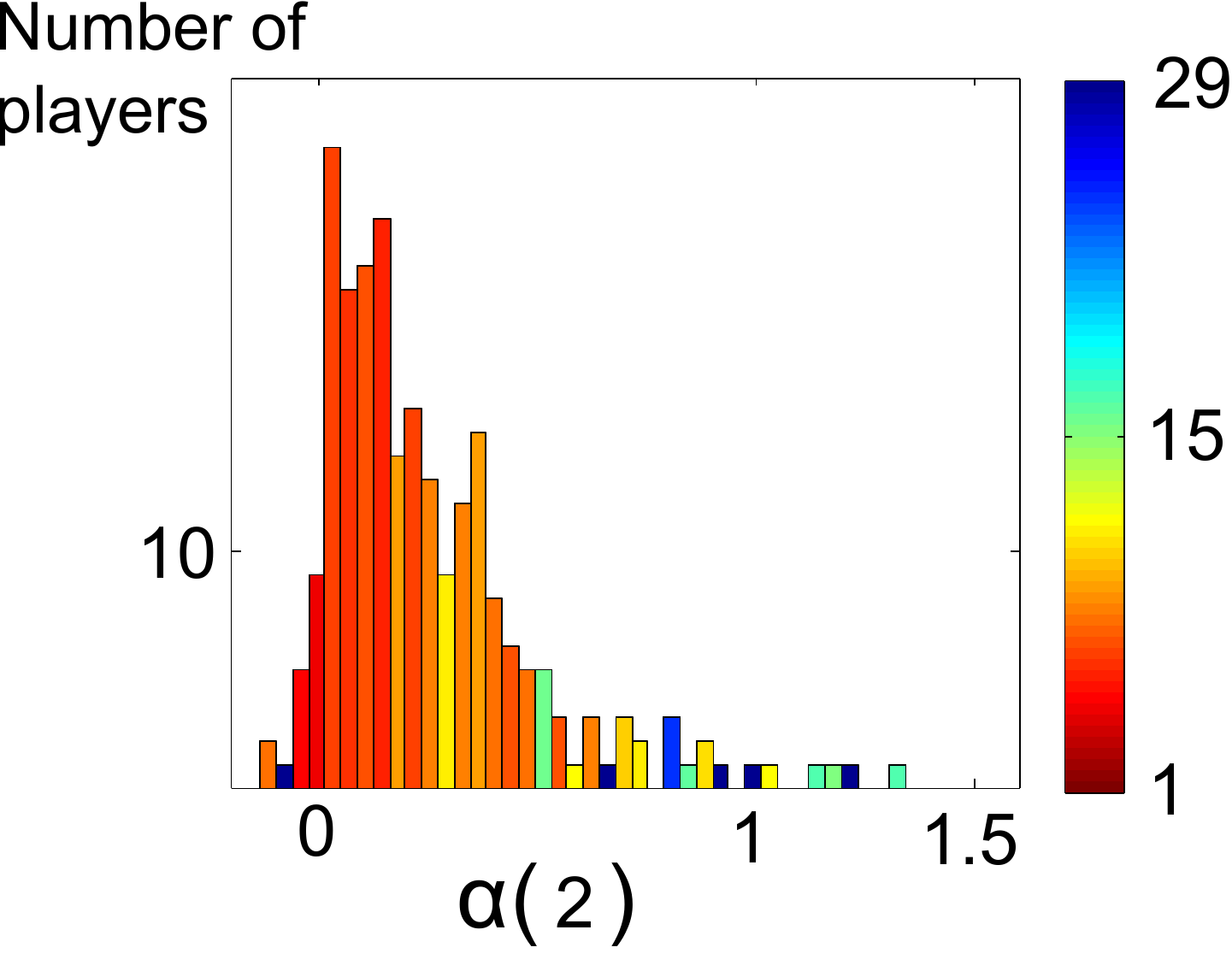}\\
\textbf{(B) Counting}\\
\vspace{0.1cm}
\includegraphics[clip=true,scale=0.3]{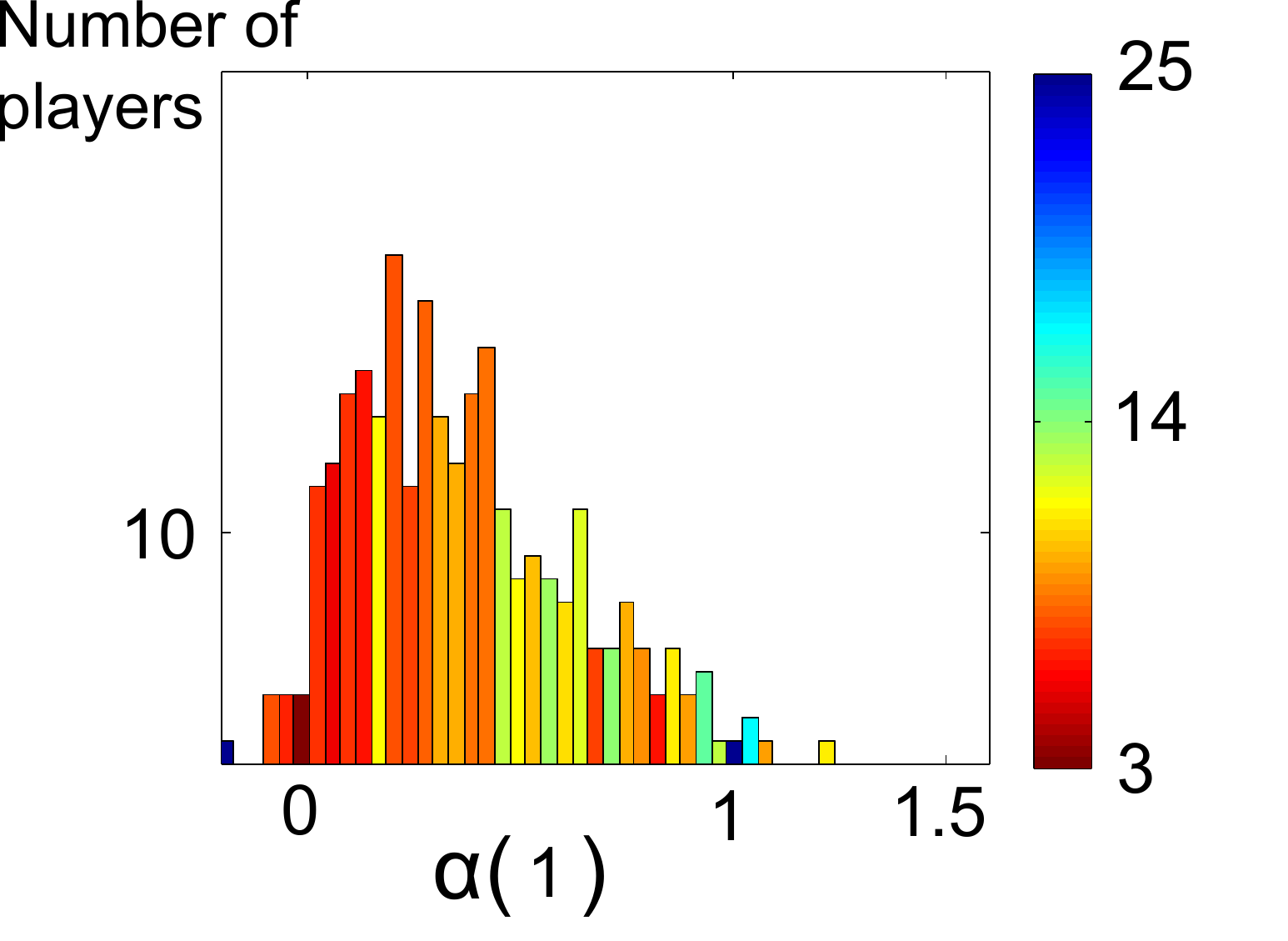}
\includegraphics[clip=true,scale=0.3]{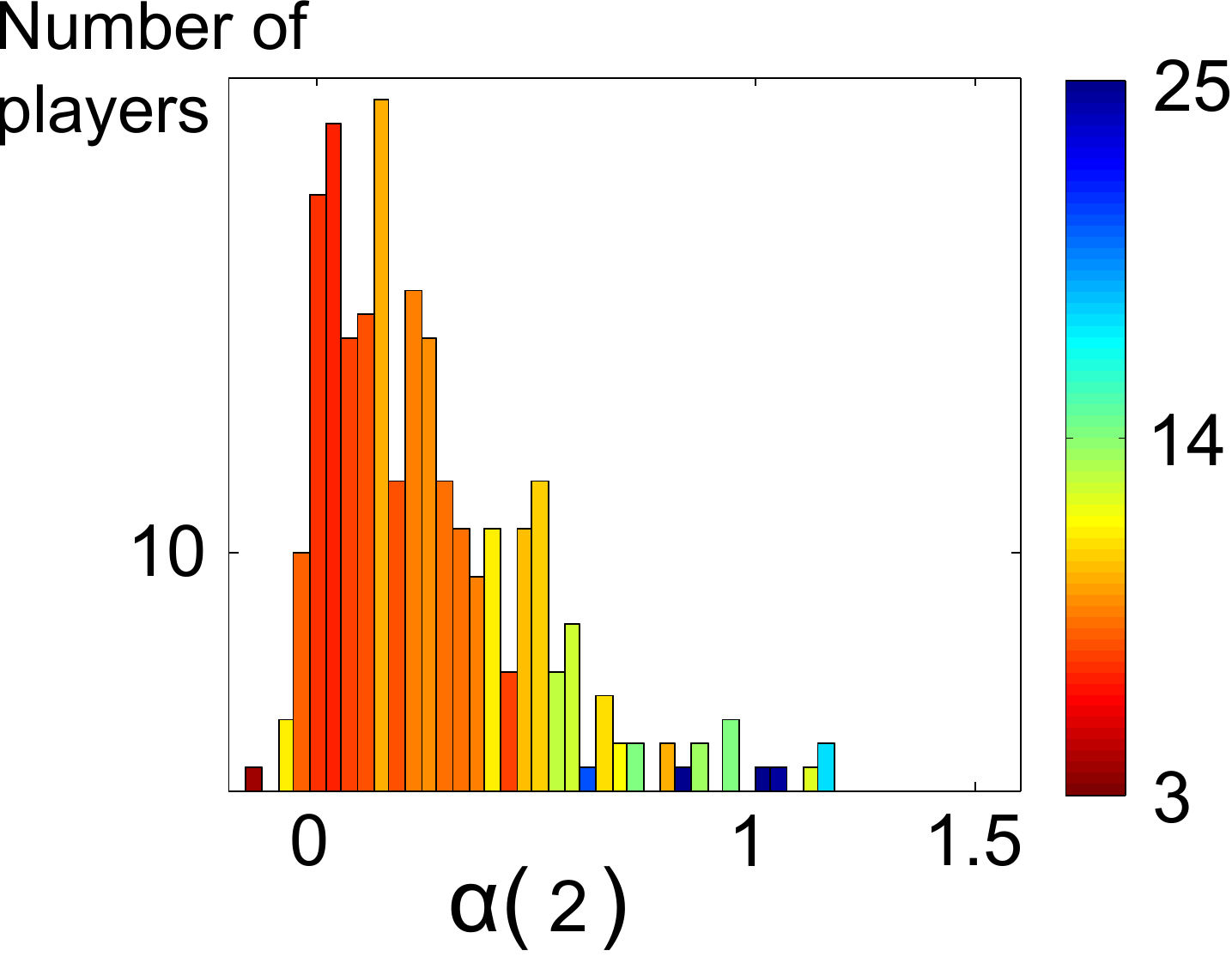}\\
\textbf{(C)}\\
\includegraphics[clip=true,scale=0.3]{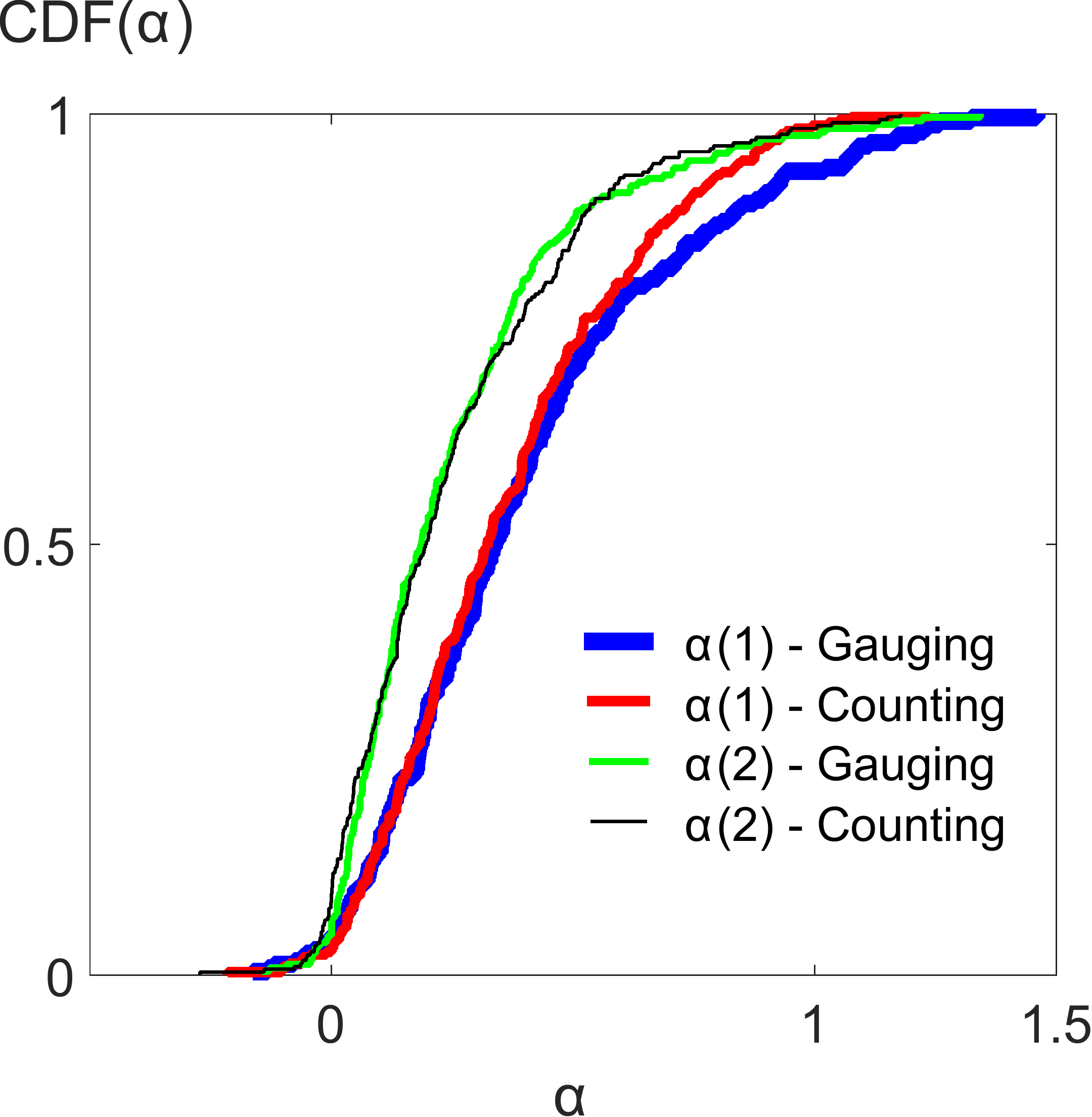}
\caption{\textbf{Influenceability over rounds and games.} (A) Gauging game, (B) Counting game : distributions of the $\alpha_i(1)$ influenceability after the first round and $\alpha_i(2)$ influenceability after the second round for the time-varying influenceability model~\eqref{eq:consensus-alpha-to-mean}. The colormap corresponds to the average prediction $RMSE$ of participants in each bin. For visualization purposes, one value at $\alpha_i(1) = -1.5$ has been removed from the histogram in the gauging game\samreview{, round $1$. (There was only one individual with $\alpha_i(1) = -1.5$. For this particular individual, the linear relationship between $x_i(2) - x_i(1)$ and $\bar{x}(1) - x_i(1)$ is not significant (p-val $= 0.38$), so that the coefficient $\alpha_i(1) = -1.5$ should not be interpreted. For the rest of the participants, the level of trust can be read from the color scale. The color of a bar corresponds to the average prediction error made for the participants with $\alpha_i$ values falling within the bar range.)} (C) Cumulated distributions of $\alpha_i(1)$ and $\alpha_i(2)$ for each type of games.
}\label{fig:alpha_1_IIr}
\end{figure}

The distribution of couples 
$(\alpha_i(1),\alpha_i(2))$
 were obtained by fitting model~\eqref{eq:consensus-alpha-to-mean} to the whole dataset via mean square minimization for each type of games independently.
The marginal distributions of 
$(\alpha_i(1),\alpha_i(2))$
are shown in Fig.~\ref{fig:alpha_1_IIr}.
Most values fall within interval $[0,1]$, meaning that the next judgment falls between one's initial judgment and the group judgment mean. Such a positive influenceability has been shown to improve judgment accuracy~\cite{yaniv2007using} (see also the \textit{Practical Implications of the Model} section). Most individuals overweight their own opinion compared to the mean opinion to revise their judgment with $\alpha_i(r)<0.5$. This fact is in accordance with the related literature on the subject~\cite{bonaccio2006advice}. 

\samreview{An interesting research direction is to link the influenceability of an individual to their personality. \samrevieww{One way to measure personality is via
the big five factors of personality~\cite{gosling2003very}.}
It turns out that influenceability is found not to be significantly correlated to the big five factors of personality, to education level and to gender. These negative findings are reported in the \textit{Influenceability and personality} supplementary section.}

\cocoreview{The \samrevieww{plots} in Fig.~\ref{fig:alpha_1_IIr} also display a small fraction of negative influenceabilities. One interpretation would be that the concerned participants recorded opinions very close to the average group opinion for multiple games. When this happens at one round, the opinion at the following round has a high probability to move away from the group average. This contributes to negative influenceabilities in the participants' influenceabilities during the fitting process.}

\samrevieww{Is the prediction error homogeneous for all influenceability values ? To see this, each bin of the influenceability distributions in Fig.~\ref{fig:alpha_1_IIr} is colored to reflect average prediction error.
The error is given in terms of root mean square error (\textit{RMSE}). The color corresponds to the prediction error regarding participants having their influenceability falling within the bin.}
\samreview{A detailed definition of RMSE is provided in paragraph \textit{Validation procedure} in \textit{Material and Methods} section}. This information shows that \samrevieww{the model makes the best predictions for participants with a small but non-negative influenceability. On the contrary, predictions for participants with a high influenceability are less accurate.}

The distribution of influenceabilities of the population is subject to evolution over time. A two-sample Kolmogorov-Smirnov test rejects equality of distributions between $\alpha_i(1)$ and $\alpha_i(2)$ 
for both types of games ($p$-val $<10^{-6}$\samreview{, with KS distance of $0.25$ and $0.23$ for the \gauginggame~and \countinggame, respectively}). 
A contraction toward $0$ occurs with a median going from $0.34$ for $\alpha_i(1)$ to $0.18$ for $\alpha_i(2)$ in the \countinggame~and $0.32$ to $0.20$ in the \gauginggame~\samreview{(p-val$<10^{-15}$ for a paired one-sided Wilcoxon sign-rank test)}. In other words, the participants continue to be influenced after round $2$ but this influence is lightened.
Fig.~\ref{fig:alpha_1_IIr}-C shows the discrepancy between the cumulated distribution functions over rounds.

\subsection*{Model Performance}

\paragraph*{Prediction scenarios} \label{sec:prediction-scenarios}

When one wishes to predict how a participant revises their opinion in a decision making process, the level of prediction accuracy will highly depend on data availability. More prior knowledge on the participant should improve the predictions. When little prior information is available about the participant, the influenceability derived from it will be unreliable and may lead to poor predictions. In this case, it may be more efficient to proceed to a classification procedure provided that data from other participants are available. These approaches are tested by computing the prediction accuracy in several situations reflecting data availability scenarios.

In the worst case scenario, no data is available on the participant and the judgment revision mechanism is assumed to be unknown. In this case, predicting constant opinions over time is the only option. This corresponds to the null model against which the consensus model~\eqref{eq:consensus-alpha-to-mean} is compared.

In a second scenario, prior data from the same participant is available. The consensus model can then be fitted to the data (individual influenceability method). 
\corentin{Data ranging from 1 to 15 prior instances of the judgment process are respectively used to learn how the participant revises their opinion.
Predictions are assessed in each of these cases to test 
\samgreen{how the predictions are impacted by the amount of prior data available.}}

\corentin{In a final scenario, besides having access to prior data from the participant, it is assumed that a large body of participants took part in a comparable judgment making process. 
\samgreen{These additional data are expected to reveal the most common behaviours in the population and enable}
to derive typical influenceabilities by classification tools (population influenceability methods).}
\samrevieww{For the population influenceability methods, there are two possibilities. First, assume that prior information on the participant is available. In this case, the influenceability class of the participant is determined using this information.
In the alternative case, no prior data is available on the targeted participant. It is then impossible to discriminate which influenceability class they belong to. Instead, the most typical influenceability is computed for the entire population and the participant is predicted to follow this most typical behaviour.}


\paragraph*{Prediction accuracy}\label{sec:prediction-performance}

\begin{figure}[!ht]
\centering
\textbf{(A) Gauging}\\
\includegraphics[scale=0.3]{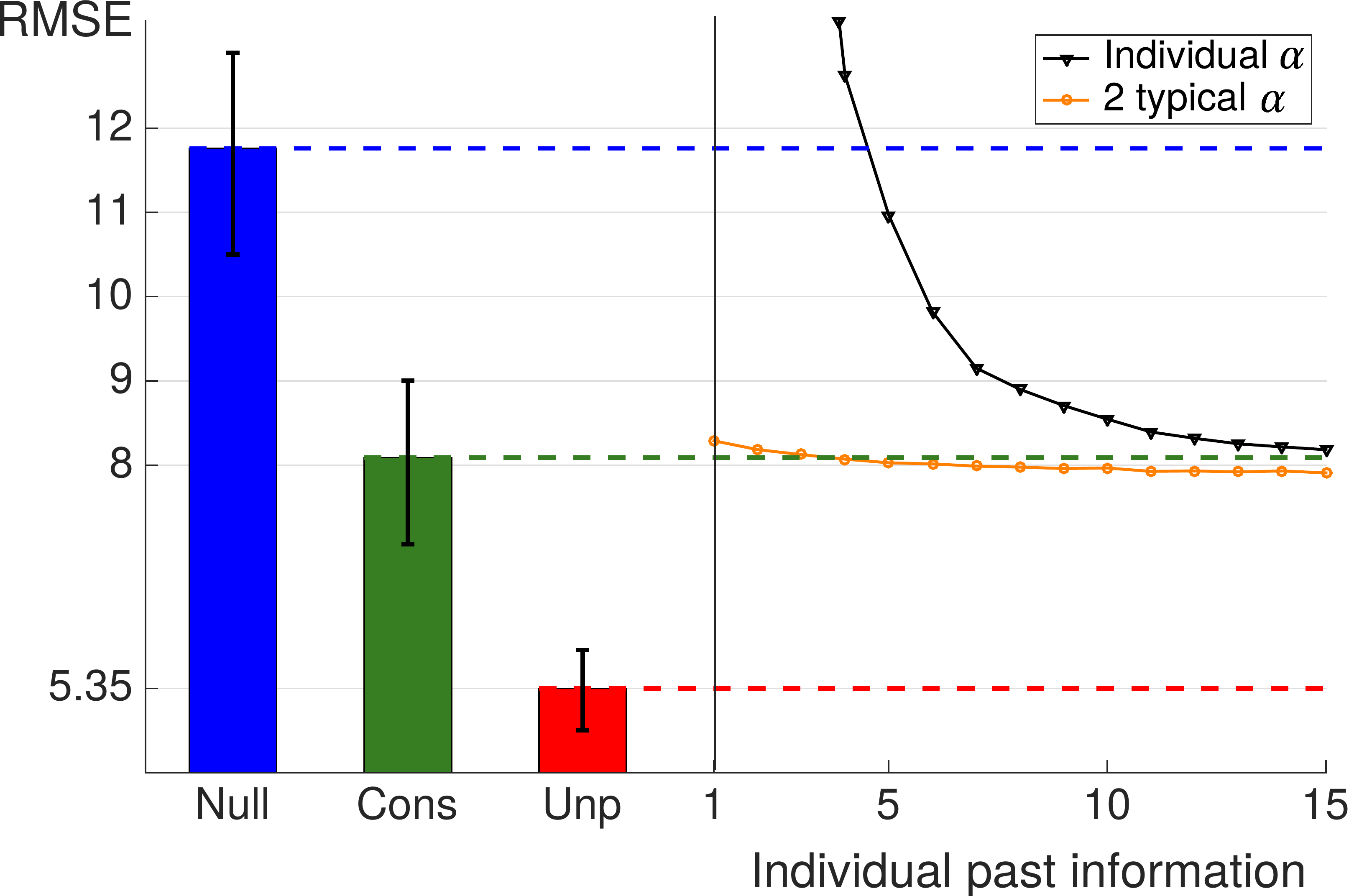}\\
\vspace{0.5cm}
\textbf{(B) Counting}\\
\includegraphics[scale=0.3]{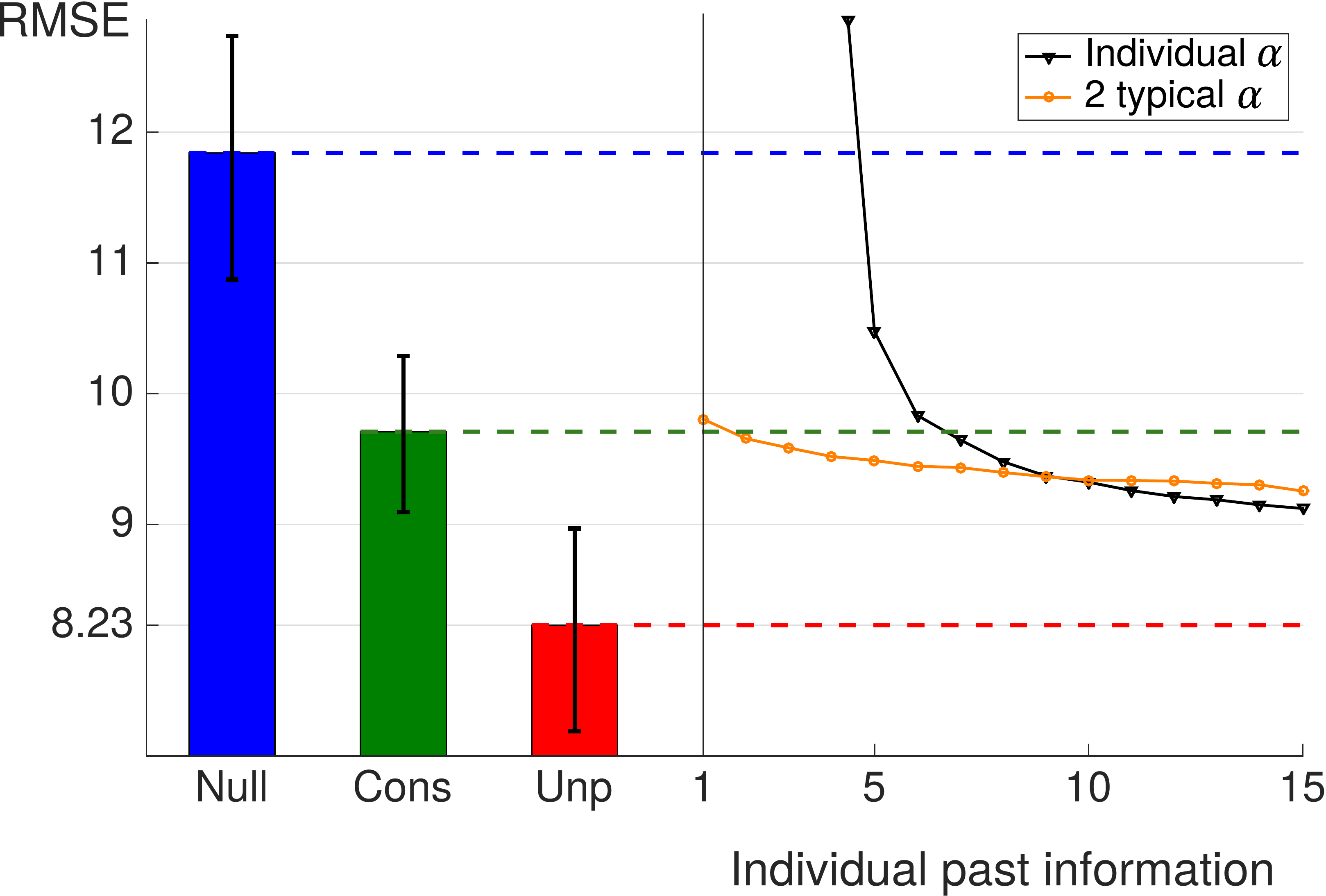}\\
\vspace{0.5cm}

\caption{\cocoreview{\textbf{Root mean square error (\textit{RMSE}) of the predictions for the final round.} The RMSEs are obtained from crossvalidation.
(A) \gauginggame, (B) \countinggame. In (B), the $RMSE$ has been scaled by a factor of $5$ to be comparable to the (A) plot.
The bar chart displays crossvalidation errors for models that does not depend on the training size.
 Top {blue} horizontal line 
corresponds to the null model of constant opinion. 
The {middle} horizontal {green} line 
corresponds to fitting using the same typical couple of influenceability for the whole population. {The bottom horizontal red} line 
corresponds to the prediction error due to intrinsic variations in judgment revision. 
The decreasing {black curve} (triangle dots) corresponds to fitting with the individual influenceability method. The slightly decreasing {orange} curve ({round} dots) corresponds to fitting choosing among $2$ typical couples of influenceability. All $RMSE$ were obtained on validation games. \samrevieww{The error bars provide $95\%$ confidence intervals for the RMSEs.} \vspace{-0.3cm}}}\label{fig:rmse}
\label{fig:compare-all-models}
\end{figure}

We assess the predictions when the number of training data is reduced, accounting for realistic settings where prior knowledge on individuals is scarce. Individual parameter estimation via the \DI~method is compared to \DIc~method. This last method uses one or two $(\alpha(1),\alpha(2))$ couples of values derived on independent experiments. Fig.~\ref{fig:compare-all-models} presents the $RMSE$ 
\corentin{(normalized to the range 0-100)}
obtained on validation sets 
\corentinc{for final round predictions}
using the model~\eqref{eq:consensus-alpha-to-mean} with parameters fitted on training data of varying size. 
\samgreen{The methodology was also assessed for second round predictions instead of final round predictions. The results also hold in this alternative case, as described in \textit{Second round predictions} section in \textit{Material and Methods}.}

The \DI~and \DIc~methods are compared to a null model assuming constant opinion with no influence, \ie $\alpha(r) = 0$ (\samreview{\textit{Null} in Fig.~\ref{fig:compare-all-models}}). The null model does not depend on training set size.
By contrast, the \DI~method which, for each individual, fits parameters $\alpha_i(1)$ and $\alpha_i(2)$ based on training data, is sensitive to training set size \samreview{(\textit{Individual $\alpha$} in Fig.~\ref{fig:compare-all-models})} : it performs better than the null model when the number of games used for training is higher than $5$ in both types of games but its predictions become poorer otherwise, due to overfitting. 

Overfitting is alleviated using the \DIc~methods which restrict the choice of $\alpha(1)$ and $\alpha(2)$, making it robust to training size variations.
The population method which uses only one typical couple of influenceability as predictor presents one important advantage. It provides a method which does not require any prior knowledge about the participant targeted for prediction. \samreview{It is thus insensitive to training set size (\textit{Cons} in Fig.~\ref{fig:compare-all-models}).} This method improves by $31\%$ and $18\%$ the prediction error for the two types of games compared to the null model of constant opinion.

\begin{figure}[!ht]
\centering
\begin{tabular}{cc}
\textbf{(A) Gauging} & \textbf{(B) Counting}\\
\includegraphics[scale=0.18]{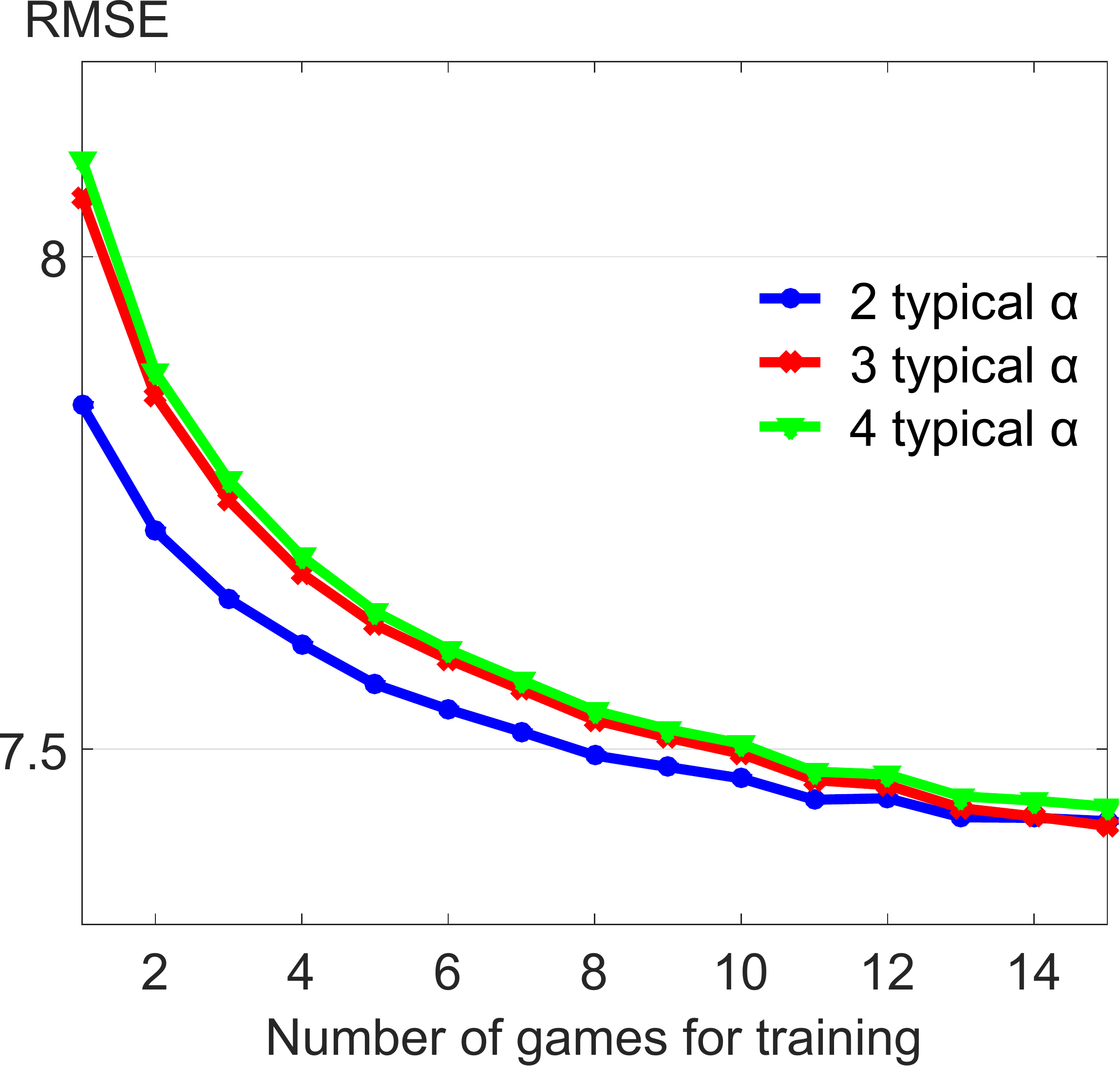} &
\includegraphics[scale=0.18]{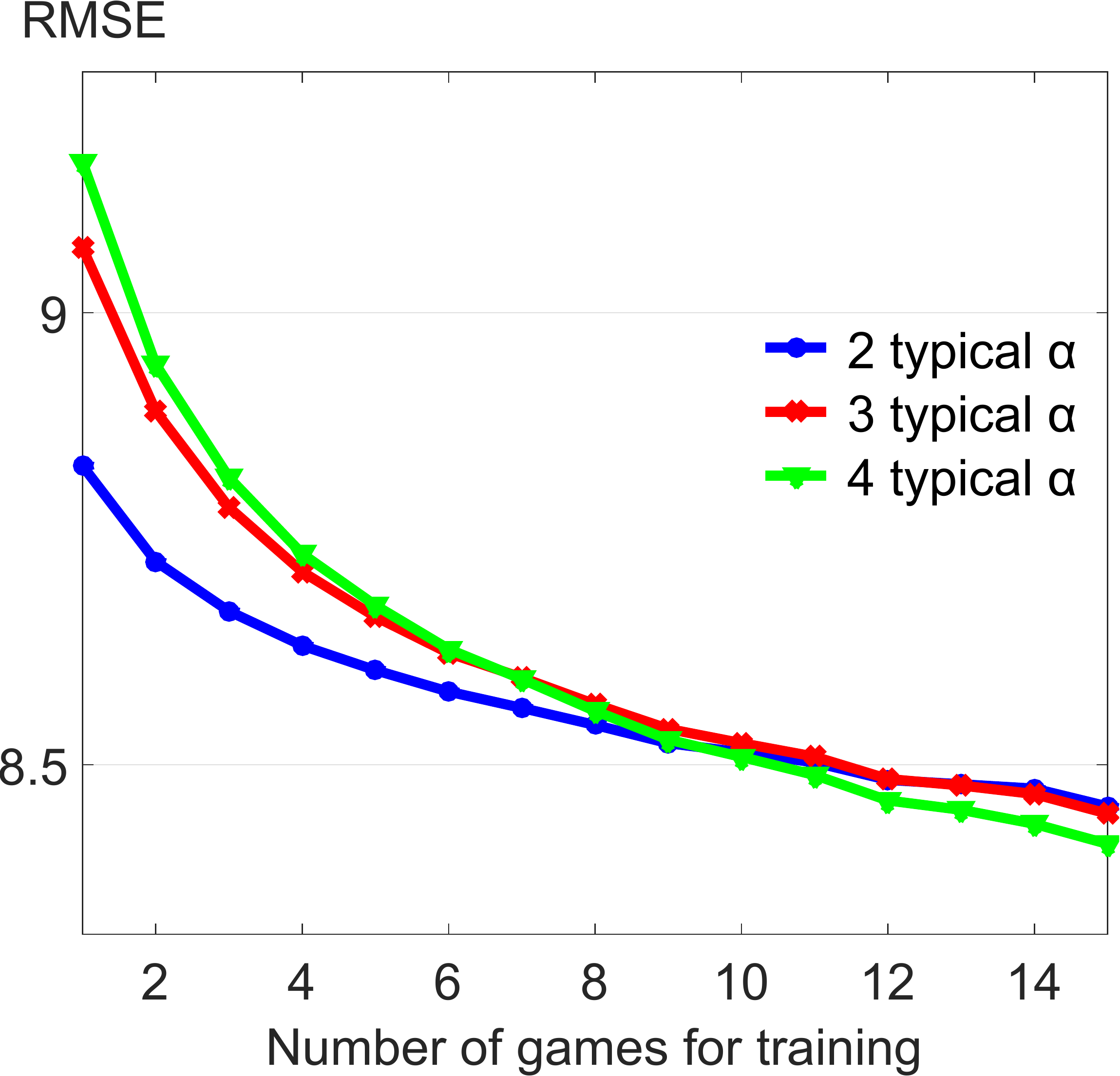}
\end{tabular}
\caption{\textbf{Predictive power of the consensus model when the number of typical couples of influenceability used for model calibration varies.} Root mean square error (\textit{RMSE}) plotted for training set size from $1$ to $15$ games. (A) \gauginggame, (B) \countinggame. In (B), \textit{RMSE} has been scaled by a factor of $5$ to be comparable to the (A) plot.\vspace{-0.3cm}}
\label{fig:II_class}
\end{figure}

The population methods based on two or more typical couples of influenceability require to possess at least one previous game by the participant to calibrate the model \samreview{(\textit{2 typical $\alpha$} in Fig.~\ref{fig:compare-all-models})}. These methods are more powerful than the former if enough data is available regarding the participant's past behaviour ($2$ or $3$ previous games depending on the type of games). The number of typical couples of influenceabilities to use depends on the data availability regarding the targeted participant. This is illustrated in Fig.~\ref{fig:II_class}. The modification obtained using more typical influenceabilities for calibration is mild. Moreover, too many typical influenceabilities may lead to poorer predictions due to overfitting. This threshold is reached for $4$ couples of influenceabilities in the \gauginggame~data. As a consequence, it is advisable to restrict the choice to $2$ or $3$ couples of influenceabilities.
This analysis shows that possessing data from previous participants in a similar task is often critical to obtain robust predictions on judgment revision of a new participant.

The results of the control experiments are displayed 
\cocoreview{by a red dashed line}
in Fig.~\ref{fig:compare-all-models}-A,B.
This \cocoreview{bottom line}
corresponds to the amount of prediction error which is due to the intrinsic unpredictability of judgment revision. No model can make better predictions than this threshold (see \textit{Control experiment} section in \textit{Material and Methods}).

The \gauginggame~obtains an unpredictable \textit{RMSE} of
 \corentinc{$5.35$}
  while the \countinggame~obtains 
  \corentinc{$8.23$}.
\corentin{By contrast, the average square variation of the judgments between first and final rounds are respectively $11.76$ and $11.84$ for both types of games (corresponding to the \textit{RMSE} of the null model).}
\corentin{Taking the intrinsic unpredictable variation thresholds as a reference}, the relative prediction \textit{RMSE} is more than halved when using the time varying influenceability model~\eqref{eq:consensus-alpha-to-mean} \samlast{with one couple of typical influenceabilities} instead of the null model with constant opinion. 
\samgreen{In other words, more than two thirds of the prediction error made by the consensus model is due to the intrinsic unpredictability of the decision revision process.}

\samrevieww{The error bars in Fig.~\ref{fig:compare-all-models} provide $95\%$ confidence intervals for the RMSEs. They confirms statistical significance of the difference between RMSEs. For clarity, the error bars are provided only for regression methods which do not depend on training set size. For completeness, supplementary Fig.~\ref{fig:errbar} provides error bars for all models.
}

\samrevieww{RMSEs were used in this study since it corresponds to the quantity being minimized when computing the influenceability parameter $\alpha$. Alternatively, reporting the \textit{Mean Absolute Errors} (MAEs) may help the reader to obtain more intuition on the level of prediction error. For this reason, MAEs are provided in supplementary Fig.~\ref{fig:mae}.
}

\subsection*{Practical Implications of the Model}\label{sec:implications-of-model}

\paragraph*{Do groups reach consensus ?}

\begin{figure}[!ht]
\centering
\includegraphics[scale=0.50, trim=1cm 0cm 0cm 0cm]{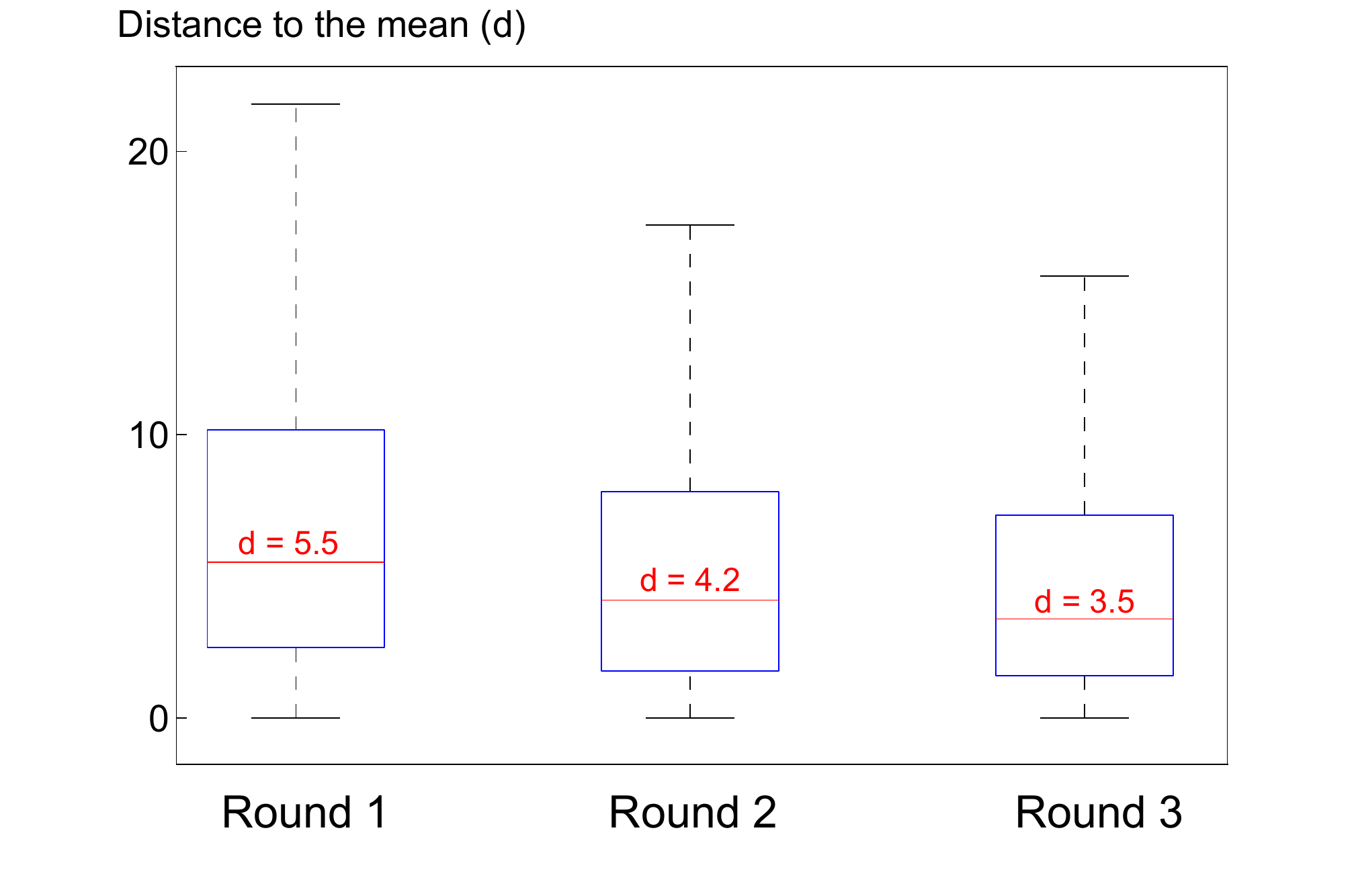}\\
\caption{\textbf{Distance of participants' judgment to mean judgment for rounds $1$, $2$ and $3$ in the \gauginggame.} Is displayed only the data coming from games where for all $3$ rounds at least 5 out of 6 participants provided a judgment. This ensures that the judgment mean and standard deviation can be compared over rounds.}\label{fig:evol_distance_to_mean_over_rounds}
\end{figure}

Because of social influence, groups tend to reduce their disagreement. However, this does not necessarily implies that groups reach consensus. To test how much disagreement remains after the social process, the distance between individual judgments and mean judgments in corresponding groups is computed at each rounds. The results are presented for the \gauginggame. The same conclusions also hold for the \countinggame. Fig.~\ref{fig:evol_distance_to_mean_over_rounds} presents the statistics summary of these distances. The median distances are respectively $5.5$, $4.2$ and $3.5$ for the three successive rounds, leading to a median distance reduction of $24\%$ from round $1$ to $2$ and $16\%$ from round $2$ to $3$. 
\samreview{In other words, the contraction of opinion diversity is less important between rounds $2$ and $3$ than between rounds $1$ and $2$ and more than $50\%$ of the initial opinion diversity is preserved at round $3$. This is in accordance to the influenceability decay observed in \textit{Influenceability of participants} section.}

\samreview{
If one goes a step further and assumes that the contraction continues to lessen at the same rate over rounds, it may be that groups will never reach consensus. This phenomenon is quite remarkable since it would explain the absence of consensus without requiring non-linearities in social influence. An experiment involving an important number of rounds would shed light on this question and is left for future work.
}

\paragraph*{Influenceability and individual performance}
Each game is characterized by a true value, corresponding to an exact proportion to guess (for \gauginggames) or an exact amount of items displayed to the participants (for \countinggames). 
Whether social influence promotes or undermines individual performance 
can be measured for the two tasks. 
\samreview{
Individual performance can also be compared to the performance of the mean opinions in each group of $6$ participants.}

\begin{figure}[!ht]
\begin{center}
\begin{tabular}{cc}
\textbf{(A) Gauging} & \textbf{(B) Counting}\\
\vspace{0.1cm}
\includegraphics[clip=true,trim=1cm 0cm 0.7cm 0cm,scale=0.51]{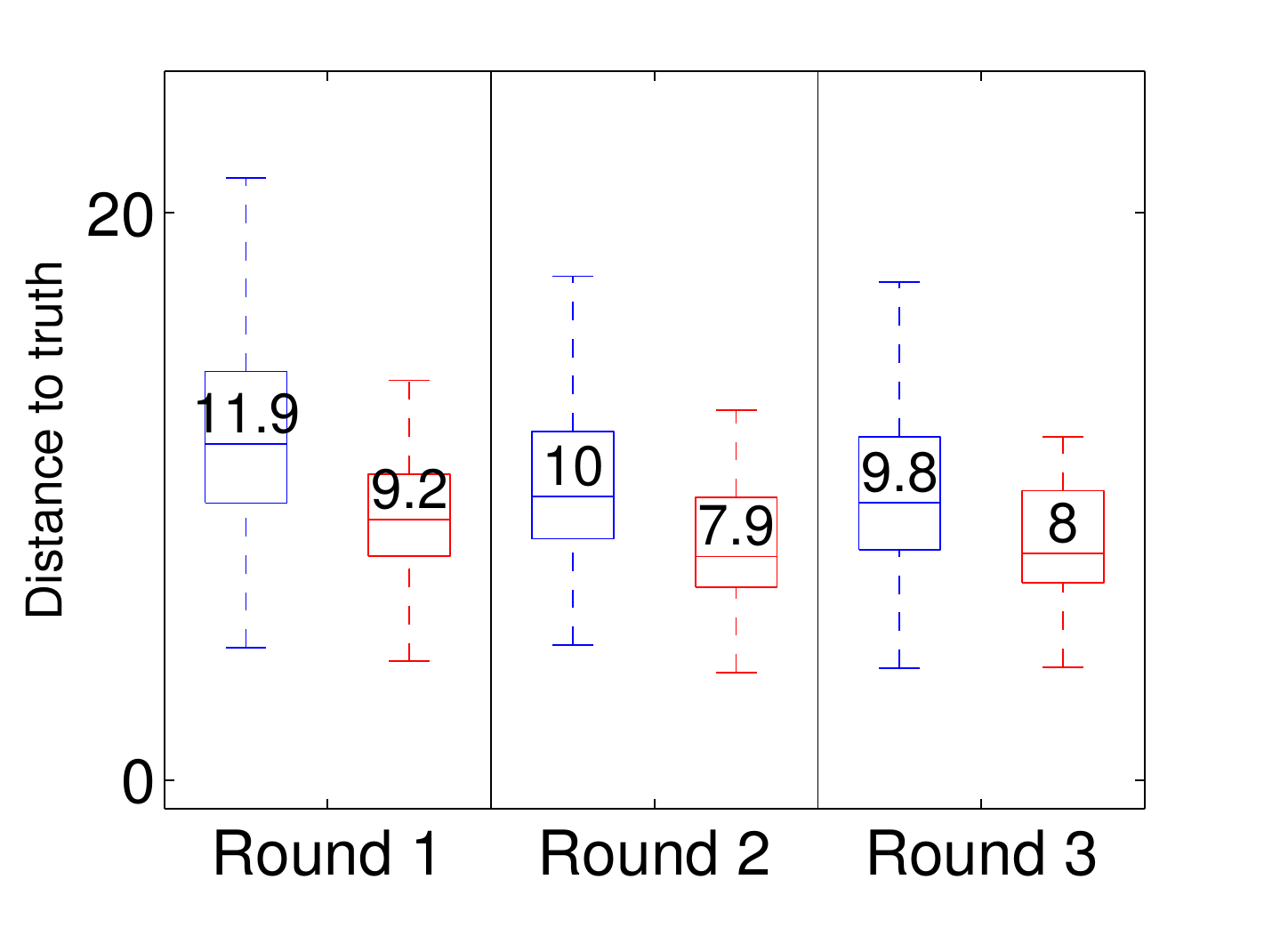} &
\includegraphics[clip=true,trim=1cm 0cm 0.7cm 0cm,scale=0.51]{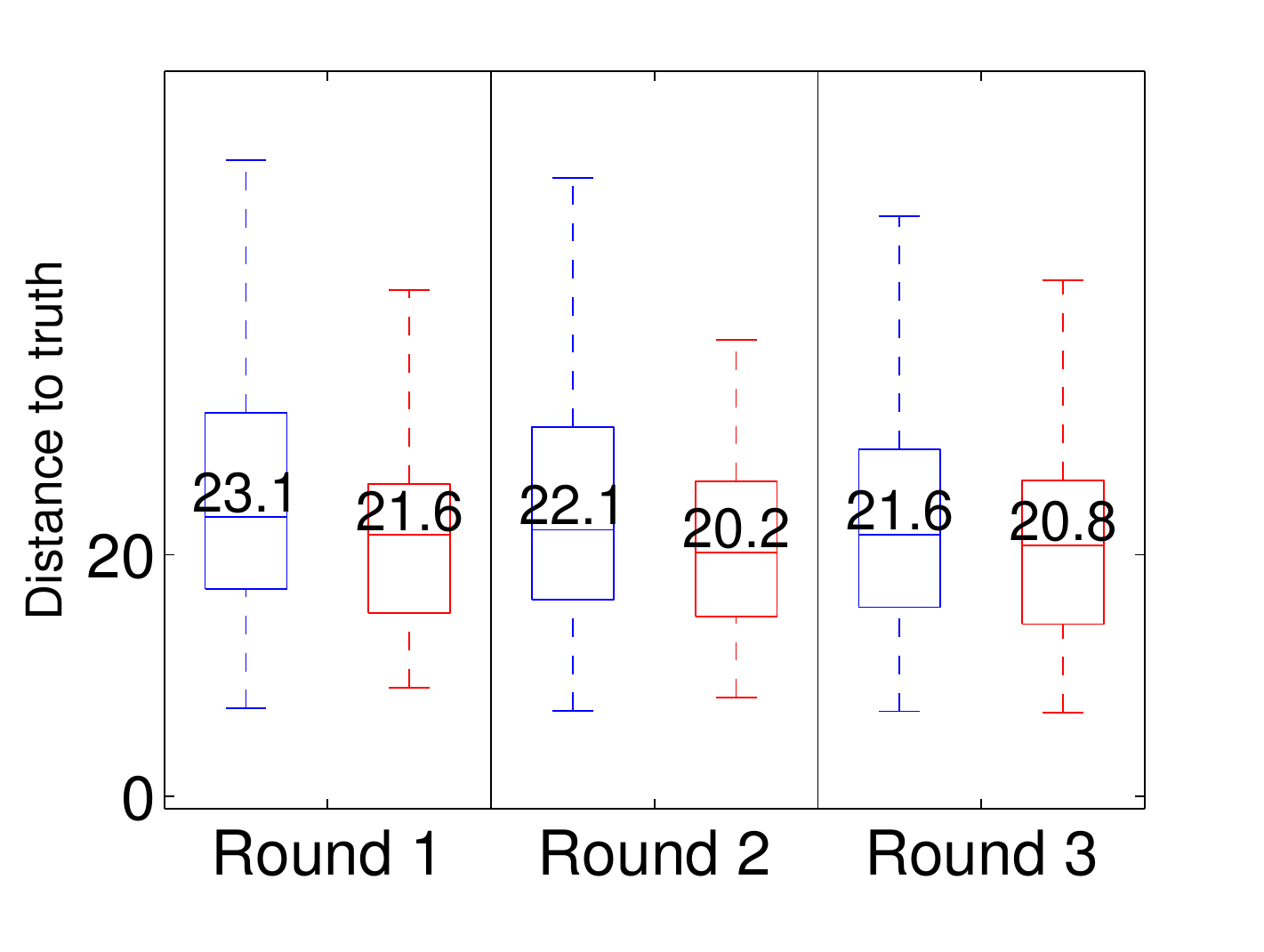}
\end{tabular}
\end{center}
\caption{\samreview{(A) \gauginggame ; (B) \countinggame. From left to right : (Boxplot 1,3 and 5, in blue) Root mean square distance from individual opinions to truth ($E_i(r)$); 
(Boxplot 2,4 and 6 in red) Root mean square distance from mean opinion to truth. Values shown for the \countinggame~have been scaled by a factor of $5$.}}\label{fig:RMSE_Xpers_distance_to_truth}
\end{figure}

At each round $r$, a participant's success is characterized by the root mean square distance to truth, denoted $E_i(r)$. The error $E_i(r)$ depicts how far a participant is to truth. Errors are normalized to fit in the range $0-100$ in both types of tasks so as to be comparable. A \samreview{global measure of individual errors} is defined as the median over participants of $E_i(r)$, and the success variation between two rounds is given by the median value of the differences $E_i(r) - E_i(r+1)$. A positive or negative success variation corresponds respectively to a success improvement or decline of the participants after social interaction. \samreview{The errors are displayed in Fig.~\ref{fig:RMSE_Xpers_distance_to_truth}.
The results are first reported for the \gauginggame.
The median error $E_i(r)$ for rounds $1$, $2$ and $3$ are respectively 
$11.9$, $10.0$ and $9.8$ (Fig.~\ref{fig:RMSE_Xpers_distance_to_truth}-(A)).
It reveals an improvement with a success variation of $1.9$ and $0.2$ for $E_i(1)-E_i(2)$ and $E_i(2)-E_i(3)$ respectively (p-values$<10E-5$, sign test), showing that most of the improvement is made between first and second round.
Regarding the \countinggame, the median error $E_i(r)$ for rounds $1$, $2$ and $3$ are respectively
$23.1$, $22.1$ and $21.6$ (Fig.~\ref{fig:RMSE_Xpers_distance_to_truth}-(B)). Note that the errors for the \countinggame~have been rescaled by a factor of $5$ to fit in the range $0-100$. This corresponds to an improvement with a success variation of $1.0$ and $0.5$ for $E_i(1)-E_i(2)$ and $E_i(2)-E_i(3)$ respectively \samrevieww{(the significance of the improvement is confirmed by a sign-test with p-values $<10E-5$).} Fig.~\ref{fig:RMSE_Xpers_distance_to_truth} also reports the aggregate performance in terms of the root mean square distance from mean opinions to truth in each group of $6$ participants. Unlike individual performance, the median aggregate performance does not consistently improve. Regarding the \gauginggame, the median aggregate error is significantly higher in round $1$ than in rounds $2$ and $3$ (p-val$<10E-4$, sign test) and this difference is not significant between rounds $2$ and $3$ (p-val$>0.05$). Regarding the \countinggame, no significant difference is found among the $3$ rounds for the median aggregate error (p-val$>0.05$). As a consequence, social influence consistently helps the individual performance but does not 
consistently promote the aggregate performance.
}

\samreview{
The reason why social influence helps the individual performance is a combination of two factors. First, at round $1$, the mean opinion is closer to the truth than the individual opinion is (p-val$<0.01$, Mann–Whitney–Wilcoxon test). Second, in accordance to the consensus model~\eqref{eq:consensus-alpha-to-mean}, individuals \samrevieww{move closer} to the mean opinion over subsequent rounds.
The fact that initially the mean opinion is closer to truth than the individual opinions corresponds to the wisdom of the crowd effect.
The wisdom of the crowd is a statistical effect stating that averaging over several \textit{independent} judgments yields a more accurate evaluation than most of the individual judgments would (see the early ox experiment by Galton in 1907~\cite{galton1907vox} or more recent work~\cite{ariely2000effects}). Since the aggregate performance does not consistently improve over rounds, it can be said that social influence does not consistently promote the wisdom of the crowd. Lorenz et al.~\cite{lorenz2011social} say that social influence \textit{undermines} the wisdom of the crowd because it ``reduces the diversity of the group without improving its accuracy''. This variance reduction is also observed in the present study and corroborates the consensus model~\eqref{eq:consensus-alpha-to-mean}.
Interestingly, in the first round, the wisdom of the crowd effect is more prominent in the \gauginggame~than in the \countinggame : the median individual error is $37\%$ higher than the median error of the mean opinion in the \gauginggame~while it is only $8\%$ higher in the \countinggame. The reason for this difference is studied in details in supplementary section~\ref{SI-sec:wisdom}.
}


\samreview{The fact that the mean opinion is more accurate than individual opinions leads to posit that participants using the mean opinion to form their own opinion, \ie those with higher influenceability $\alpha_i$, will increase their performance.} We examine relationships between success variation and the model parameters $\alpha_i$ by computing partial Pearson correlations $\rho$ controlling for the effect of the rest of the variables.
Only significant Pearson correlations are mentioned (p-val $<0.05$). All corresponding p-values happen to be smaller than $0.001$ \samreview{except for one as explicitly mentioned. Pearson correlations are given by pairs : the first value corresponds to the \gauginggame~while the second to the \countinggame.}
Influenceability $\alpha_i(1)$ between round $1$ and $2$ and improvement are positively related with $\rho(\alpha_i(1),E_i(1) - E_i(2)) = 0.41 / 0.22$.
\samreview{This is in accordance to the posited hypothesis. The \textit{wisdom of the crowd effect} found at round $1$ implies that 
participants who improve more from round $1$ to $2$ are those who give more weight to the average judgment. Since the wisdom of the crowd effect is more prominent in the \gauginggame~than in the \countinggame, it is consistent that the correlation is higher in the former than in the latter.}
A similar effect relates success improvement and the influenceability $\alpha_i(2)$ between round $2$ and $3$ with $\rho(\alpha_i(2) \,,\, E_i(2) - E_i(3)) = 0.36 / 0.32$.
As may be expected, higher initial success leaves less room for improvement in subsequent rounds, which explains that $\rho(E_i(1) , E_i(1) - E_i(2)) = 0.68 / 0.21$ and $\rho(E_i(1) , E_i(2) - E_i(3)) = 0.25 / 0.16$ (where $0.16$ is significant for p-val $<0.01$). This also means that initially better participants are not better than average at using external judgments.

\paragraph*{Modelling influenceability across different types of games}

The assessment of the predictive power of model~\eqref{eq:consensus-alpha-to-mean} on both types of games provides a generalisability test of the prediction method.
The two types of games vary in difficulty. The root mean square relative distance 
\corentin{$E_i(1)$}
between a participant's first round judgment and truth is taken as the measure of inaccuracy for each participant. The median inaccuracy for the \countinggame~is 
\corentind{$23.1$ while it is $11.8$} 
 for the \gauginggame~(Mood's median test supports the rejection of equal median, $p=0$). 
 Moreover,  a Q-Q plot shows that inaccuracy is more dispersed for the \countinggame, 
 \corentin{suggesting that estimating quantities is more difficult than gauging proportion of colors.}

The accuracy of model~\eqref{eq:consensus-alpha-to-mean} is compared for the two datasets in Fig.~\ref{fig:compare-all-models}. Interestingly, the model prediction ranks remains largely unchanged for the two types of games. As depicted in Fig.~\ref{fig:alpha_1_IIr}-C, influenceability distributions do not vary significantly between the two games. A two-sample Kolmogorov-Smirnov test fails to reject the equality of distribution null hypothesis of equal median with $p>0.65$ \samreview{and a KS distance of $0.06$} for both $\alpha_i(1)$ and $\alpha_i(2)$. This means that although the participants have an increased difficulty when facing the \countinggame, they do not significantly modify how much they take judgments from others into account. 
Additionally, the relationships between participants' success and influenceability are preserved for both types of games. The preserved tendencies corroborate the overall resemblance of behaviours across the two types of games. These similarities indicate that the model can be applied to various types of games with different level of difficulty.

\section*{\cocoreview{Conclusions}}

The way online social systems are designed has an important effect on judgment outcome~\cite{muchnik2013social}. Operating or acting on these online social systems provides a way to significantly impact our markets, politics~\cite{bond2012nature} and health. 
Understanding the social mechanisms underlying opinion revision is critical to plan successful interventions in social networks. It will help to promote the adoption of innovative behaviours (e.g., quit smocking~\cite{christakis2008collective}, eat healthy)~\cite{valente2012network}.
The design and validation of models of opinion revision will enable to create a bridge between system engineering and network science~\cite{liu2011controllability}.

The present work shows that it is possible to model opinion evolution in the context of social influence in a predictive way. When the data regarding a new participant is available, parameters best representing their influenceability are derived using mean-square minimization. When the data is scarce, the data from previous participants is used to predict how the new participant will revise their judgments. To validate our method, results were compared for two types of games varying in difficulty. The model performs similarly in the two experiments, indicating that our influenceability model can be applied to other situations.

The decaying influenceability model after being fit to the data 
\corentin{suggests}
that despite opinion settlement, consensus will not be reached within groups and disagreement will remain. This suggests that there needs to be incentives for a group to reach a consensus. The analysis also reveals that participants who improve more are those with highest influenceability, this independently of their initial success.

The degree to which one may successfully intervene on a social system is directly linked to the degree of predictability of opinion revision. Because there must always be factors which fall out of the researcher's reach (changing mood or motivations of participants), part of the process cannot be predicted. The present study provides way to assess the level of unpredictability of an opinion revision mechanism. This assessment is based on a control experiment with hidden replicated tasks.

The proposed experiment type and validation method can in principle be generalized to any sort of continuous judgment revision.
The consensus model can also serve as a building block to more complex models when collective judgments rely on additional information exchange.


\section*{Material and Methods}

\subsection*{Experiment}

Our research is based on an experimental website that we built, which received participants from a crowdsourcing platform.
When a participant took part in an experiment, they joined a group of $6$ participants. Their task was to successively play $30$ games of the same sort related to $30$ distinct pictures. 

\paragraph*{Criteria for online judgment revision game}

The games were designed to reveal how opinions evolve as a result of online social influence. Suitable games have to satisfy several constraints. First, to finely quantify influence, the games ought to allow the evolution of opinion to be gradual. Numbers were chosen as the way for participant to communicate their opinion. Multiple choice questions with a list of unordered items (e.g., choosing among a list of holiday locations) were discarded.
Along the same lines, the evolution of opinion requires uncertainty and diversity 
\corentinb{of a sufficient magnitude}
in the initial judgments. The games were chosen to be sufficiently difficult to obtain this diversity.
Thirdly, to encourage serious behaviours, the participants were rewarded based on their success in the games. This required the accuracy of a participant to be computable. Games were selected to have an ideal opinion or \textit{truth} which served as a reference. Subjective questions involving for instance political or religious opinions were discarded.

Additionally, the game had to satisfy two other constraints related to the online context where, unlike face-to-face experiments, the researcher cannot control behavioural trustworthiness.
Since the educational and cultural background of participants is a priori unknown, the game had to be \textit{accessible}, i.e., any person which could read English had to be able to understand and complete the game. As a result, the games had to be as simple as possible. For instance, games could not involve high-level mathematical computations. Despite being simple to understand our games were still quite difficult to solve, in accordance with the first constraint. Lastly, to anticipate the temptation to cheat, the solution to the games had to be absent from the Internet. Therefore, questions such as estimating the population of a country were discarded.

\paragraph*{Gauging and counting games}

\corentin{Each game was associated with a picture.}
In the \gauginggame, the pictures were composed of $3$ colors and participants estimated the percentage as a number between $0$ and $100$ of the same given color in the picture. In the \countinggame, the picture was composed of between $200$ and $500$ many small items, so that the participant could not count the items one by one. The participants had then to evaluate the total number of these items as a number between $0$ and $500$.
A game was composed of $3$ rounds. The picture was kept the same for all $3$ rounds.
In each round, the participant had to make a judgment.
During the first round,
each of the $6$ participants provided their judgment, independently of the other participants. During the second round, each participant anonymously received all other judgments from the first round and provided their judgment again. The third round was a repetition of the second one.
Accuracy of all judgments were converted to a monetary bonus to encourage participants to improve their judgment at each round. Screenshots of the games' interface are provided in the \textit{Design of the Experiment} section. 

\paragraph*{Design of the experiment}

\samgreen{The present section describes the experiment interface. 
A freely accessible single player version of the games was also developed to provide a first hand experience of the games. In the single player version, participants are exposed to judgments stored on our database obtained from real participant in previous games. The single player version is freely accessible at \textit{http://collective-intelligence.cran.univ-lorraine.fr/eg/login}. The interface and the timing of the single player version is the same as the version used in the control experiment. The only difference is that the freely accessible version does not involve redundant games and provides accuracy feedback to the participants.}

In the multi-player version which was used for the uncontrolled experiment, 
the participants came from the \textit{CrowdFlower}\textsuperscript{\textregistered} external crowdsourcing platform where they received the URL of the experiment login page along with a keycode to be able to login. The \cocoreview{ad} we posted on CrowdFlower was as follows :
\noindent\blockquote{\textit{
\textbf{Estimation game regarding color features in images} \\
You will be making estimations about features in images. Beware that this game is a 6-player game. If not enough people access the game, you will not be able to start and get rewarded. To start the game : click on <estimation-game> and login using the following information : \\
login : XXXXXXXX \\
password: XXXXXXXX \\
You will receive detailed instruction there. At the end of the game you will receive a reward code which you must enter below in order to get rewarded : \\
<  > \\
}}
The participants were told they will be given another keycode at the end of the experiment which they had to use to get rewarded on the crowdsourcing platform, this forced the participants to finish the experiment if they wanted to obtain a payment. Secondly, the participants arrived on the experiment login page, chose a login name and password so they could come back using the same login name if they wanted to for another experiment (see supplementary Fig.~\ref{fig:login}). Once they had logged in, they were requested to agree on a consent form mentioning the preservation of the anonymity of the data (see the \textit{Consent and privacy} section below for details).
Thirdly, the participants were taken to a questionnaire regarding personality, gender, highest level of education, and whether they were native English speaker or not (all the experiment was written in English). The questions regarding personality come from a piece of work by Gosling and Rentfrow~\cite{gosling2003very} and were used to estimate the five general personality traits.
The questionnaire page is reported in supplementary Fig.~\ref{fig:questionnaire}. Once the questionnaire submitted, the participants have access to the detailed instructions on the judgment process (supplementary supplementary Fig.~\ref{fig:instructions}). After this step, they were taken to a waiting room 
until 6 participants had arrived at this step. At this point, they started the series of 30 games which appeared 3 at a time, with one lone round where they had to make judgments alone and two social rounds where the provided judgment being aware of judgments from others. An instance of the lone round is given in supplementary Fig.~\ref{fig:lone-round}-(A) for the \countinggame~while a social round is shown in supplementary Fig.~\ref{fig:social-round}. Instances of pictures for the \gauginggame~are provided in supplementary Fig.~\ref{fig:lone-round}-(B). In the \gauginggame, the question was replaced by ``\textit{What percentage of the following color do you see in the image ?}''. For this type of games, a sample of the color to be gauged for each pictures was displayed between the question and the $3$ pictures.
At the end of the 30 games, the participants had access to a debrief page where was given the final score and the corresponding bonus. They could also provide a feedback in a text box. They had to provide their email address if they wanted to obtain the bonus (see supplementary Fig.~\ref{fig:debrief}).

\paragraph*{Consent and privacy}

Before starting the experiment, participants had to agree electronically on a consent form mentioning the preservation of the anonymity of the data :

\blockquote{\textit{Hello! Thank you for participating in this experiment. You will be making estimations about features in images. The closer your answers are to the correct answer, the higher reward you will receive. Your answers will be used for research on personality and behaviour in groups. We will keep complete anonymity of participants at all time. If you consent you will first be taken to a questionnaire. Then, you will get to a detailed instruction page you should read over before starting the game. Do you understand and consent to the terms of the experiment explained above ? If so click on \emph{I agree} below.}}

In this way, participants were aware that the data collected from their participation were to be be used for research on personality and behaviour in groups.
IP addresses were collected. Email addresses were asked. Email addresses were only used to send participants bonuses via Paypal\textsuperscript{\textregistered} according to their score in the experiments. IP addresses were used solely to obtained the country of origin of the participants. Behaviours were analyzed anonymously. Information collected on the participants were not used in any other way than the one presented in the manuscript and were not distributed to any third party. Personality, gender and country of origin presented no correlation with influenceability or any other quantity reported in the manuscript.
The age of participants was not collected. Only adults are allowed to carry out microtasks on the CrowdFlower platform : CrowdFlower terms and conditions include : "you are at least 18 years of age".
The experiment was declared to the Belgian Privacy Comission (https://www.privacycommission.be/) as requested by law. 
\samrevieww{The French INSERM IRB read the consent procedure and confirmed that their approval was not required for this study since the data were analyzed anonymously.
}

\paragraph*{Control experiment}

Human judgment is such a complex process that no model can take all its influencing factors into account. The precision of the predictions is limited by the intrinsic variation in the human judgment process. 
\corentin{To represent this degree of unpredictability, we consider}
the variation in the judgment revision process that would occur if a participant were exposed to two replicated games
in which the set of initial judgments happened to be identical. A control experiment served to measure this degree of unpredictability.

\begin{figure}[h!]
    \centering
    \includegraphics[trim=0cm 0cm 0cm 0cm,scale = 0.3]{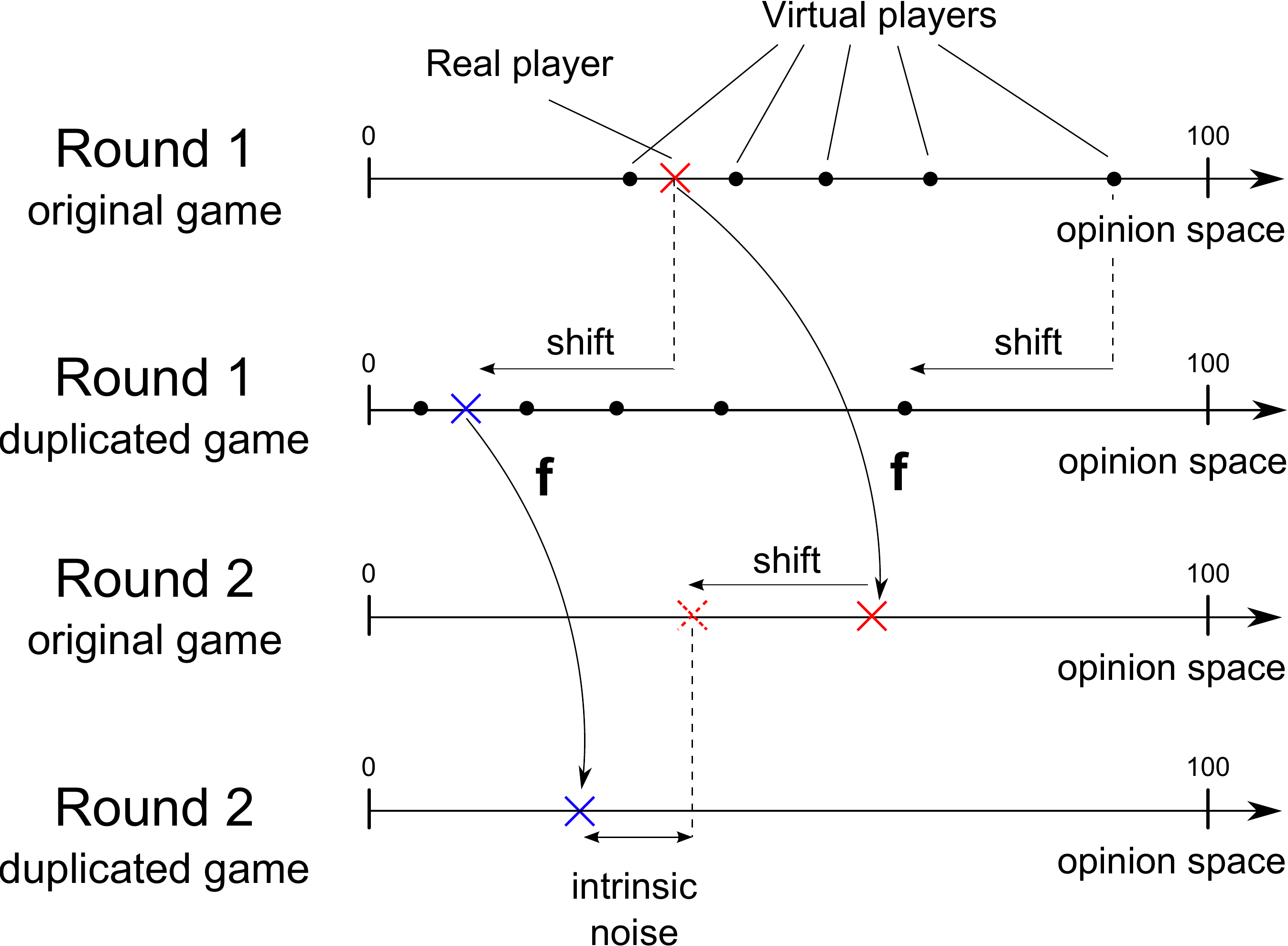}
    \caption{\textbf{Illustration of the shift applied to the synthetic judgment in order to preserve the distances between initial judgments.} \samlast{The shift made by the participant in the first round was synthetically applied to all other initial judgments. The intrinsic variation in judgment revision is the distance between second round judgments, reduced by the shift in initial judgments. The dotted cross is not a judgment and is only displayed to show how the second round intrinsic variation is computed using the initial shift. All shifts are equal. The same shift is also applied to all other second round judgments, and used to compute final round intrinsic variations.}}
    \label{fig:new_exp}
\end{figure}

To create replicated experimental conditions, the judgments of five out of the six participants were synthetically designed. The only human participant in the group was not made aware of this, so they would act as in the uncontrolled experiments. Practically, participants took part in 30 games, among these, 20 games had been designed to form 10 pairs of replicated games with an identical picture used in both games of a pair. To make sure the participants did not notice the presence of replicates, the remaining 10 games were distributed between the replicates. The order of appearance of the games with replicates is as follows :
$\numhi{1} , \numhi{2} , 11 | \numhi{3}, \numhi{4}, 12 | \numhi{6}, 13, \numhi{5}| 14 , \numhi{7}, \numhi{8} |\numhi{9}, \numhi{1},15 | \numhi{4}, \numhi{10}, 16| \numhi{2}, \numhi{6},17 | \numhi{7}, 18, \numhi{3} | \numhi{8}, \numhi{5}, 19 | \numhi{10}, \numhi{9}, 20$, where games 1 to 10 are the replicated games. The games successively appeared three at a time from left to right.
The 15 synthetic judgments (5 participants over 3 rounds) which appeared in the first instance of a pair of replicates were copies of past judgment made by real participants in past uncontrolled experiments.
The copied games collected in uncontrolled experiments were randomly selected among the games in which more than 5 participants had provided judgments.
Since the initial judgment of the real participant could not be controlled, the 15 synthetic judgments in the second replicate had to be shifted in order to maintain constant the initial judgment distances in each replicate. The shift was computed in real time to match the variation of the real participant initial judgments between the two replicates. The same shift was applied to all rounds to keep the synthetic judgments consistent over rounds (see Fig.~\ref{fig:new_exp} for the illustration of the shifting process). This provided exactly the same set of initial judgments up to a constant shift in each pair of replicated games. Such an experimental setting allowed assessing the degree of unpredictability in judgment revision (see the \textit{Prediction accuracy} section in \textit{Results} for details).

\subsection*{Participants}

\paragraph*{Uncontrolled experiment}

\begin{figure}[!ht]
\centering
\begin{tabular}{cc}
\textbf{(A) Gauging} & \textbf{(B) Counting}\\
\includegraphics[scale=0.27]{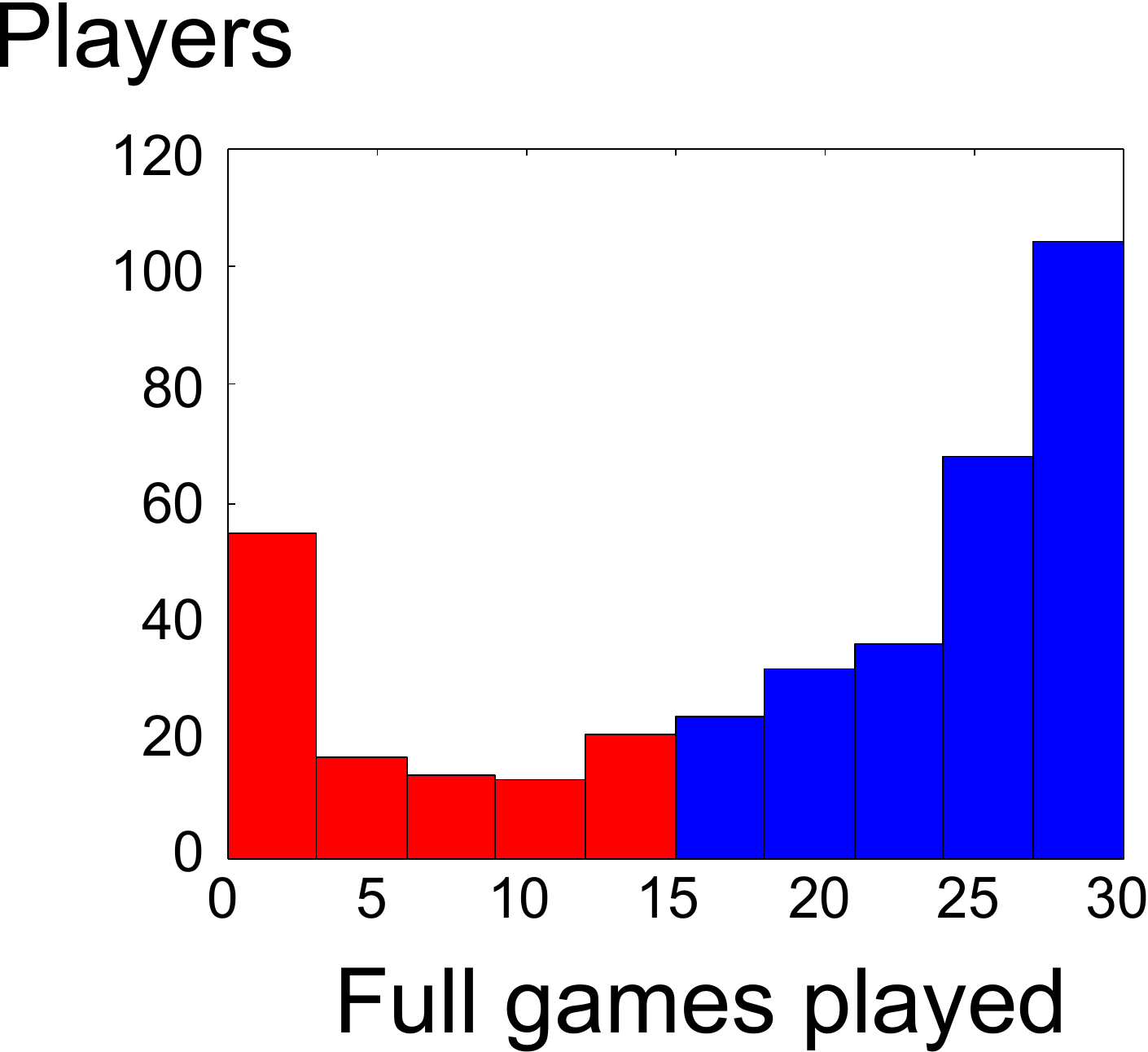}&
\includegraphics[scale=0.27]{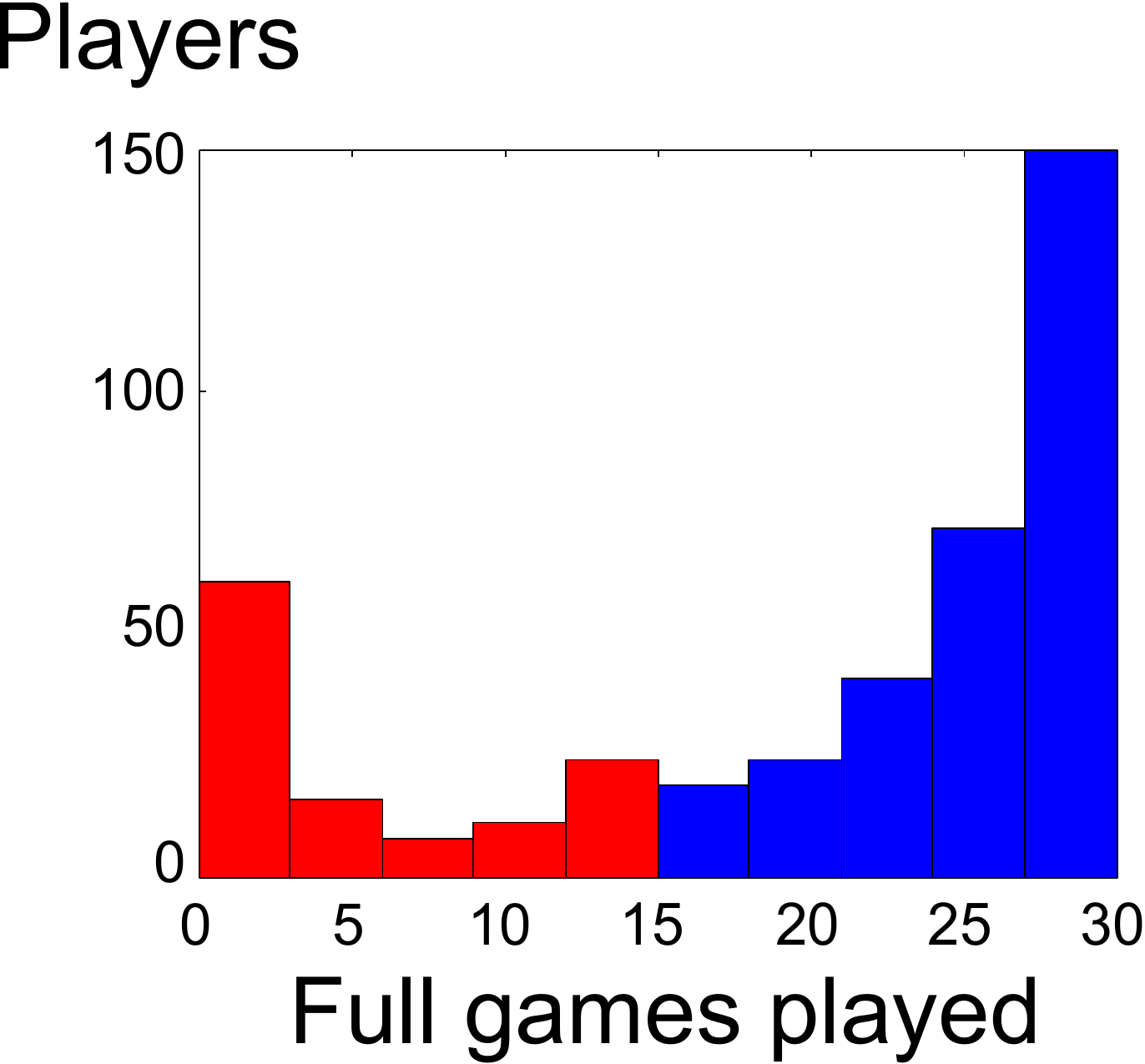}
\end{tabular}

\textbf{(C) Gauging}\\
\includegraphics[scale=0.60]{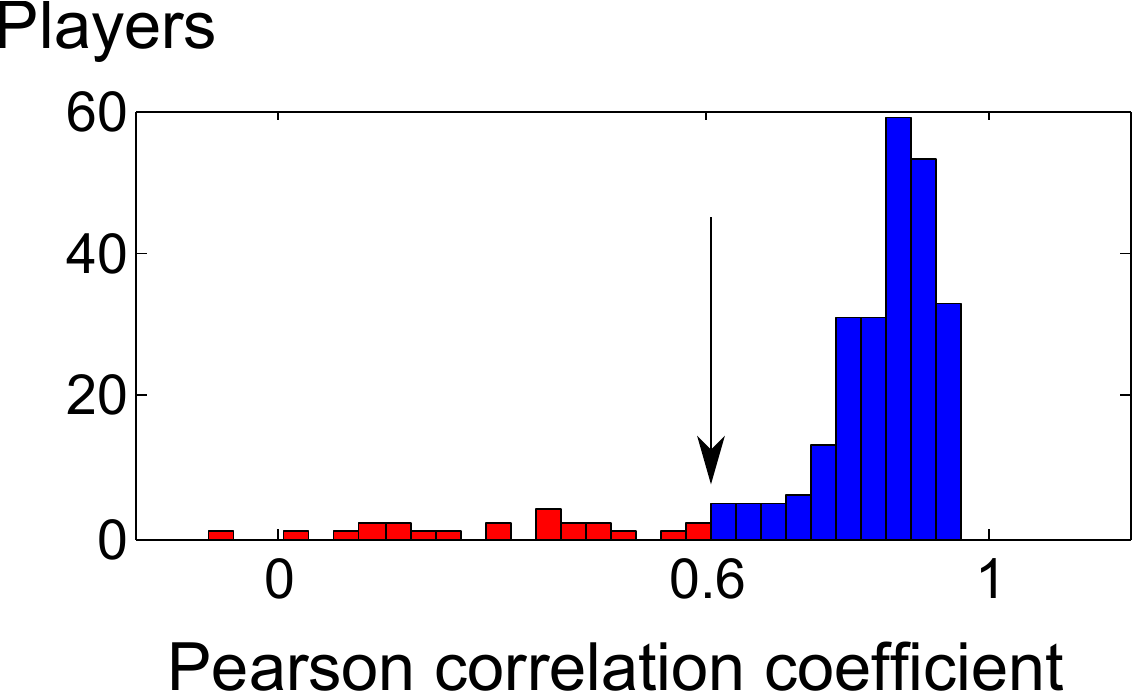}\\
\textbf{(D) Counting}\\
\includegraphics[scale=0.60]{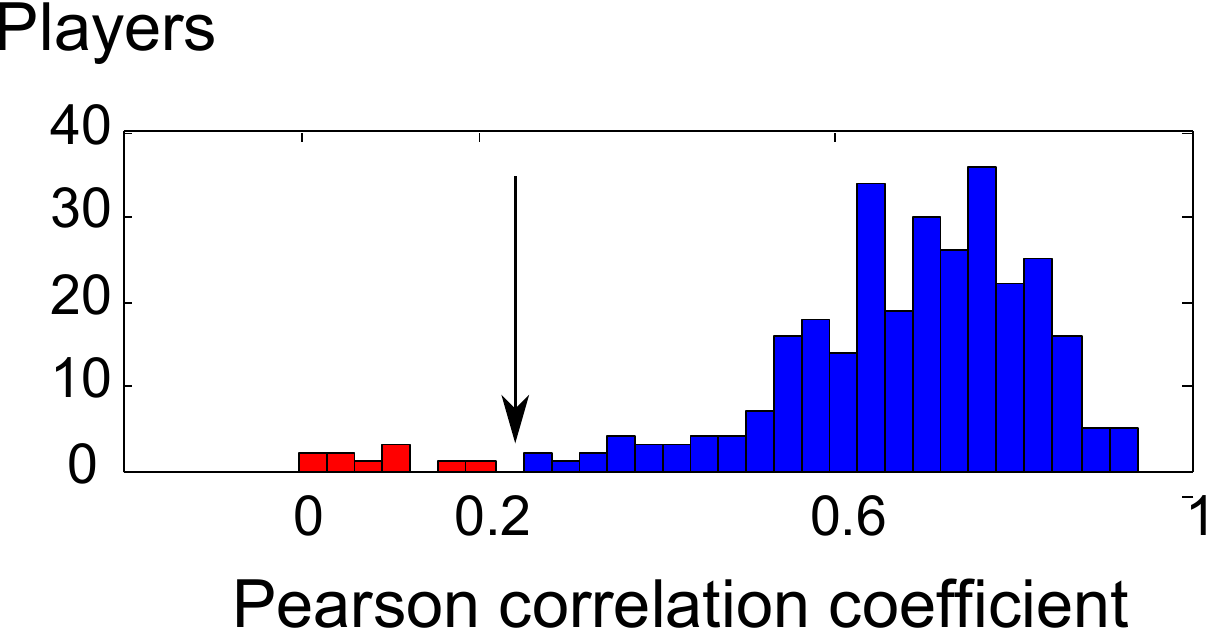}\\
\caption{\textbf{Filters for participants' trustworthiness.} (A),(C) \gauginggame, 
(B),(D) \countinggame.
(A,B) : Histograms of the number of full games played by participants. 
(C,D) : Histograms of the correlations between judgments and true values for each participant. Rejected participants are displayed in red while kept participants are in blue, to the right of the black arrow.
}\label{fig:fractals_filter_full_games}
\end{figure}

The data were collected during July, September and October 2014. Overall, $654$ distinct participants took part in the study ($310$ in the \gauginggame~only, $308$ in the \countinggame~only and $36$ in both). In total, $64$ groups of $6$ participants completed a \gauginggame, 
while $71$ groups of $6$ participants completed a $\countinggame$. According to their IP addresses, participants came from $70$ distinct countries.
Participants mostly originated 
from $3$ continents : $1/3$ from Asia, $1/2$ from Europe and $1/6$ from South America.
As detailed at the end of the paragraph, most participants completed most of the $30$ games and played trustworthfully. The others were ignored from the study via two systematic filters. First, since the prediction method was tested using up to $15$ games in the model parameter estimation process, the predictions reported in present study concern only the participants who completed more than $15$ out of the $30$ games. This ensures that the number of games used for parameter estimations is homogeneous over all participants. The prediction performance can then be compared among participants. The median number of fully completed games per participants was $24$ with $10.5$ std for the \gauginggame~and $27$ with $10.5$ std for the \countinggame. Lower numbers are possibly due to loss of interest in the task or connexion issues. The first filter lead to keep $68\%$ of the participants for the \gauginggame~and $71\%$ for the \countinggame~(see Fig.~\ref{fig:fractals_filter_full_games}--A,C for details).
Secondly, the prediction were only made on judgments of trustworthy participants. Trustworthiness was computed via correlation between participant's judgments and true answers. Most participants carried out the task truthworthfully with a median correlation of 0.85 and median absolute deviation (MAD) of 0.09
for the \gauginggame~and 0.70 median and 0.09 MAD for the \countinggame. A few participants either played randomly or systematically entered the same aberrant judgment. A minimum Pearson correlation thresholds of $0.61$ for the \gauginggame~and $0.24$ for the \countinggame~were determined using Iglewicz and Hoaglin method based on median absolute deviation~\cite{iglewicz1993detect}. The difference between the two thresholds is due to the higher difficulty of the \countinggame~as expressed by the difference between median correlations. This lead to keep $91\%$ and $96\%$ of the participants which had passed the first filter~(see Fig.~\ref{fig:fractals_filter_full_games}--B,D for details).

\samreview{
It should be acknowledged that the selection procedure may have lead to sample bias. This could be due to self-selection :
some people choose to participate in the experiment and others do not.
The fact that the study was carried out online is another factor that could bias the sample. The \textit{a posteriori} filters may be another source of bias. These are common issues in the behavioural sciences.
Possibly, the nature of the study will have appealed to certain types of people and not others. Although that could have biased the characteristics of the sample, we are unaware of any empirical evidence suggesting that people \samrevieww{who like participating in this kind of tasks} are more or less susceptible to social influence.
}

\paragraph*{Control experiment}

The data were collected during May and June 2015. 
Overall, $207$ distinct participants took part in the study ($113$ in the \gauginggame~only, $87$ in the \countinggame~only and $7$ in both). The $120$ \gauginggame~participants took part in $139$ independent games while the $94$ \countinggame~participants were involved in $99$ independent games. Each independent game was completed by $5$ synthetic participants to form groups of $6$. The same filters as those used in the uncontrolled experiment were applied to the participants in the control experiment. This lead to keep $88\%$ of the \countinggames~and $80\%$ of the \gauginggames.


\subsection*{Opinion revision model}

\corentin{To capture the way individual opinions evolve during a collective judgment process, a \textit{consensus model} is used. These models have a long history in social science~\cite{French1956}
and their behaviour has been thoroughly 
analyzed in a theoretical way~\cite{olfati2007,SamAntoine_Persistent_SICON2013}.
Our model~\eqref{eq:consensus-alpha-to-mean}
assumes that when an individual sees a set of opinions, their opinion changes linearly in the distance between their opinion and the mean of the group opinions. There is recent evidence supporting this assumption~\cite{mavrodiev2013quantifying}. \samreview{See also supplementary section~\ref{SI-sec:linearity-test} for a test of the validity of the linearity assumption.}
The rate of opinion change as a result of social influence is termed the \textit{influenceability} of a participant. The model also assumes that this influenceability may vary over time. The decrease of influenceability represents opinion settling.
\samgreen{The model is described in mathematical terms in equation~\eqref{eq:consensus-alpha-to-mean}, with $\alpha_i(r)$ being the influenceability of participant $i$ after round $r$.}
\cocoreview{When $\alpha_i(1)$ is nonzero,} the ratio $\alpha_i(2)/\alpha_i(1)$ represents the decaying rate of influenceability. Parameters $\alpha_i(1)$ and $\alpha_i(2)$ are to be estimated to fit the model to the data. It is expected that $\alpha_i(1), \alpha_i(2) \in [0,1]$.
If $\alpha_i(r) = 0$, the participant does not take into account the others and their opinion remains constant. If $\alpha_i(r+1)=\alpha_i(r)$, the influenceability does not change in time and the opinion $x_i(t)$ eventually converges to the mean opinion $\bar{x}$. Instead, if $\alpha_i(r+1)/\alpha_i(r) < 1$, \influenceability starts positive but decays over time, which represents opinion settling.}

\samreview{
There exist several variations to the linear consensus model presented above. In particular, the bounded confidence models~\cite{Deffuant2000,Hegselmann2002} assume that the influenceability $\alpha_i(t)$ also depends on the distance between one's opinion and the influencing opinion. Alternatively, the model by Friedkin and Johnsen~\cite{friedkin1990social} assumes that individuals always remain influenced by their initial opinion or \textit{prejudice} over time. Rather than providing an exhaustive assessment of the alternative models found in the literature, the objective of the present study is to show how the predictive power of a simple model of opinion dynamics can be assessed and to estimate the minimal prediction error that one can expect for any opinion dynamics model.
}

\samreview{Consensus models of opinion dynamics are well adapted to represent opinions evolving in a continuous space. This corresponds to many real world situations in which the opinion represents the inclinaison between two extreme opposite options such as left and right in politics.
Alternatively, part of the literature consider models with binary choices or actions (\eg voting for one candidates, buying a car or going on strikes). These discrete choice models include the rumour~\cite{dodds2004universal} and threshold models~\cite{Granovetter1978,watts2002simple}. In the latter, an individual changes their action when a certain proportion of its neighbours does so. These models directly link the discrete action of an individual to the actions in their neighbourhood. This allows to describe cascade propagation of behaviours. Presumably, before someone changes their actions, their opinion had to change as a result of the social stimuli. 
A recent model bridges these two bulks of work, considering the social influence of discrete actions on continuous opinions (the \textit{CODA} model) which itself results in an individual action~\cite{martins2008continuous}. It appears that this model also naturally leads to cascade of behaviours over a social network~\cite{chowdhury2016continuous}. It would be interesting to see how the threshold parameter in Granovetter's model may be expressed in terms of the initial opinion of individual in the CODA model : individuals with an opinion close to the boundary leading to either of the two discrete actions would have a lower threshold for action change.
}

\samreview{
The past two decades have witnessed a few attempts to confront simple models of opinion dynamics to real world data on collective human behaviours. These include conflicts and controversies on Wikipedia~\cite{iniguez2014modeling,torok2013opinions} or how voters distribute their votes among candidates in elections~\cite{bernardes2002election,caruso2005opinion,fortunato2007scaling}. However, since individual opinions involved in real world processes are not directly available, researchers had to calibrate their models on global measures such as level of controversy or distribution of votes. Moreover, the predictive power of these models was not assessed : the data used to calibrate the models also served to validate them. \textit{In vitro} studies such as the present one have the advantage of providing the micro-level data driving the collective dynamics. Another advantage of \textit{in vitro} studies is the possibility to differentiate social influence from other confounding factors such as homophily thanks to the anonymity of influencing individuals (see also~\cite{aral2009distinguishing,shalizi2011homophily}).
}

\subsection*{Prediction procedure}

\paragraph*{Regression procedure}

The goal of the procedure is to predict the future judgment of a given participant in a given game. The set of first round judgments of this game are supposed to be known to initialize the model~\eqref{eq:consensus-alpha-to-mean}, this includes the initial judgment from the participant targeted for prediction and the initial judgments from the five other participants who possibly influenced the former.
  
To tune the model, prior games from the same participant are assumed to be available. These data serve to estimate the influenceability parameters $\alpha_i(1)$ and $\alpha_i(2)$. In one scenario, the influenceability parameters of the participant are estimated independently of the data from other participants (\DIbf~method). This is the only feasible method when no prior data on other participants is available (see the \textit{Prediction scenarios} section in \textit{Results} for details on the data availability scenarios). In this first case, parameters $\alpha_i(1)$ and $\alpha_i(2)$ are determined using a mean square error minimization procedure (this procedure amounts to likelihood maximization when the errors between model predictions and actual judgments are normally distributed and have zero mean, see for instance \cite[p27]{bishop2006pattern}). 

In situations where the number of prior games available to estimate the influenceability parameters is small, little information is available on the participant's past behaviour. This may result in unreliable parameter estimates.
To cope with such situations, another scenario is considered : besides having access to prior data from the targeted participant, part of the remaining participants (half of them, in our study) are used to derive the typical influenceabilities in the population. The expectation-maximization (EM) algorithm is used to classify the population into groups of similar influenceabilities~\cite{bishop2006pattern}. These typical influenceabilities serve as a pool of candidates. The prior games of the targeted participant are used to determine which candidate yields the smallest mean square error (\DIcbf~method).
The procedure to determine which typical candidate best suits a participant requires less knowledge than the one to accurately estimate their influenceability without a prior knowledge on its value (see the results in the \textit{Prediction accuracy} section in \textit{Results}).

\paragraph*{Validation procedure}
\samgreen{The three scenarios presented in the \textit{Prediction scenarios} section in \textit{Results} are validated via crossvalidation.}
The validation procedure 
\corentin{of the last scenario (i.e., access to prior data from participants, existence and access to typical influenceabilities)}
starts by 
\corentin{randomly}
splitting the set of participants into two
\corentin{equal parts.}
The prediction focuses on one of the two halves while the remaining half is used to derive the typical influenceabilities in the \DIc~method. In the half serving for prediction, our model is assessed via repeated random sub-sampling crossvalidation : for each participant, a training subset of the games is used to
\samgreen{assign the appropriate typical influenceabilities to each participant.}
The rest of the games serves as the validation set to compute the root mean square error (\textit{RMSE}) between the observed data and predictions. The error specific to participant $i$ is denoted as $RMSE_i$. The results are compared for various training set sizes. The \textit{RMSE} is obtained from averaging errors over $300$ iterations using a different randomly selected training set each time. 
\corentin{To compare scenarios 1 and 2 with scenario 3, we only consider the half serving for prediction. Scenario 1 (no data available) does not require any training step. For scenario 2 (access to prior data from participants), instead of learning 
\samgreen{affectation of typical influenceabilities to each participant, the learning process directly estimates the influenceabilities $\alpha_i(1)$ and $\alpha_i(2)$ for each participant without prior assumption on their values}.
\samgreen{The whole validation process is also carried out reversing the role of the two halves of the population and a global \textit{RMSE} is computed out of the entire process}.}


\subsection*{Intrinsic unpredictability estimation}

\paragraph*{Unpredictability of the second round judgment}\label{SI-sec:unpredictability-second-round}

\samgreen{
Even though the study mainly focuses on the predictions of third round judgments, we first focus in the present section on the two first rounds of the games, for the sake of clarity. The prediction procedure can easily be adapted to predict second round rather than third round judgments. Results for second round predictions based on the consensus model~\eqref{eq:consensus-alpha-to-mean} are presented in \textit{Second round predictions} section below}.

\begin{figure}[!ht]
\centering
\begin{tabular}{cc}
\textbf{(A) Gauging}& \textbf{(B) Counting}\\
\includegraphics[scale=0.22]{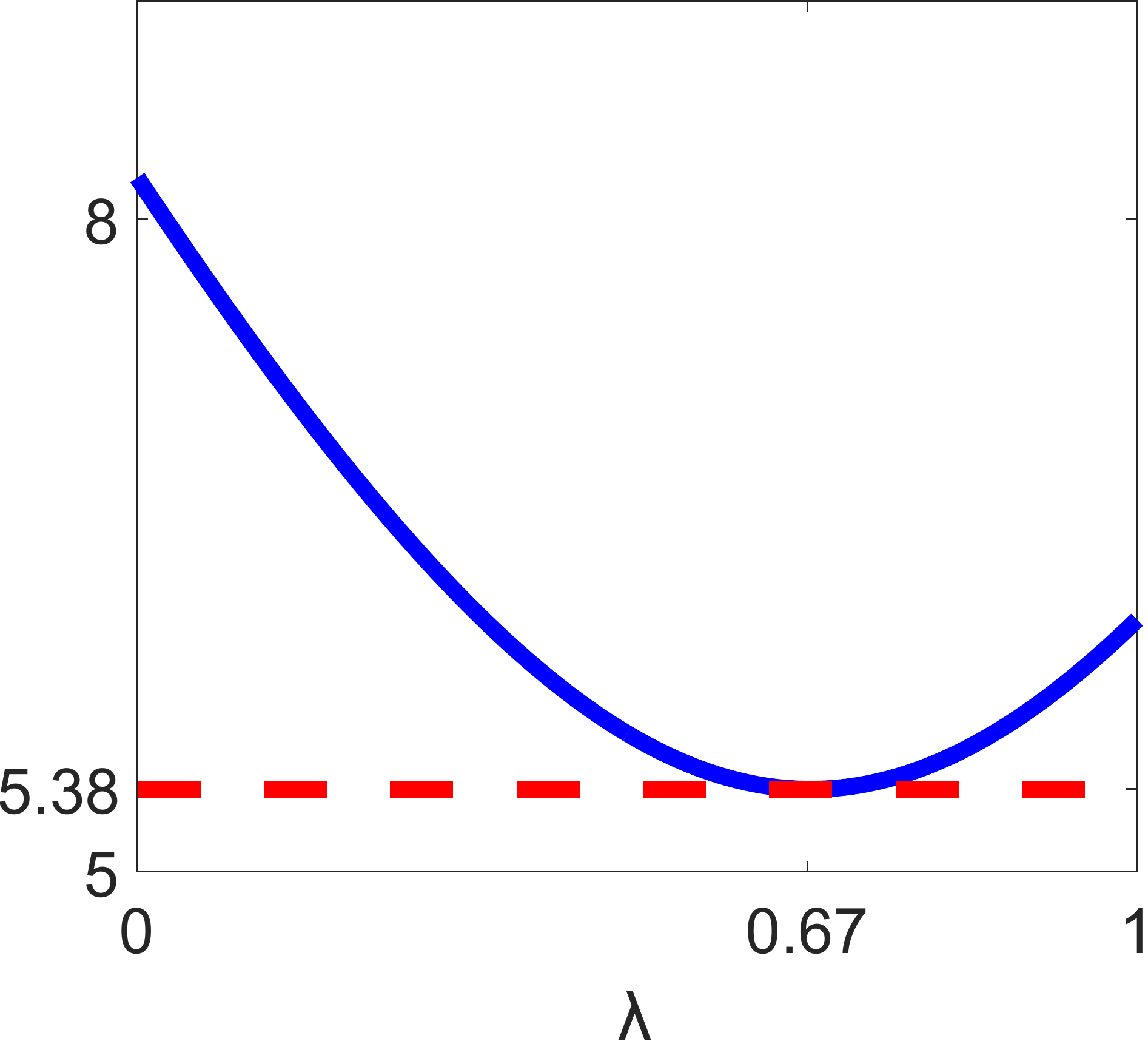}&
\includegraphics[scale=0.22]{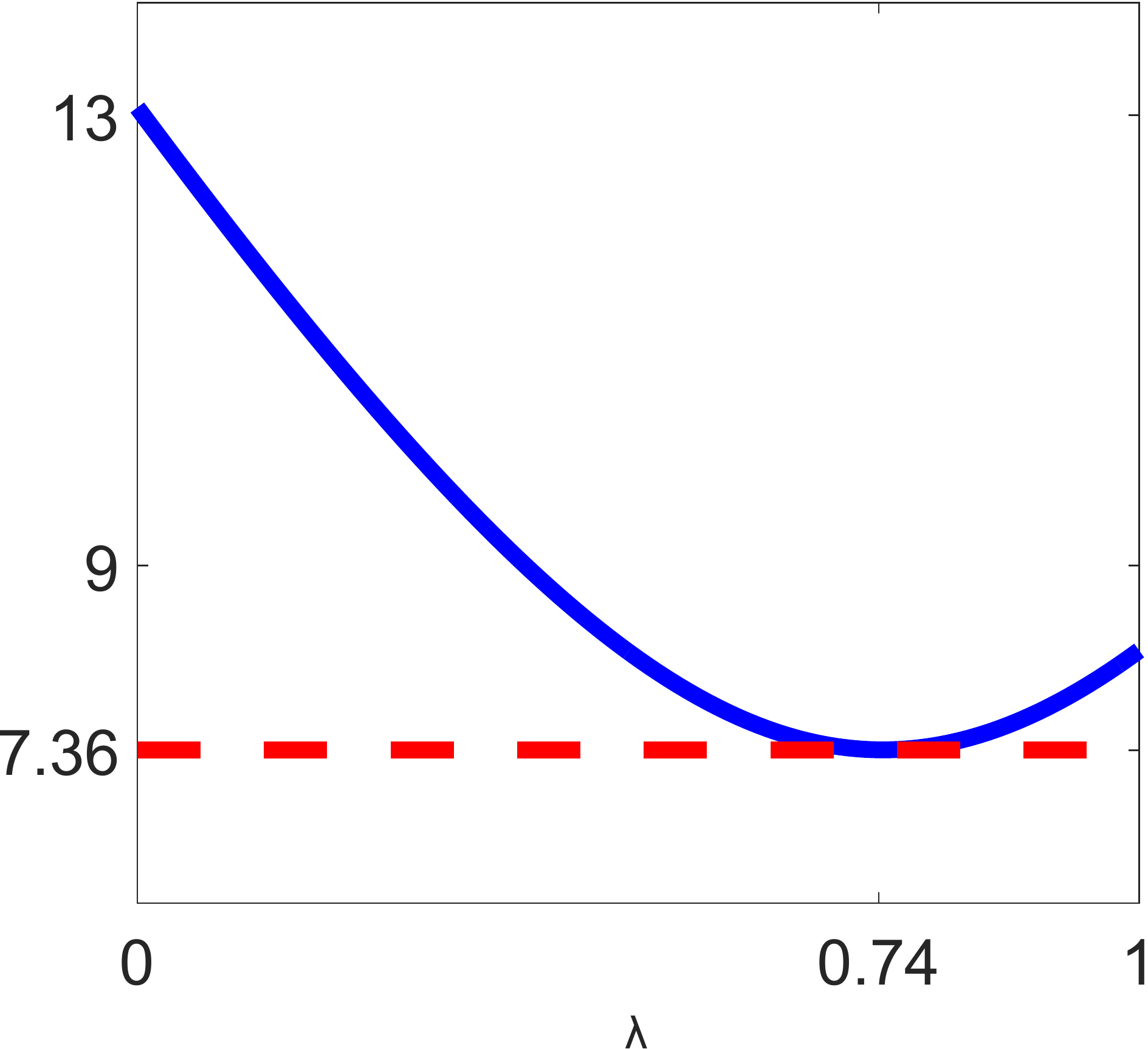}\\
\textbf{(C) Gauging}& \textbf{(D) Counting}\\
\includegraphics[clip=true,trim=4cm 8.6cm 4.9cm 8.6cm, scale=0.33]{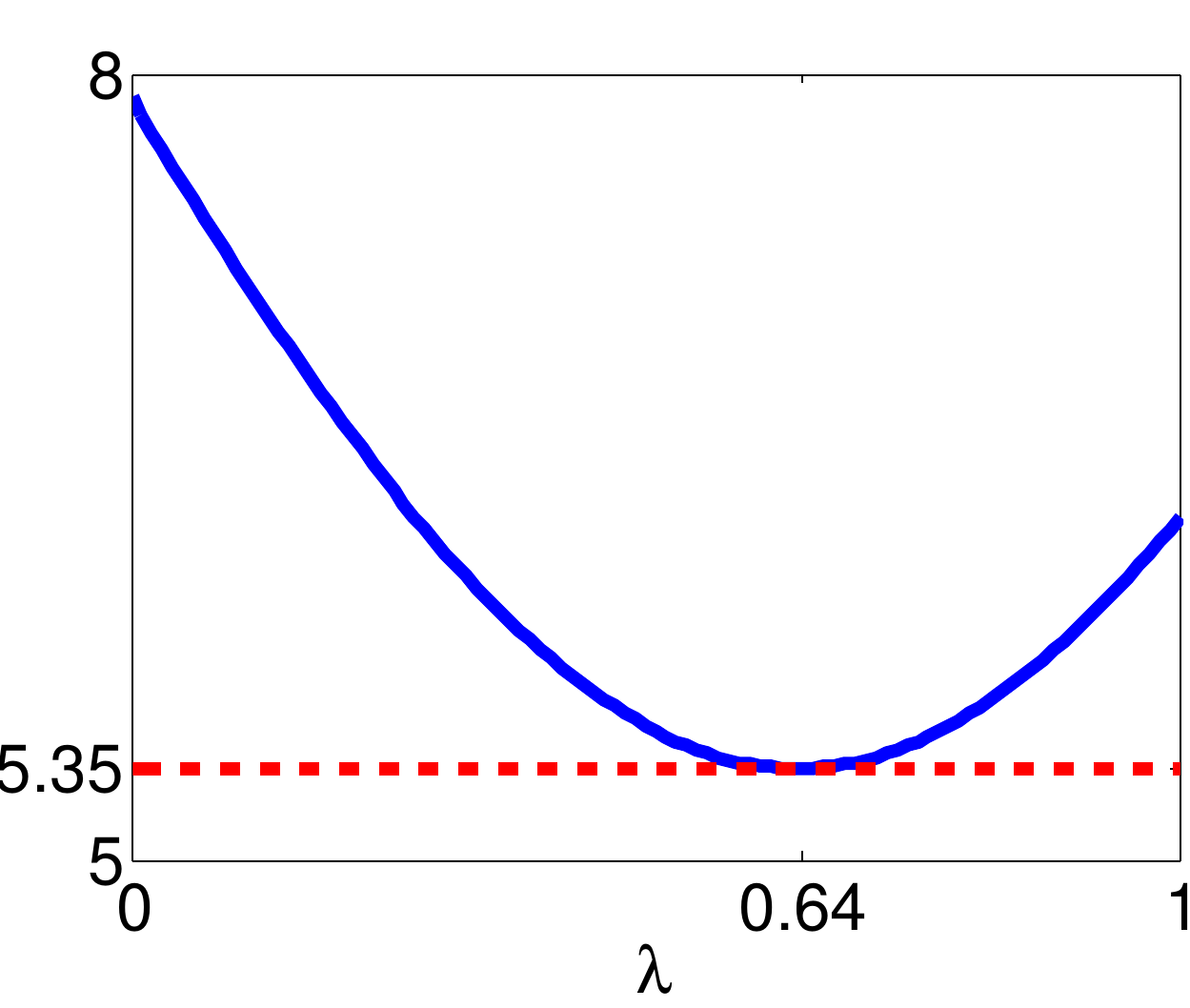}&
\includegraphics[clip=true,trim=4cm 8.6cm 4.9cm 8.6cm,scale=0.33]{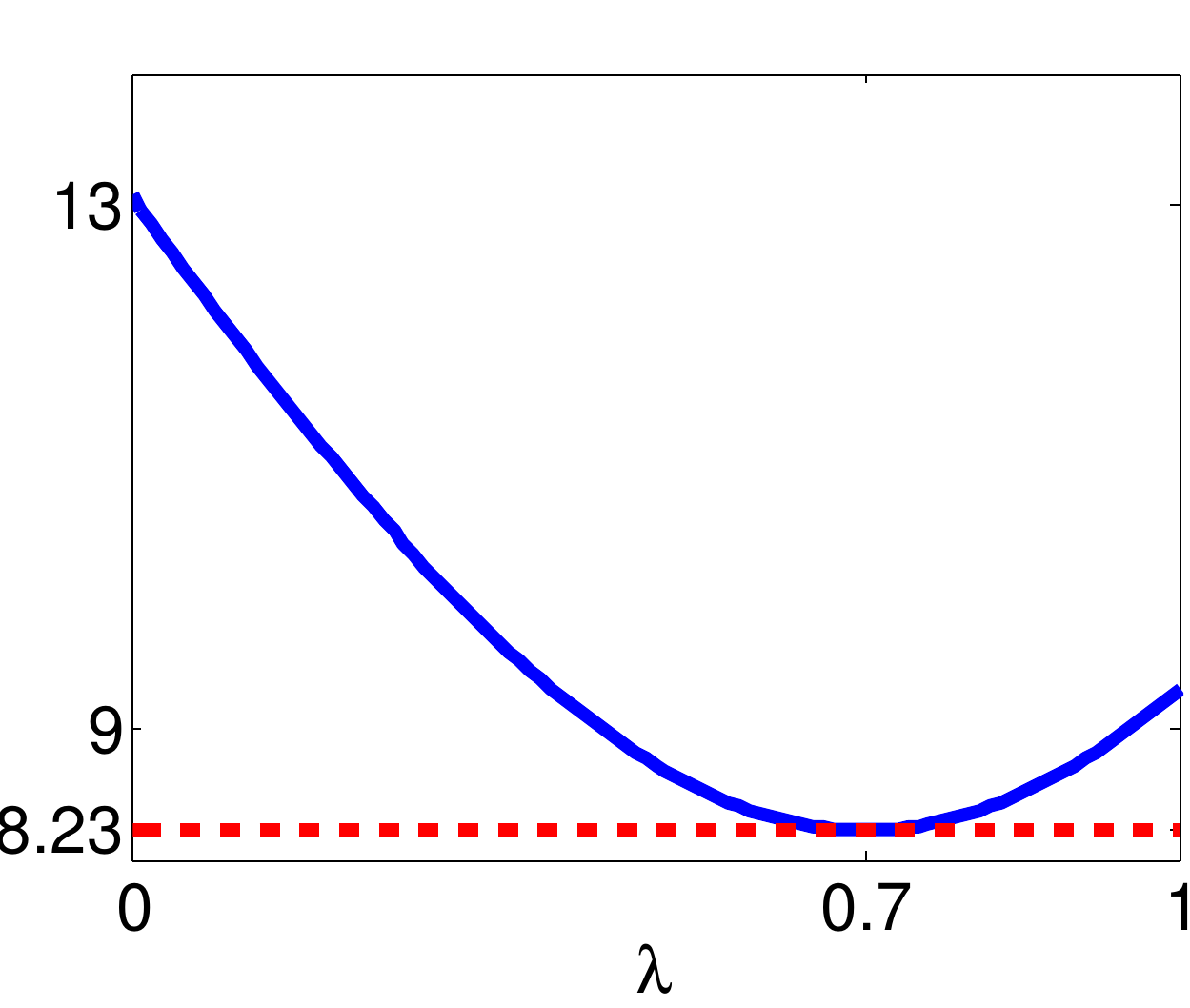}
\end{tabular}
\caption{\textbf{Root mean square intrinsic variation as a function of the $\lambda$ parameters}. (A), (B) : the second round judgments, as given in equation~\eqref{eq:unpredictability}. (C), (D) : the third round judgments, as given in equation~\eqref{eq:unpredictability2}. (A), (C) \gauginggame, (B), (D) \countinggame.
}\label{fig:intrinsic_variation}
\end{figure}

The control experiment described in \textit{Control experiment} section 
provides a way of estimating the intrinsic variations in the human judgment process.
When a participant takes part in a game, their actual second judgment depends on several factors : their own initial judgment $x_i(1)$, the \corentind{vector of} initial
judgments from other participants \samgreen{denoted as}
\corentind{$x_{others}(1)$}
and the displayed picture. As a consequence, the second round judgment of a participant can always be written as
\begin{equation}\label{eq:1rmodel}
x_i(\corentinb{2}) = f^{1}_i(x_i(1),\corentind{x_{others}(1)},picture) + \eta,
\end{equation}
where $f^{1}_i$ describes how a participant revises their judgment in average depending on their initial judgment and external factors. 
The term $picture$ is the influence of the picture on the final judgment. The quantity $\eta$ captures the intrinsic variation made by a participant when making their \samgreen{second round} judgment despite having made the same initial judgment and being exposed to the same set of judgments and identical picture. Formally $\eta$ is a random variable with zero mean. \samreview{This is shown in supplementary section~\ref{SI-sec:eta-zero-mean}}. The standard deviation of $\eta$ is assumed to be the same for all participants to the same game, denoted as $\text{std}(\eta)$. 
\samgreen{The standard deviation $\text{std}(\eta)$ measures the root mean square error between $f_i^1$ and the actual judgment $x_i(2)$ ; \samlast{this error measures} by definition the intrinsic unpredictability of the judgment revision process.}
\corentin{If it was known, the function $f^{\corentinc{1}}_i$} would provide the best prediction regarding judgment revision. By definition, no other model can be more precise.
The function $f^{\corentinc{1}}_i$ is unknown, but it is reasonable to make the following assumptions. First, the function $f^{\corentinc{1}}_i$ \samlast{is assumed to be} a sum of \corentind{(i)} the influence of \samlast{the initial judgments} and \corentind{and (ii)} the influence of the picture. Thus $f^{\corentinc{1}}_i$ splits into two components :
$$
f^{\corentinc{1}}_i(x_i(1),\corentind{x_{others}(1)},picture) = 
$$
$$
\lambda g^{\corentinc{1}}_i(x_i(1),\corentind{x_{others}(1)}) + (1-\lambda) h^{\corentind{1}}_i(picture),
$$
where $g^{\corentinc{1}}_i$ represents the dependence of the second round judgment on past judgments while $h^{\corentind{1}}_i$ contains the dependency regarding the picture. The parameter $\lambda \in [0,1]$ weights the relative importance of the first term compared to the second
\corentin{and is considered unique for each particular type of game.} 
It is further assumed that if the initial judgment $x_i(1)$ and the others judgment at round $1$ are shifted by a constant shift, the $g^{\corentinc{1}}_i$ component in the second round judgment will on average be shifted in the same way, in other words, it is possible to write
$$
g^{\corentinc{1}}_i(x_i(1) + s, \corentind{x_{others}(1)}+s) = g^{\corentinc{1}}_i(x_i(1),\corentind{x_{others}(1)}) + s,
$$
where $s$ is a constant shift applied to the judgments. Under this assumption, the control experiment provides a way of measuring the intrinsic variation. The intrinsic variation estimation $\text{std}(\eta)$ can be empirically measured as the root mean of
\begin{equation}\label{eq:unpredictability}
\frac{1}{2}(\corentinb{x'_i(2)} - \corentinb{x_i(2)} - \lambda(x'_i(1) - x_i(1)) )^2 
\end{equation}
over all repeated games and all participants, where the prime notation is taken for judgments from the second replicated game in the control experiment (see \textit{Participants} section in \textit{Material and Methods}.
The derivation of equation~\eqref{eq:unpredictability} is provided in supplementary section~\ref{SI-sec:unpredictability-derivation}.
\samgreen{
Since $\eta$ is assumed to have zero mean, the function $f_i$ properly describing the actual judgment revision process is the one minimizing $\text{std}(\eta)$. Correspondingly, the constant $\lambda \in [0,1]$ is set so as to satisfy this minimization. The intrinsic variation estimation $\text{std}(\eta)$ is displayed in Fig.~\ref{fig:intrinsic_variation}. \samgreen{The optimal $\lambda$ values are found to be $0.67$ and $0.74$ and correspond to intrinsic unpredictability estimations of $5.38$ and $7.36$ for the \gauginggame~and the \countinggame, respectively.}
These thresholds can be used to assess the quality of the predictions for the second round (see \textit{Second round predictions} section).}


\paragraph*{Unpredictability of the final round judgment}\label{SI-sec:unpredictability-third-round}

\samgreen{
The procedure to estimate third round intrinsic unpredictability, whose results are shown in Fig.~\ref{fig:compare-all-models}, varies slightly from second round estimation procedure. For third round judgments, a function with the same input as $f_i^1$, depending only on the initial judgments and the picture, cannot properly describe the judgment revision of one participant independently of the other participant's behaviour. In fact, the third round judgments depend on the second round judgments of others, which results from the revision process of other players. In other words, in a situation where a participant were to be faced successively to two groups of other participants, who by chance had given an identical set of initial judgments, the second round judgments of the two groups could vary due to \samlast{distinct ways of revising their judgments}.
}

\samgreen{
Since the initial judgments does not suffice in describing third round judgments, function $f_i^1$ is modified to take the second round judgments of others $\corentind{x_{others}(2)}$ as an additional input. Then, the control experiment provides a way to estimate the intrinsic variations occurring in judgment revision up to the third round. This is described formally in the rest of the section.
However, it should be noted that this description of judgment revision does not strictly speaking provide the exact degree of intrinsic variation included in the final round prediction error made in the uncontrolled experiment, it is rather an under-estimation of it : The predictions via the consensus model presented in the \textit{Prediction performance} section in \textit{Results}, are based solely on the initial judgments, whereas, the second round judgment from other participants are also provided in the present description. This additional piece of information necessarily makes the description of third round judgment more precise. As a consequence, the intrinsic variation estimated here (see details below) is an under-estimation of the actual intrinsic variation included in the prediction error of the consensus model. From a practical point of view, this means that the actual intrinsic unpredictability threshold is actually even closer to the predictions made by the consensus model than displayed in Fig.~\ref{fig:compare-all-models}. In other words, there is even less space to improve the predictions provided by the consensus model, since more than two thirds of the error comes from the intrinsic variation rather than from the model imperfections.
}

Formally, the deterministic part of the third round judgment of a participant $x_i(3)$ is fully determined as a function of their initial judgment $x_i(1)$, the initial judgment of others $\corentind{x_{others}(1)}$, the second round judgment of others $\corentind{x_{others}(2)}$ and the picture. So, the third round judgment can be written as
\begin{equation}\label{eq:2rmodelb}
\corentinc{x_i({3}) = f^2_i(x_i(1),\corentind{x_{others}(1)},\corentind{x_{others}(2)},picture) + \bar{\eta}},
\end{equation}
where $\bar{\eta}$ is the estimation of the intrinsic variation occurring after three rounds under the same initial judgments, same picture and same second round judgments from others.
Under the same assumptions on function $f^2_i$ as those made on $f_i^1$, and analogously to equation~\eqref{eq:unpredictability}, $\text{std}(\bar{\eta})$ is measured by the root mean of 
\begin{equation}\label{eq:unpredictability2}
\frac{1}{2}(\corentinb{x'_i(3)} - \corentinb{x_i(3)} - \lambda(x'_i(1) - x_i(1)) )^2
\end{equation}
over all repeated games and all participants. This intrinsic variation is provided as a function of parameter $\lambda$ in Fig.~\ref{fig:intrinsic_variation}, (C)-(D).

\section*{Second round predictions}\label{SI-sec:second-round-predictions}

\begin{figure}[!ht]
\centering
\textbf{(A) Gauging}\\
\includegraphics[scale=0.3]{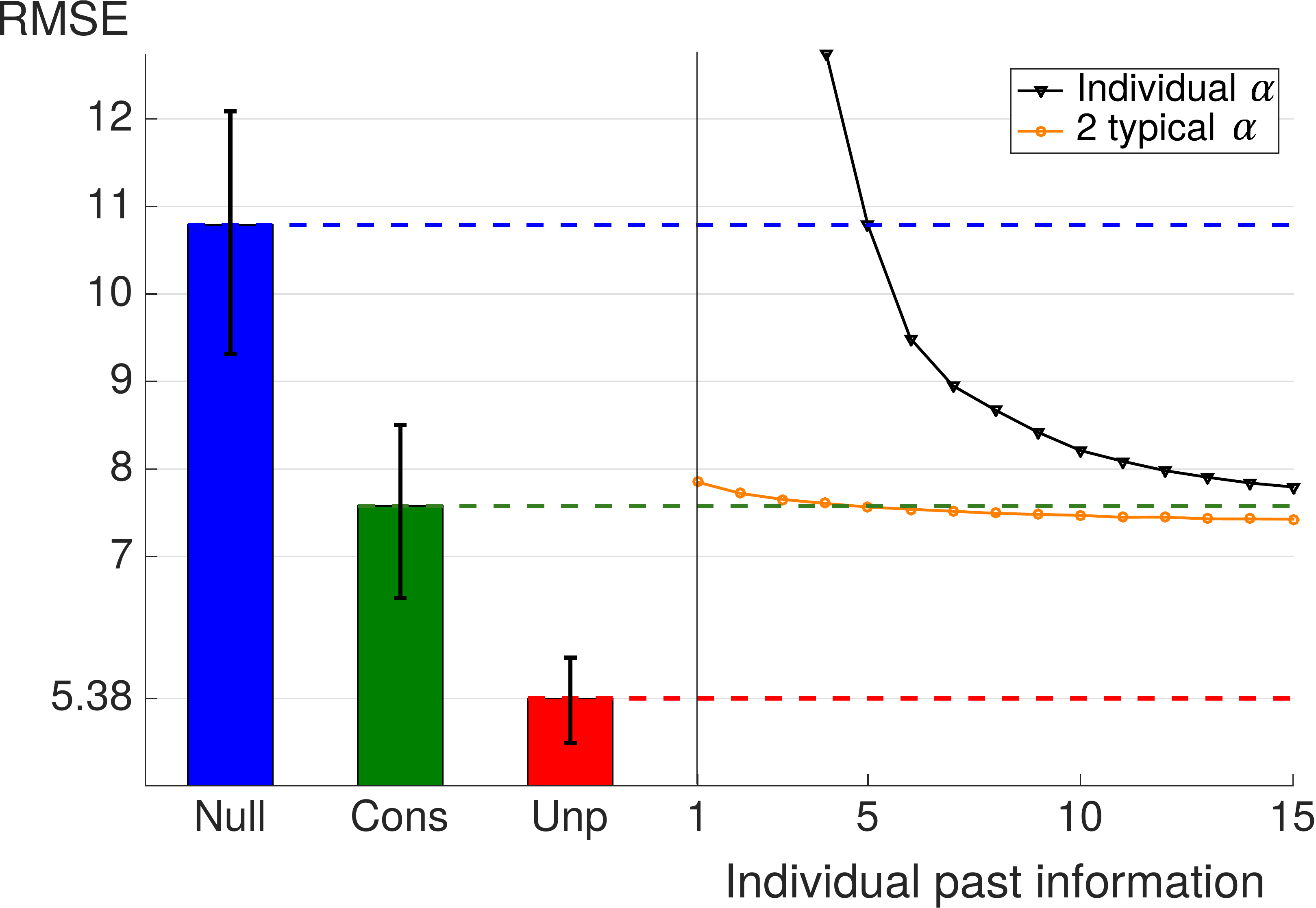}\\
\vspace{0.5cm}
\textbf{(B) Counting}\\
\includegraphics[scale=0.3]{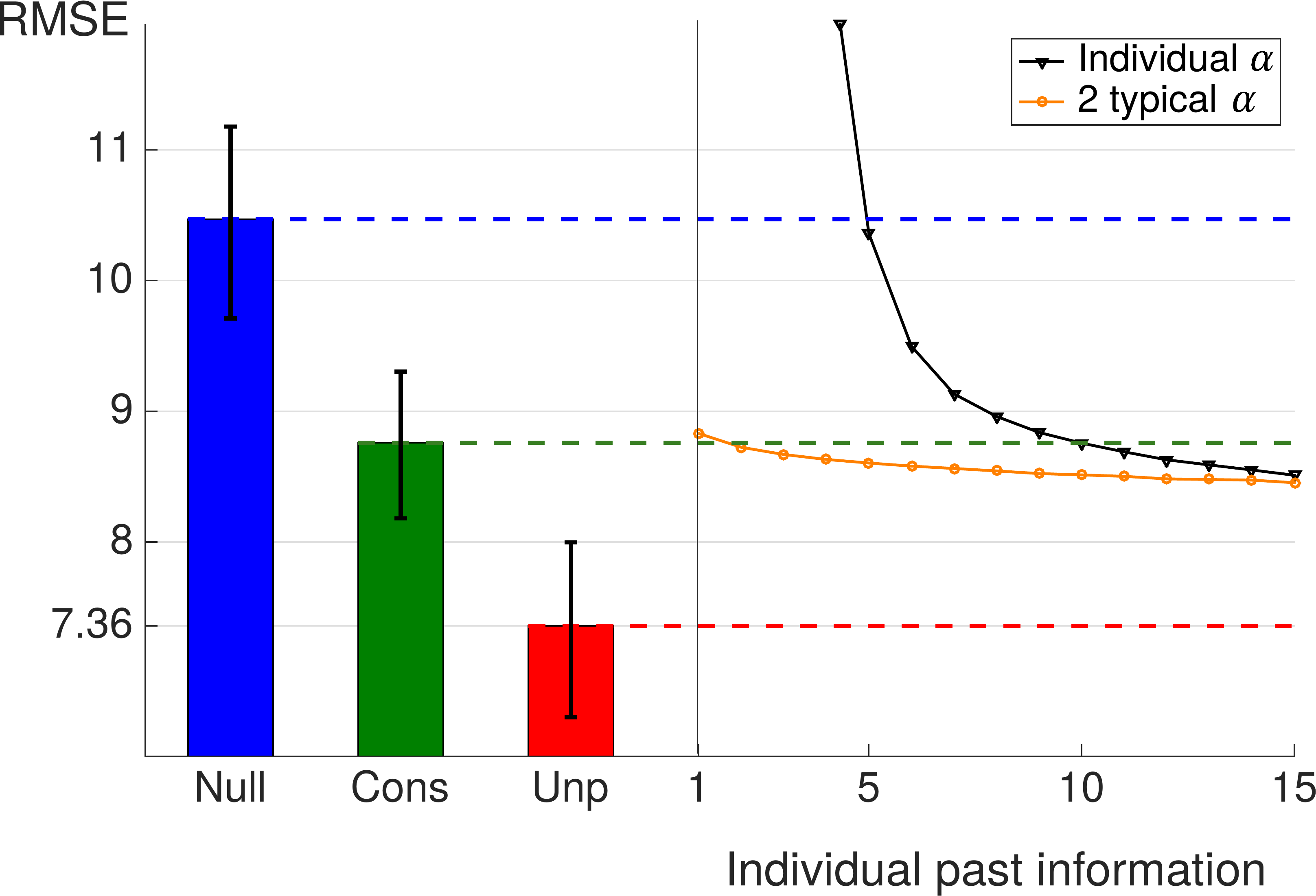}\\
\vspace{0.5cm}

\caption{\cocoreview{\textbf{Root mean square error (\textit{RMSE}) of the predictions for the \textit{second} round.} The RMSEs are obtained from crossvalidation.
(A) \gauginggame, (B) \countinggame. In (B), the $RMSE$ has been scaled by a factor of $5$ to be comparable to the (A) plot.
The bar chart displays crossvalidation errors for models that does not depend on the training set size.
 Top {blue} horizontal line 
corresponds to the null model of constant opinion. 
The {middle} horizontal {green} line 
correspond to fitting using the same typical couple of influenceability for the whole population. {The bottom horizontal red} line 
correspond to the prediction error due to intrinsic variations in judgment revision. 
The decreasing {black curves} (triangle dots) correspond to fitting with the individual influenceability method. The slightly decreasing {orange} curves ({round} dots) correspond to fitting choosing among $2$ typical couples of influenceability. All $RMSE$ were obtained on validation games.\vspace{-0.3cm}}}\label{fig:rmse-bis}

\label{fig:compare-all-models-bis}
\end{figure}

The prediction procedure based on the consensus model~\eqref{eq:consensus-alpha-to-mean} is applied to predict the second round judgments. Crossvalidation allows to assess the accuracy of the model. Results are presented in Fig.~\ref{fig:compare-all-models-bis}.
These results are qualitatively equivalent to the prediction errors for the third round as shown in Fig.~\ref{fig:rmse}, although the second round predictions lead to a lower RMSEs, as expected since they correspond to shorter term predictions.

\FloatBarrier

\section*{Acknowledgments}
C. V. K. is a F.N.R.S./FRIA research fellow. Research partly supported by the Belgian Interuniversity Attraction Poles (\textit{Dynamical systems, control and optimization network}), by an Actions incitatives 2014 of the Centre de Recherche en Automatique de Nancy, by project \textit{Modélisation des dynamiques dans les réseaux d'échange de semences (PEPS MADRES)} funded by the Centre National pour la Recherche Scientifique and project \textit{Computation Aware Control Systems (COMPACS)}, ANR-13-BS03-004 funded by the Agence National pour la Recherche. Contact : samuel.martin@univ-lorraine.fr. The funders had no role in study design, data collection and analysis, decision to publish, or preparation of the manuscript.

\section*{Author contributions statement}

C. V. K., S. M., J. M. H., P. J. R. and V. D. B. conceived the experiment(s).  C. V. K., S. M. and P. G. conducted the experiments. C. V. K. and S. M. analysed the results. All authors reviewed the manuscript.

\section*{Additional information}

\textbf{Competing financial interests}: The authors declare no competing financial interests.

\nolinenumbers

%
%
%

\bibliography{references}




\newpage
\pagestyle{empty}
\newgeometry{legalpaper}

\setcounter{table}{0}
\renewcommand{\thetable}{S\arabic{table}}%
\setcounter{figure}{0}
\renewcommand{\thefigure}{S\arabic{figure}}%

\begin{center}
\LARGE{Supplementary Information}\\
\Large{Modelling influence and opinion evolution in online collective behaviour}\\
\vspace{0.2cm}
\large{Corentin Vande Kerckhove, Samuel Martin, Pascal Gend, Peter J. Rentfrow, Julien M. Hendrickx, Vincent D. Blondel}
\end{center}

\renewcommand*{\thesection}{S\arabic{section}}
\setcounter{secnumdepth}{2}


\section{Derivation of the measure of unpredictability}\label{SI-sec:unpredictability-derivation}

\subsection{\samreview{Proof that $\eta$ has zero mean}}\label{SI-sec:eta-zero-mean}

\samrevieww{In section~\textit{Intrinsic unpredictability estimation}, we used that $\eta$ has zero mean. This fact is proven in the sequel.}
\samreview{
Assume, to show a contradiction that $m = \E(\eta) \neq 0$. Denote $\tilde{x}_i(2) = f_i^1(x_i(1),x_{others}(1),picture)$. Then, we can show that
$\tilde{x}_i(2) + m$ would be a better model for $x_i(2)$ than $\tilde{x}_i(2)$. This would contradict the definition of $f_i^1$ as the best model. Indeed, the expected square prediction error would be
\begin{eqnarray*}
 \E\left(  \left(x_i(2) - (\tilde{x}_i(2) + m) \right)^2  \right) &=& 
 \E\left(\left(x_i(2) - \tilde{x}_i(2)  \right)^2  \right)
- 
2m\E\left(x_i(2) - \tilde{x}_i(2)  \right) 
+
 m^2 \\
 &=&
 \E\left(\left(x_i(2) - \tilde{x}_i(2)  \right)^2  \right)
-m^2 \\
&<& \E\left(\left(x_i(2) - \tilde{x}_i(2)  \right)^2  \right),
\end{eqnarray*}
where we used the fact that $\eta = x_i(2) - \tilde{x}_i(2)$.
The same reasoning allows to show that the prediction error $\bar{\eta}$ at round $3$ also has zero mean.
}

\subsection{\samreview{Derivation of equation~\eqref{eq:unpredictability}}}

Equation~\eqref{eq:unpredictability} is derived using the following reasoning.
The judgments made in two replicated games \samlast{of a control experiment} by a same participants are described as 
\begin{eqnarray*}
 \corentinb{x_i(2)} &=& f^{\corentinc{1}}_i(x_i(1), \corentind{x_{others}(1)},picture) + \eta,\\
 \corentinb{x'_i(2)} &=& f^{\corentinc{1}}_i(x'_i(1), \corentind{x'_{others}(1)},picture) + \eta',
\end{eqnarray*}
where the prime notation is taken for judgments from the second replicated game and $\eta$ and $\eta'$ are two independent draws of the random intrinsic variation. By design, the set of judgments are all shifted by the same constant :
\begin{eqnarray*}
x_i'(1) = x_i(1) + s, \\
\corentind{x'_{others}(1)} = \corentind{x_{others}(1)} + s,
\end{eqnarray*}
where $s = x'_i(1) - x_i(1)$ is known.
According the assumption made on function $f^{\corentinc{1}}_i$, 
\begin{eqnarray*}
\corentinb{x_i(2)} &=& \lambda g^{\corentinc{1}}_i(x_i(1), \corentind{x_{others}(1)}) + (1-\lambda) h^{\corentind{1}}_i(picture) + \eta,\\
x_i'(2) &=& \lambda g^{\corentinc{1}}_i(x_i'(1), \corentind{x'_{others}(1)}) + (1-\lambda) h^{\corentind{1}}_i(picture) + \eta',
\end{eqnarray*}
the second round judgment made in the second replicate is then
$$
\corentinb{x'_i(2)} = \lambda \left(g^{\corentinc{1}}_i(x_i(1),\corentind{x_{others}(1)})+s\right) + (1-\lambda) h^{\corentind{1}}_i(picture) + \eta',
$$
where the invariance by translation of $g_i^1$ was used. Taking the difference makes the unknown terms $g^{\corentinc{1}}_i(x_i(1),\corentind{x_{others}(1)})$ and $h^{\corentind{1}}_i(picture)$ vanish to obtain
\begin{equation}\label{eq:eta-prime-minus-eta}
\corentinb{x'_i(2)} - \corentinb{x_i(2)} = \lambda s + \eta' - \eta.
\end{equation}
Since $\eta$ and $\eta'$ have zero mean and are assumed to have equal variance, 
the theoretical variance of $\eta$ is 
\begin{equation}\label{eq:var-eta-plus-cov}
\begin{split} 
\E(\eta^2) &= \frac{1}{2} \left(\E(\eta^2) + \E(\eta'^2)\right) \\
&= \frac{1}{2} (\E(\eta'^2)  -2\E(\eta'\eta) + \E(\eta^2) + 2\E(\eta'\eta))\\
&= \frac{1}{2} (\E(\eta'^2  -2\eta'\eta + \eta^2) ) + \E(\eta'\eta)\\
&= \frac{1}{2} \E((\eta'-\eta)^2) + \E(\eta'\eta).
\end{split}
\end{equation}

Moreover $\eta$ and $\eta'$ are assumed to be independent with zero mean, \ie, $\E(\eta)=\E(\eta')=0$, therefore, their covariance is null :
$
\E(\eta\eta') = \E(\eta)\E(\eta') = 0.
$
Consequently, $\E(\eta^2) = \frac{1}{2} \E((\eta'-\eta)^2)$ and using equation~\eqref{eq:eta-prime-minus-eta}, the variance of $\eta$ is empirically measured as the average of
$$
\frac{1}{2}(\corentinb{x'_i(2)} - \corentinb{x_i(2)} - \lambda s)^2
$$
over all repeated games and all participants. This corresponds to equation~\eqref{eq:unpredictability}.


\subsection{\samreview{Discussion on the assumptions on $\eta$ and $\eta'$}}\label{SI-sec:eta-equal-variance}

\samreview{
The only assumption used to derive equation~\eqref{eq:eta-prime-minus-eta} is that $\eta$ and $\eta'$ have the same variance and are independent for each participant. Since function $f^1_i$ is unknown, it is not possible to directly test these assumptions. However, since pairs of replicates in the control experiment are related to the same picture, it is unlikely that the covariance between $\eta$ and 
$\eta'$ would be negative. If the covariance was positive, the quantity given in equation~\eqref{eq:eta-prime-minus-eta} would become a lower bound on the unpredictability threshold, as shown through equation~\eqref{eq:var-eta-plus-cov}. Finally, if $\eta$ and $\eta'$ did not satisfy the assumption of equal variance, the quantity in equation~\eqref{eq:eta-prime-minus-eta} would still correspond to the average variance $\frac{1}{2} \left(\E(\eta^2) + \E(\eta'^2)\right)$ which also represents the average intrinsic unpredictability, as seen in equation~\eqref{eq:var-eta-plus-cov}. 
}



\section{\samreview{Circumstances of the wisdom of the crowd}}\label{SI-sec:wisdom}

\samreview{
The wisdom of the crowds may not always occur. The present section recalls one important hypothesis underlying the wisdom of the crowds. The hypothesis is then tested against the empirical data from the study.
In the context of the present study, the wisdom of the crowd \samrevieww{corresponds to the following fact:} the mean opinion is most often \samrevieww{\textit{much}} closer to the true answer than the individual opinions are.
Denoting $\bar{x}$ the mean of $n$ opinions $x_i$ and $T$ the corresponding true answer, this is formally expressed as
\begin{equation}\label{eq:wisdom_of_crowds}
|\bar{x} - T| << \frac{1}{n} \ssum_{i=1}^n |x_i - T|,
\end{equation}
where $<<$ stands for \textit{significantly smaller than}.
The wisdom of the crowd given by equation~\ref{eq:wisdom_of_crowds} does not always take place. \samrevieww{It only occurs if the opinions $x_i$ are distributed sufficiently symmetrically around the true answer. When the distribution is largely biased above or below the true answer, equation~\ref{eq:wisdom_of_crowds} fails to hold.}
To understand this fact, the group of individual is split in two : $i \in \NN^+$ if $x_i(1) > T$ and $i \in \NN^-$ if $x_i(1) < T$. Then, the distance of the mean opinion to truth rewrites as
\begin{eqnarray*}
|\bar{x} - T| &=& \left|\left(\frac{1}{n} \ssum_{i=1}^n x_i\right) - T \right|
= \frac{1}{n}\left| \ssum_{i=1}^n (x_i - T) \right|
= \frac{1}{n}\left| \ssum_{i\in \NN^+} (x_i - T) + \ssum_{i\in \NN^-} (x_i - T) \right|\\
&=& \frac{1}{n}\left| D^+ - D^- \right|
\end{eqnarray*}
where $D^+ = \sum_{i\in \NN^+} |x_i - T| \ge 0$ is the contribution from opinions above truth and $D^- = \sum_{i\in \NN^-} |x_i - T| \ge 0$ is the contribution from opinions below truth. Using these notation, the average distance to truth is
$\frac{1}{n} \sum_{i=1}^n |x_i - T| = \frac{1}{n}(D^+ + D^-)$. As a consequence, the wisdom of the crowd described in equation~\ref{eq:wisdom_of_crowds} translates to
\begin{equation}\label{eq:wisdom_of_crowds-D}
\left| D^+ - D^- \right|  << (D^+ + D^-).
\end{equation}
Two extreme cases are possible :
\begin{itemize}
 \item \textbf{Perfect wisdom of the crowd :} opinions are homogeneously distributed around the true answer and $D^+ = D^-$ so that $|\bar{x}(1) - T| = 0$.
 \item \textbf{No wisdom of the crowd :} opinions either totally overestimate or totally underestimate the correct answer and either $D^- = 0$ or $D^+ = 0$, so that $|\bar{x} - T| = \frac{1}{n} \ssum_{i=1}^n |x_i - T|$.
\end{itemize}
}

\begin{figure}[!ht]
\begin{center}
\begin{tabular}{cc}
\textbf{(A) Gauging} & \textbf{(B) Counting}\\
\vspace{0.1cm}
\includegraphics[clip=true,trim=0.2cm 0cm 0.7cm 0cm,scale=0.5]{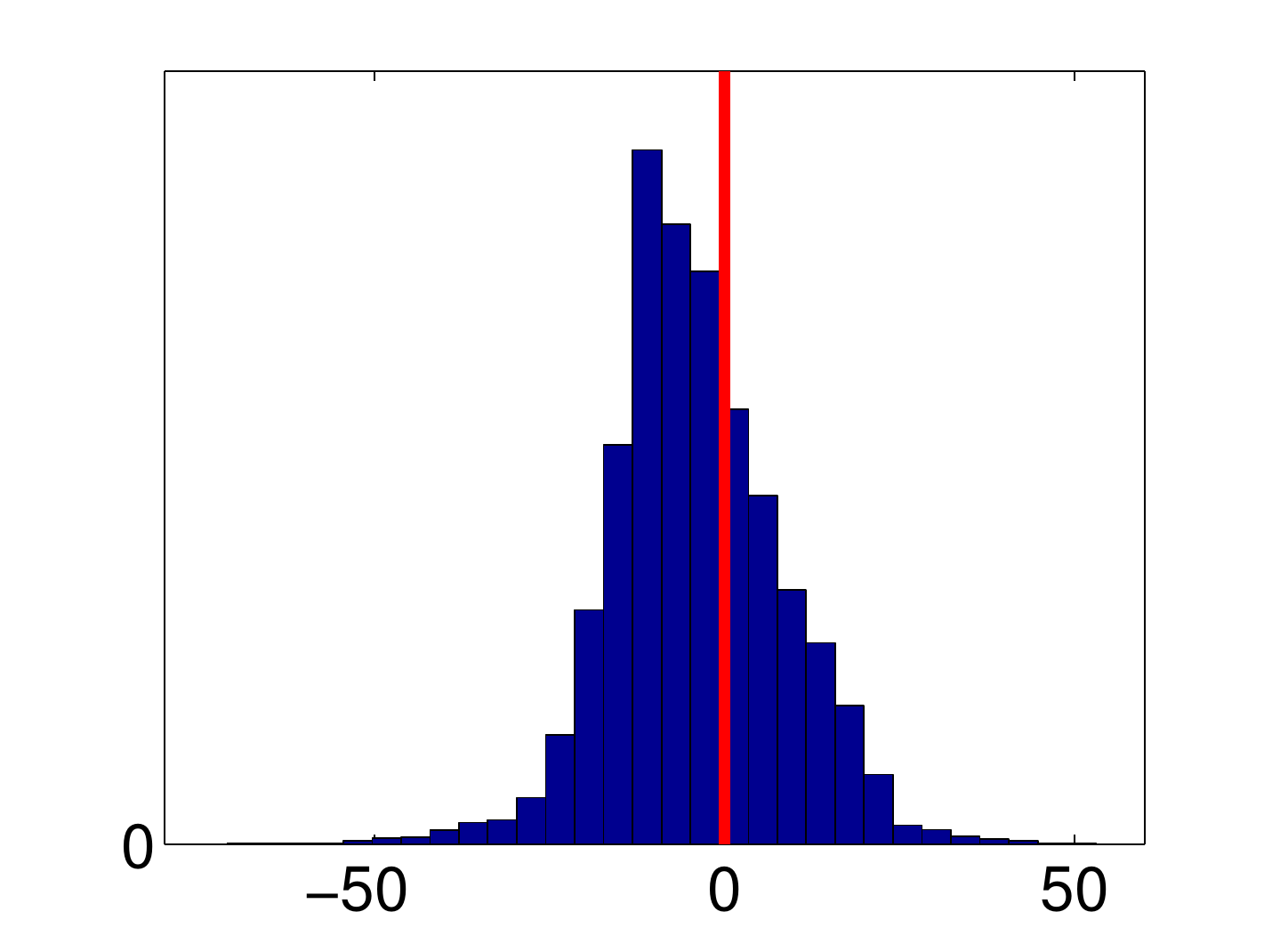} &
\includegraphics[clip=true,trim=0.2cm 0cm 0.7cm 0cm,scale=0.5]{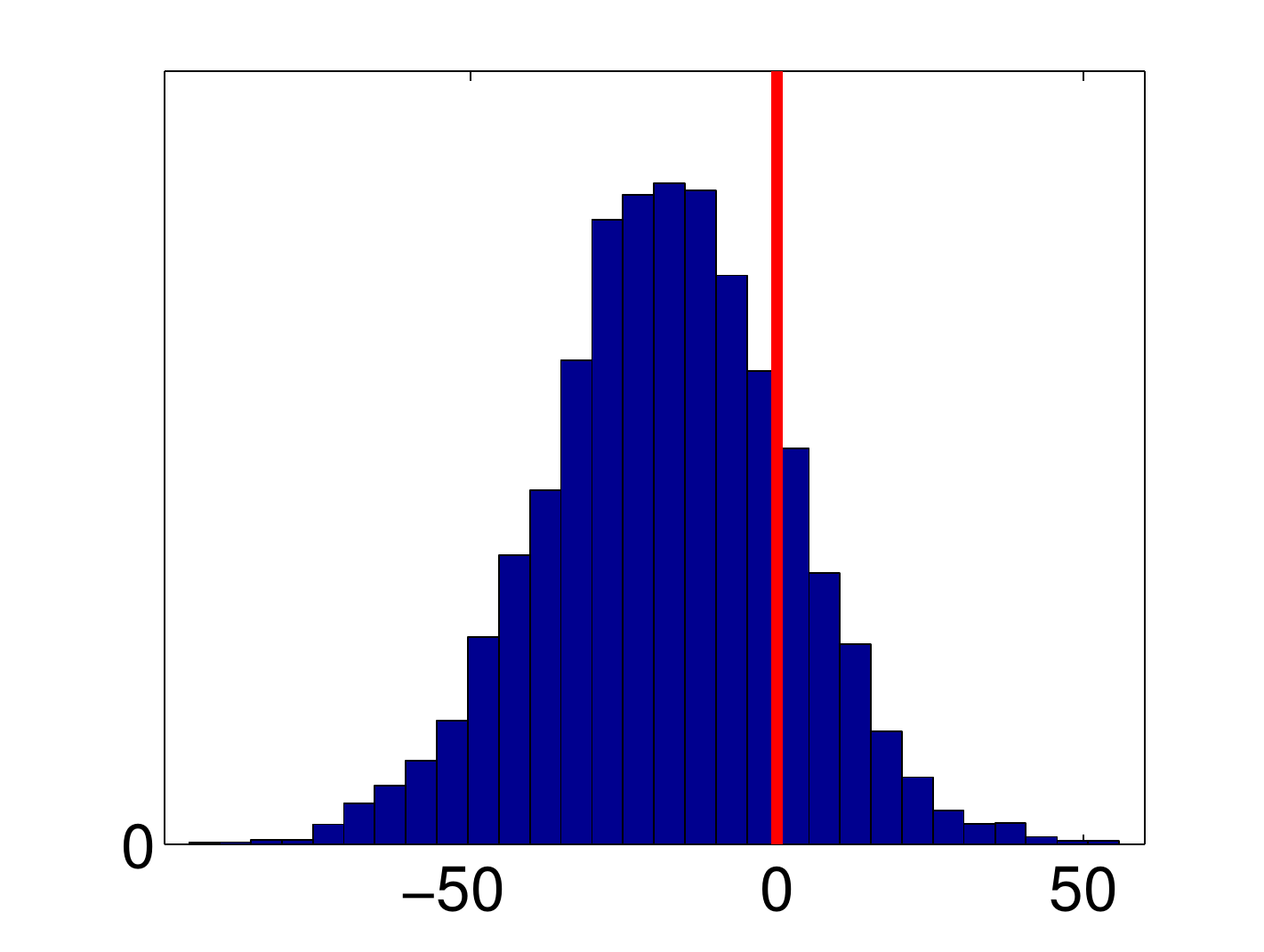}
\end{tabular}
\end{center}
\caption{Deviation of opinions to truth during the lone round $1$. Red vertical lines split the histogram into the opinions contributing negatively to the distance between mean opinion and truth (left) and the opinions contributing positively to the distance between mean opinion and truth (right). (A) \gauginggame ; (B) \countinggame.}\label{fig:Xpers_distance_to_truth}
\end{figure}

\samreview{
We now turn to the empirical data. Only the first round is discussed here because, in the subsequent rounds, the opinions are no longer independent, a criterion required for the wisdom of the crowd to occur.
Fig. \ref{fig:Xpers_distance_to_truth} displays how opinions are distributed around the true value for the \gauginggame~(A) and the \countinggame~(B). Both distributions fall between the two extreme cases with most opinions underestimating the true value. However, the bias is more important in the \countinggame~which explains that the wisdom of the crowd is more prominent in the \gauginggame~in the first round. This explains the differences between mean opinion errors and individual errors observed in Fig.~\ref{fig:RMSE_Xpers_distance_to_truth}.
}

\section{\samreview{Testing the linearity of the consensus model}}\label{SI-sec:linearity-test}

\begin{figure}[!ht]
\begin{center}
\begin{tabular}{ccc}
&\textbf{(A) Gauging} & \textbf{(B) Counting}\vspace{0.1cm}\\
\multicolumn{2}{l}{Opinion change} &\vspace{0cm}\\
\rotatebox[origin=c]{90}{\hspace{5cm} from round 1 to 2} &
\includegraphics[clip=true,trim=1cm 0cm 0.7cm 0.5cm,scale=0.5]{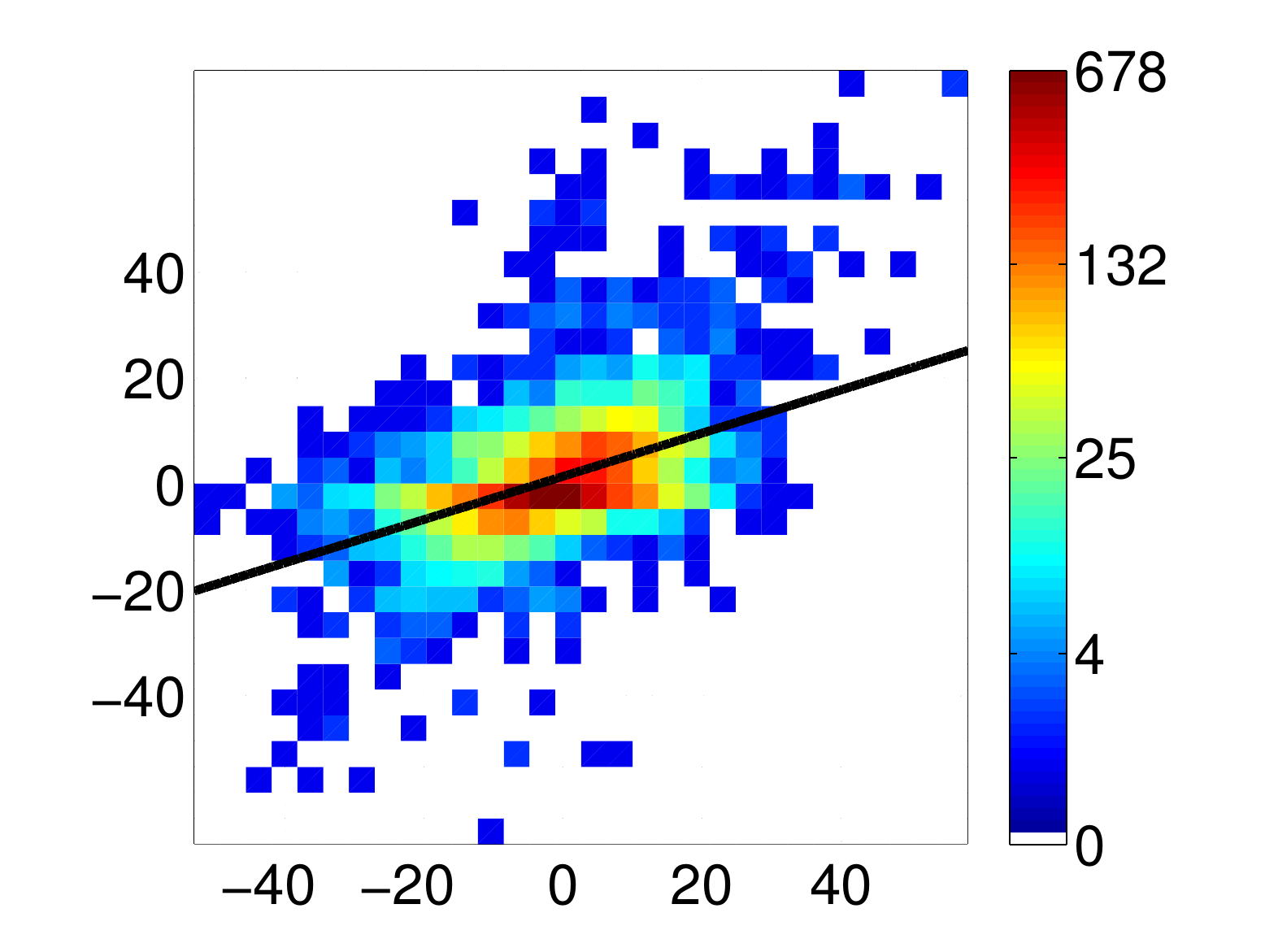} &
\includegraphics[clip=true,trim=1cm 0cm 0.7cm 0.5cm,scale=0.5]{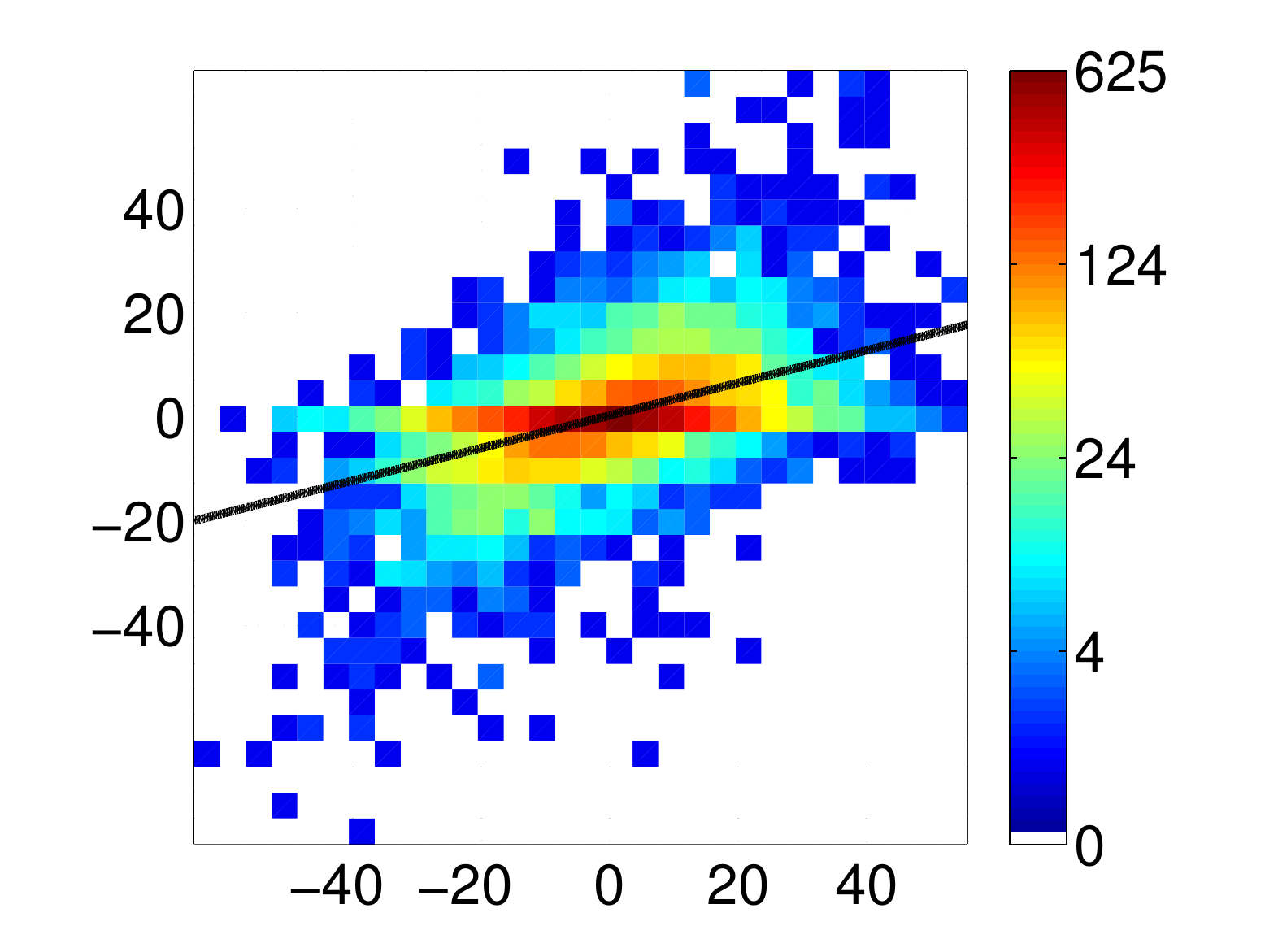}
\vspace{-3.5cm}
\\
\rotatebox[origin=c]{90}{\hspace{5cm} from round 2 to 3} &
\includegraphics[clip=true,trim=1cm 0cm 0.7cm 0.5cm,scale=0.5]{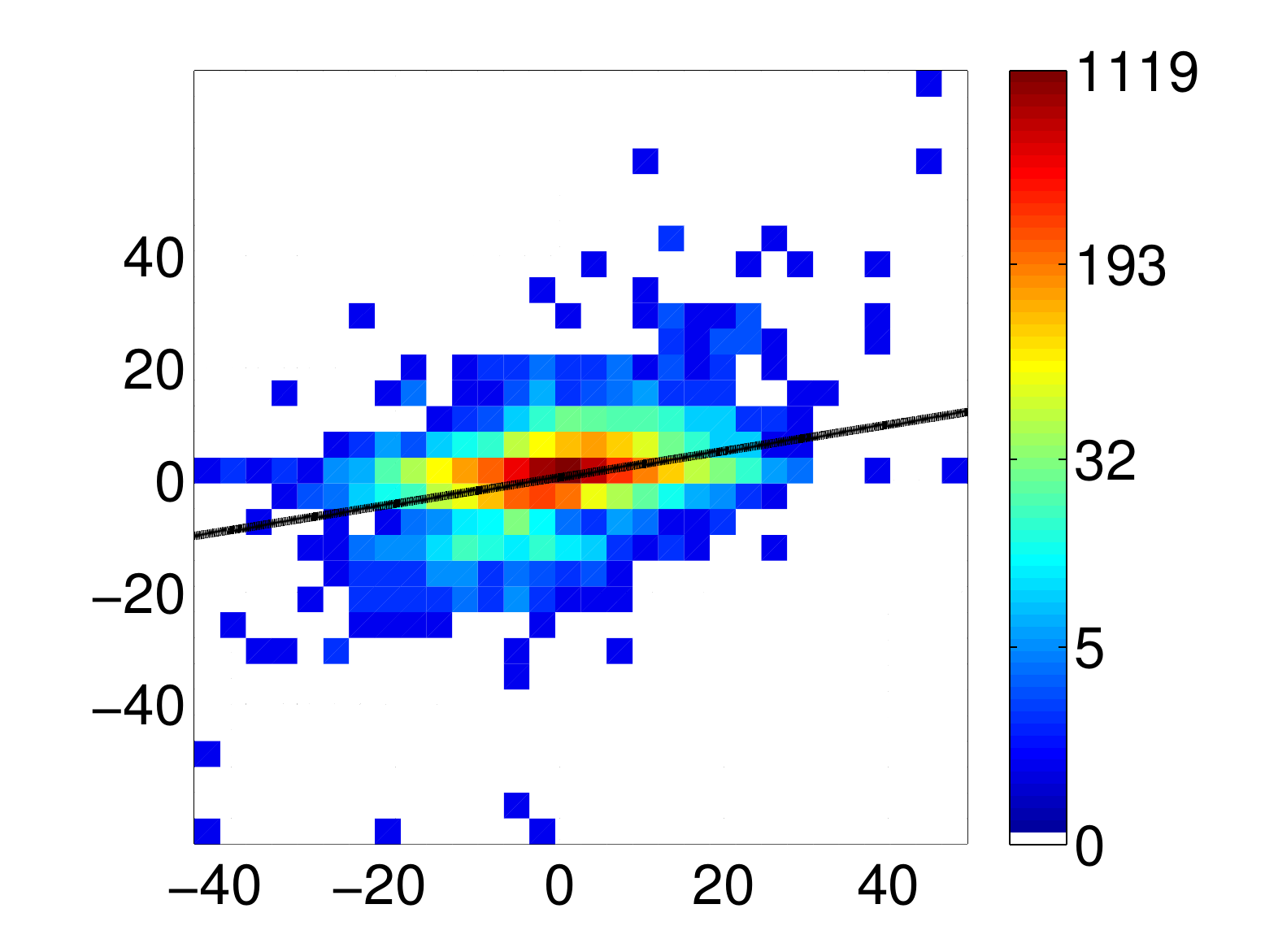} &
\includegraphics[clip=true,trim=1cm 0cm 0.7cm 0.5cm,scale=0.5]{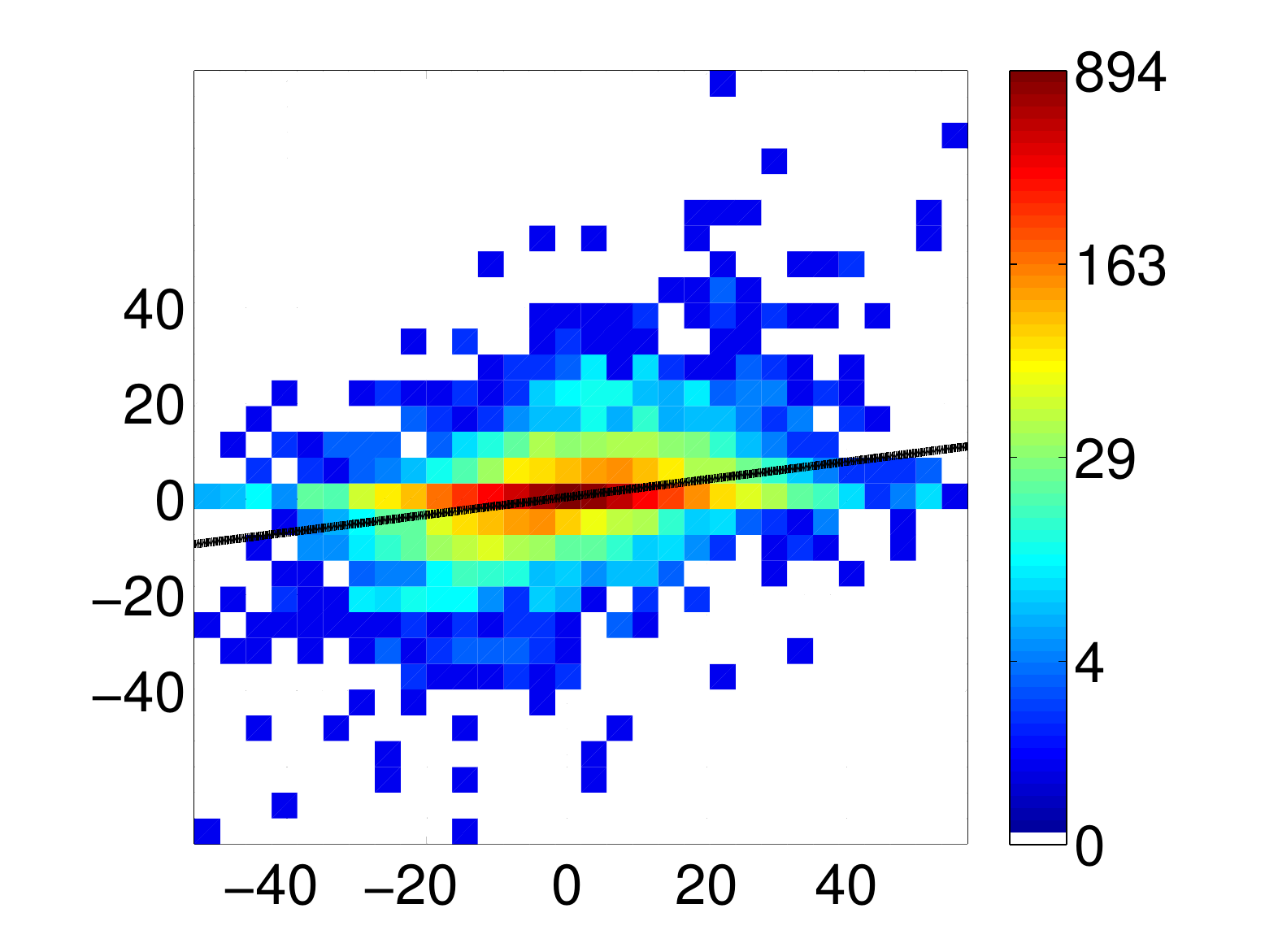}\vspace{-3.8cm}\\
\multicolumn{3}{c}{Difference between mean and own opinion}\\
\end{tabular}
\end{center}
\caption{\samreview{Opinion change $x_i(t+1) - x_i(t)$ versus difference between mean and individual opinion $\bar{x}(t) - x_i(t)$. The color in each cell corresponds to the number of data points falling in the cell. The color scale is logarithmic. The black strait line represents the linear regression $x_i(t+1) - x_i(t) = \beta_0 + \alpha(t)(\bar{x}(t) - x_i(t))$. Top : opinion change from round $1$ to $2$ ; bottom : opinion change from round $2$ to $3$. (A) \gauginggame ; (B) \countinggame.}}\label{fig:linear-model}
\end{figure}

\samreview{
The consensus model~\eqref{eq:consensus-alpha-to-mean} assumes that the opinion change $x_i(t+1) - x_i(t)$ grows linearly with the distance between $x_i(t)$ and the mean opinion $\bar{x}(t)$. This assumption is tested against the alternative 
$$x_i(t+1) - x_i(t) = \beta_0 + \alpha_1 (\bar{x}(t) - x_i(t))^\gamma$$
with $\gamma \neq 1$. 
The numerical statistics values are reported for the opinion change between rounds $1$ and $2$ for the \gauginggame. The same conclusions hold for the \countinggame~and for the opinion change between rounds $2$ and $3$.
The linearity test provided in~\cite{niermann2007testing} applied to our data gives a statistics $P = -1.4 \text{\sc{e}} 7$ with empirical variance $\text{var}(P) = 4 \text{\sc{e}} 14$ so that we fail to reject the null hypothesis $\gamma = 1$ (p-val=0.5). Fig.~\ref{fig:linear-model} displays the evolution $x_i(t+1) - x_i(t)$ against the distance to the mean $\bar{x}(t) - x_i(t)$ along with the result of the linear regression assuming $\gamma=1$.
}

\section{\samreview{Influenceability and personality}}\label{SI-sec:unpredictability-derivation}

\samreview{Is influenceability related to personality ? To answer this question, we required the participants to provide information regarding their personality, gender, highest level of education, and whether they were native English speaker. The questionnaire regarding personality comes from a piece of work by Gosling and Rentfrow~\cite{gosling2003very} and was used to estimate the five general personality traits. The questionnaire page is reported in supplementary Fig.~\ref{fig:questionnaire}. For each of the five traits, the participants rated how well they feel in adequacy with a set of synonyms (rating $s \in \{1,\ldots,7\})$ and with a set of antonyms (rating $a \in \{1,\ldots,7\}$. This redundancy allows for testing the consistency of the answer of each participants. The participants who had a distance $(8-a) - s$ too far away from $0$ were discarded (threshold values were found using Iglewicz and Hoaglin method based on median absolute deviation~\cite{iglewicz1993detect}). Partial Pearson's linear correlations are first reported between the individual traits measured by the questionnaire (see table~\ref{table:self-big-5}). The correlation signs are found to be consistent with the related literature on the topic~\cite{van2010general}. This indicates that our measure of the big five factors is trustworthy. Partial correlations are then provided to link the personal traits to influenceability. As shown in Table~\ref{table:big-5-to-incluenceability}, none of the measured personal traits is able to explain the variability in the influenceability parameter. The only exceptions concern gender and being English native speaker, with weak level of significance (p-$val \in [0.01,0.05]$). However, these relations are consistent neither between types of tasks nor over rounds, so that they cannot be trusted. We conclude that the big five personality factors and the other measured individual traits are not relevant to explain the influenceability parameter. Finding appropriate individual traits to explain the influenceability remains an open question.
}

\begin{table}
\begin{center}
\textbf{(A) Participants from the \countinggame}\\
\begin{tabular}{|l|l|l|l|l|l|l|l|}
\hline
 &C&E&A&N&Gen&Eng&Edu\\\hline
O&0.45***&0.38***&0.31***&-0.35***&0.03&-0.08&0.09\\\hline
C&&0.11&0.29***&-0.41***&-0.02&-0.06&0.29***\\\hline
E&&&0.03&-0.15*&0.02&-0.05&0\\\hline
A&&&&-0.48***&0.11&-0.16*&0.15*\\\hline
N&&&&&0.2**&0.08&-0.23***\\\hline
Gen&&&&&&0&0.01\\\hline
Eng&&&&&&&-0.12\\\hline
\end{tabular}\\
\textbf{(B) Participants from the \gauginggame}\\
\begin{tabular}{|l|l|l|l|l|l|l|l|}
\hline
 &C&E&A&N&Gen&Eng&Edu\\\hline
O&0.37***&0.24***&0.29***&-0.4***&0.06&0.15*&0.16**\\\hline
C&&0.2***&0.22***&-0.29***&0.15**&0.12*&0.11\\\hline
E&&&-0.12*&-0.12*&0.04&0.04&-0.03\\\hline
A&&&&-0.34***&0.01&-0.07&0.08\\\hline
N&&&&&0.18**&-0.02&-0.16**\\\hline
Gen&&&&&&-0.03&0\\\hline
Eng&&&&&&&-0.02\\\hline
\end{tabular}
\end{center}
\caption{\samreview{Partial Pearson's linear correlations among the big five factors of personality (O: openness, C : calmness, E : extroversion, A : agreeableness, N : neuroticism), gender (Gen), native English speaker (Eng) and highest level of education (Edu). Significance : *p-val$<0.05$, **p-val$<0.01$,***p-val$<0.001$.}
}\label{table:self-big-5}
\end{table}

\begin{table}
\begin{center}
\hspace{0.3cm} \textbf{(A) Gauging} \hspace{0.7cm} \textbf{(B) Counting}\vspace{0.1cm}\\
\begin{tabular}{|l|l|l|l|}
\hline
 &$\alpha(1)$&$\alpha(2)$\\\hline
O&-0.1&-0.04\\\hline
C&0.08&-0.06\\\hline
E&-0.1&0.02\\\hline
A&0&-0.02\\\hline
N&-0.06&0.03\\\hline
Gen&-0.14*&-0.05\\\hline
Eng&-0.09&0.02\\\hline
Edu&0.05&-0.1\\\hline
\end{tabular}
\begin{tabular}{|l|l|l|l|}
\hline
$\alpha(1)$&$\alpha(2)$\\\hline
0.07&-0.05\\\hline
-0.05&-0.09\\\hline
-0.01&-0.01\\\hline
0.1&0.07\\\hline
-0.1&0.01\\\hline
0.03&0.12*\\\hline
-0.03&-0.12*\\\hline
-0.04&-0.1\\\hline
\end{tabular}
\end{center}
\caption{\samreview{Partial Pearson's linear correlations linking influenceability to the big five factors of personality (O: openness, C : calmness, E : extroversion, A : agreableness, N : neuroticism), gendre (Gen), native English speaker (Eng) and highest level of education (Edu). Significance : *p-val$<0.05$, **p-val$<0.01$,***p-val$<0.001$.}
}\label{table:big-5-to-incluenceability}
\end{table}

\newgeometry{left=1.5cm,bottom=1cm}

\section{\cocoreview{Additional figures for prediction accuracy}}\label{SI-sec:error bars}

\subsection{\cocoreview{Confidence intervals for prediction errors}}
\begin{figure*}[!h]
    \centering
    \begin{subfigure}[b]{0.475\textwidth}
        \centering
        \textbf{(A) Gauging - RMSE, detailed error bars}\\
        \vspace{0.3cm} 
        \includegraphics[width=\textwidth]{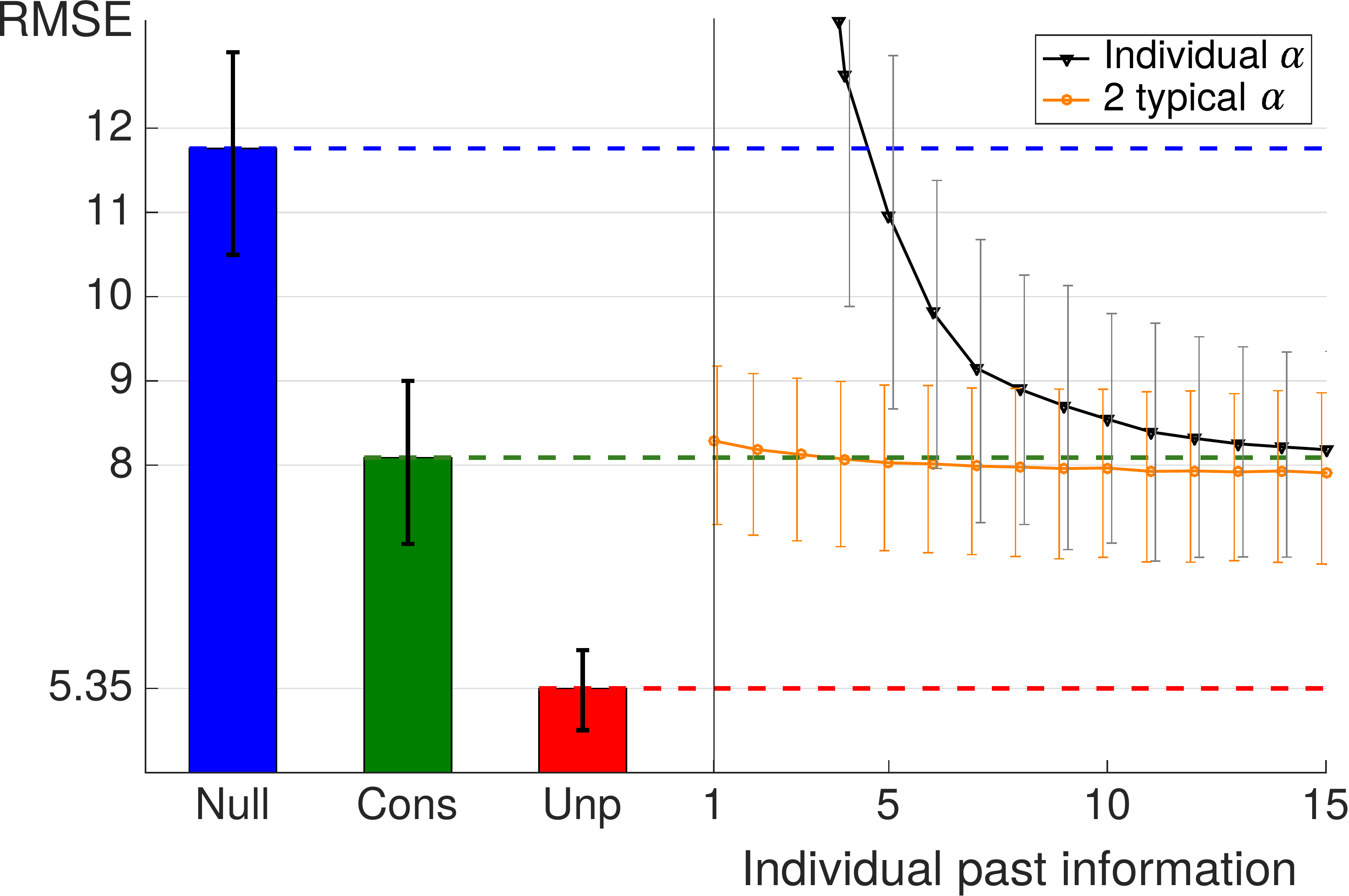}
   
        \label{fig:gaug-err}
    \end{subfigure}
    \hfill
    \begin{subfigure}[b]{0.475\textwidth}  
        \centering 
        \textbf{(B) Counting - RMSE, detailed error bars}\\
        \vspace{0.3cm}  
        \includegraphics[width=\textwidth]{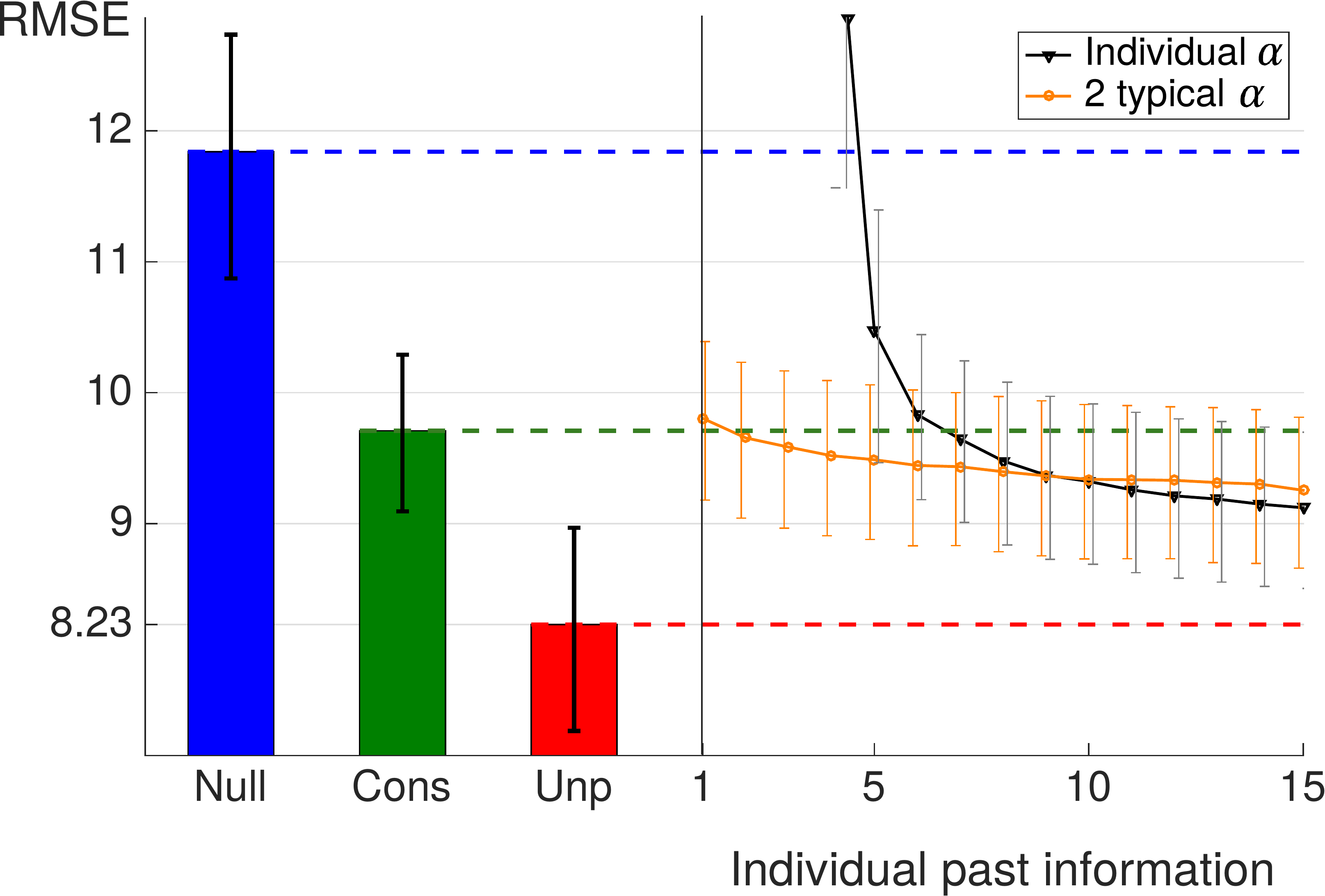}
  
        \label{fig:count-err}
    \end{subfigure}
    \vspace{0.5cm}
    \caption[]
    {\textbf{Root mean square error (\textit{RMSE}) of the predictions with detailed error bars for the final round.}}
      \label{fig:errbar}
\end{figure*}

\samrevieww{Fig. \ref{fig:errbar} displays error bars for $95\%$ confidence interval of the RMSEs. This figure reveals that the two methods depending on training set size do not perform significantly better than the consensus model with one couple of typical influenceabilities, even for large training set sizes. This is an argument to favor the model in which the whole population has a unique couple of influenceabilities $(\alpha(1),\alpha(2))$.}

\subsection{\cocoreview{Prediction accuracy in terms of Mean Absolute Errors}}
\begin{figure*}[!h]
    \begin{subfigure}[b]{0.475\textwidth}   
        \centering 
        \textbf{(C) Gauging - MAE}\\

        \includegraphics[width=\textwidth]{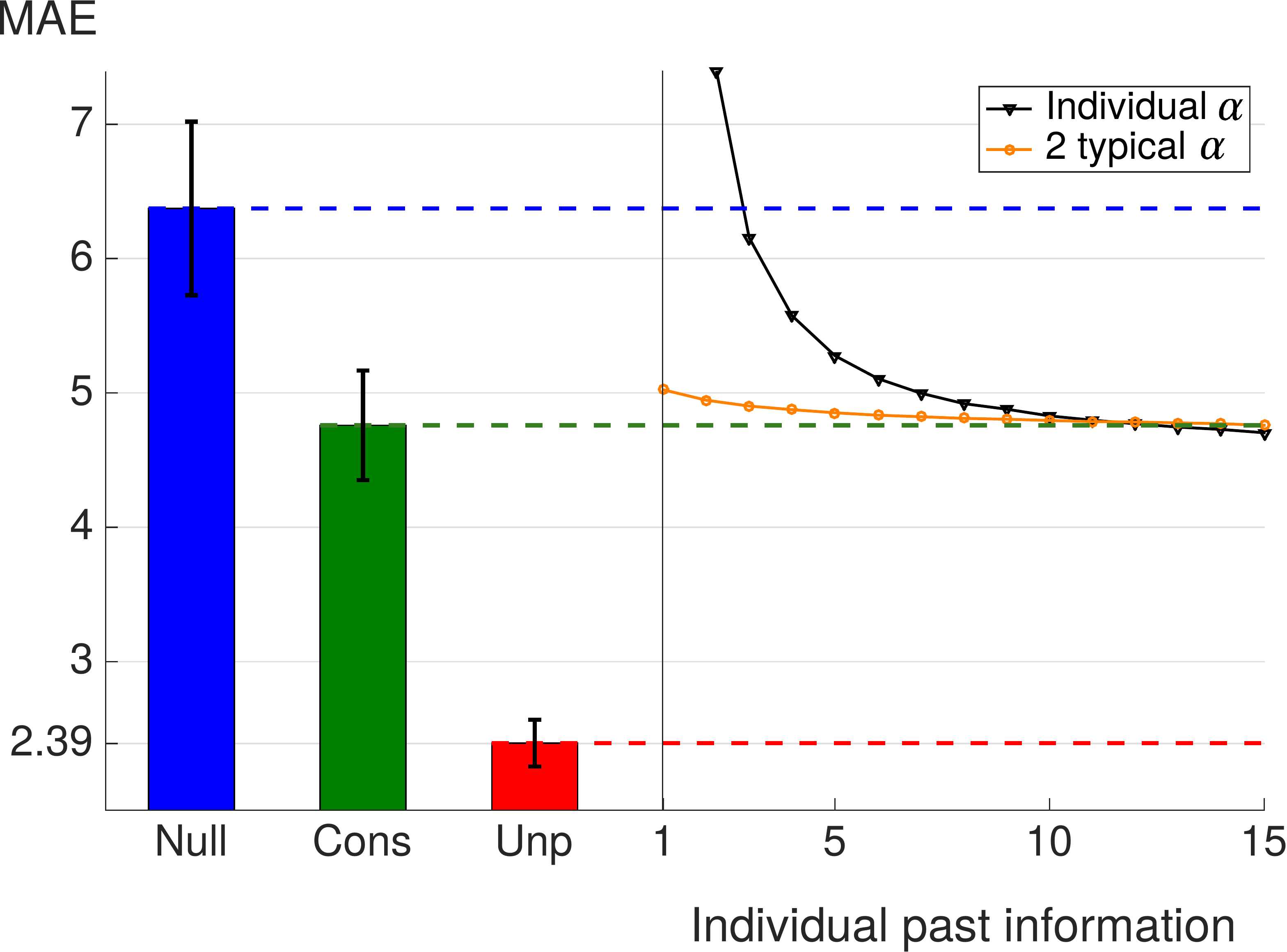}
  
        \label{fig:gaug-mae}
    \end{subfigure}
    \quad
    \begin{subfigure}[b]{0.475\textwidth}   
        \centering 
        \textbf{(D) Counting - MAE}\\

        \includegraphics[width=\textwidth]{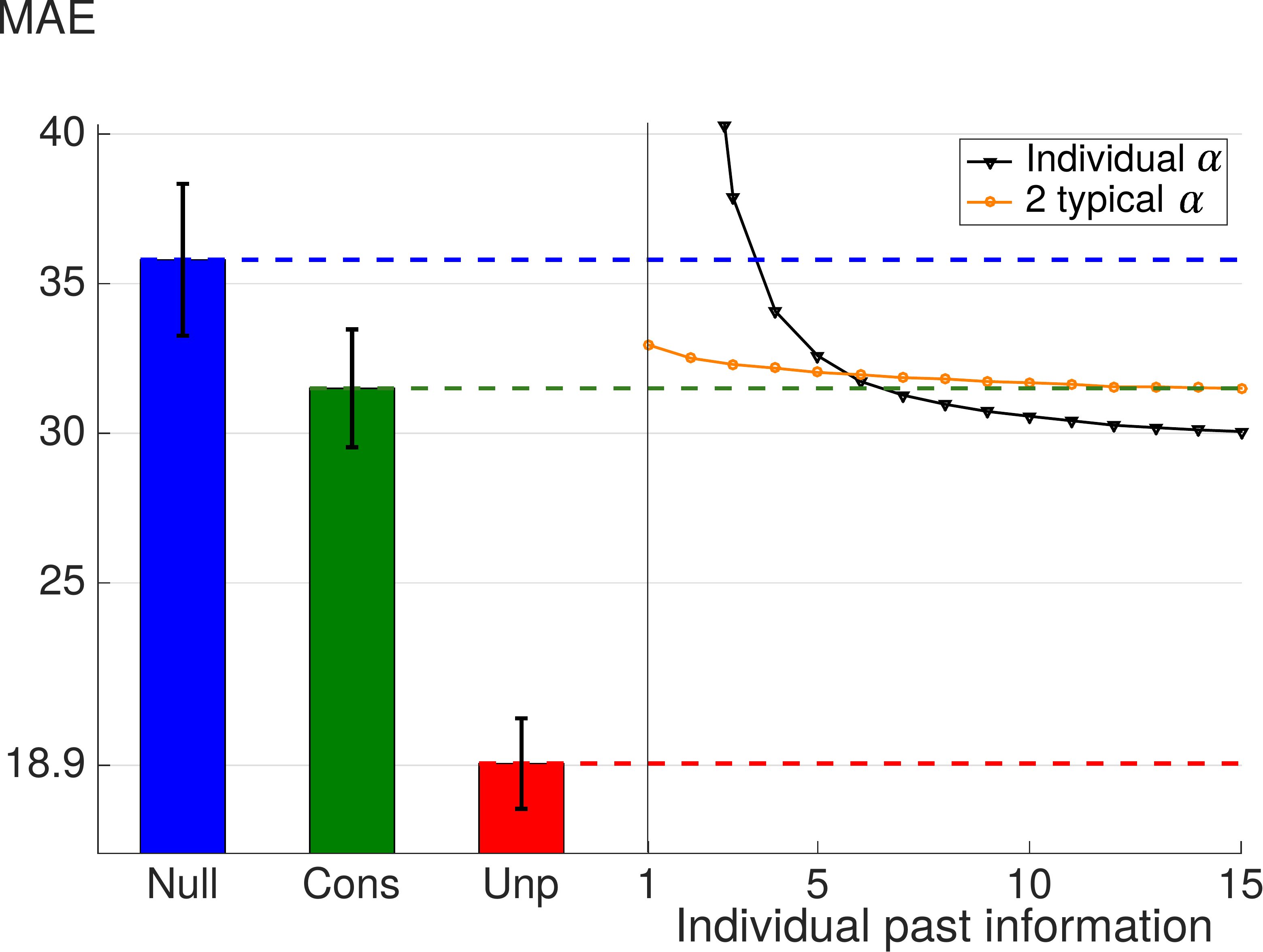}
  
        \label{fig:count-mae}
    \end{subfigure}
    \vspace{0.5cm}
    \caption[]
    {\textbf{Mean absolute error (\textit{MAE}) of the predictions (unscaled) for the final round.}}
    \label{fig:mae}
\end{figure*}

\samrevieww{Measuring prediction accuracy in terms of MAEs may appear more intuitive for comparing prediction methods. Fig. \ref{fig:mae}. assesses the models using an absolute linear scale, where the errors are deliberately unscaled for the \countinggame. The prediction methods rank equally when measured in terms of MAE or RMSE. Notice that, due to nonlinear relation between RMSE and MAE, on this alternative scale, the consensus models errors are now closer from the null model than from the unpredictability error. For comparison, recall that for the \gauginggame, the judgments range between $0$ and $100$ while they range between $0$ and $500$ for the \countinggame.}

\newpage

\begin{figure}[!ht]
\centering
\includegraphics[clip=true,trim=0cm 0cm 0cm 0cm,scale=0.5]{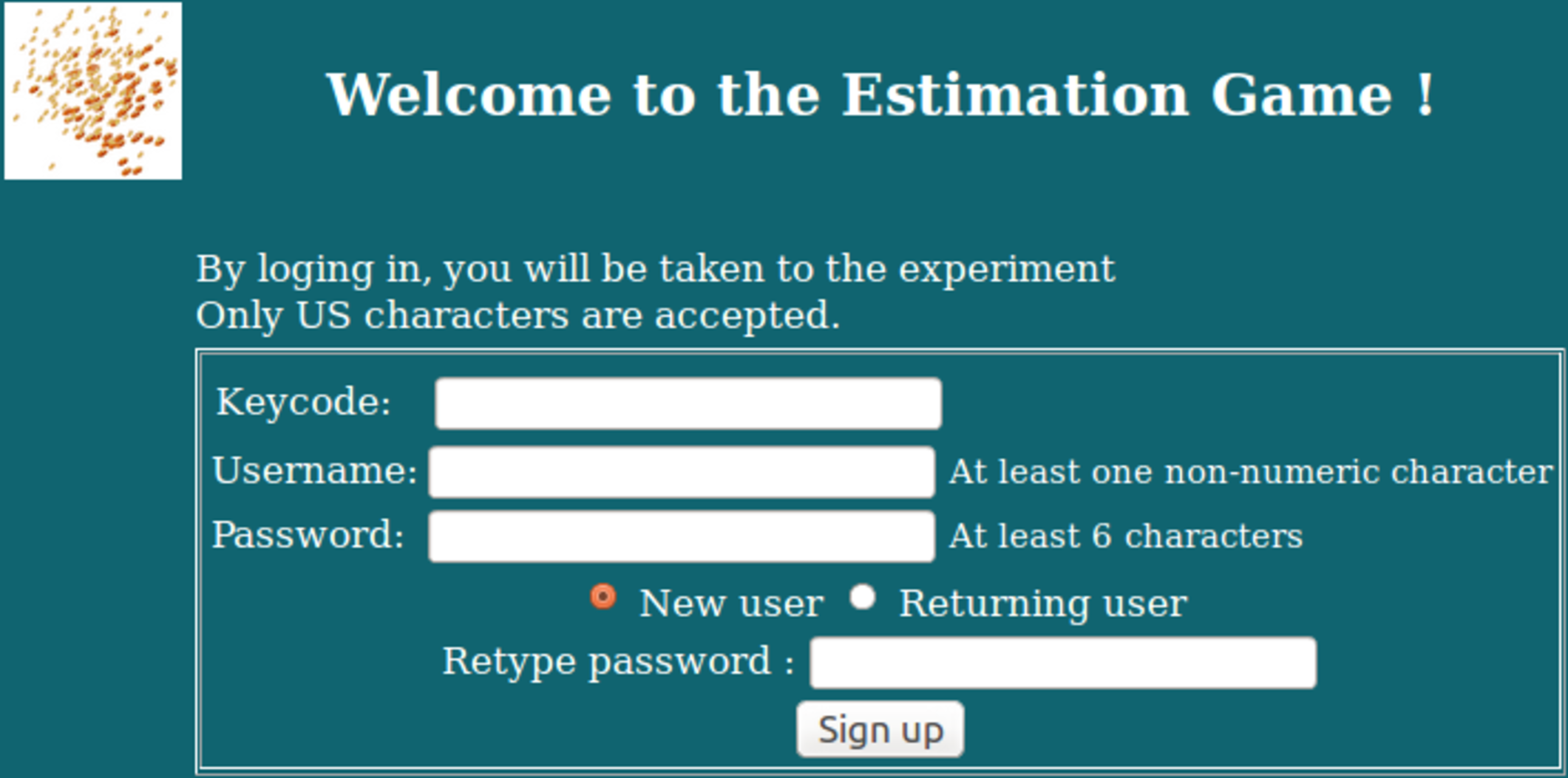}\\
\caption{Login page.}\label{fig:login}
\end{figure}


\begin{figure}[!ht]
\centering
\includegraphics[clip=true,trim=0cm 0cm 0cm 0cm,scale=0.7]{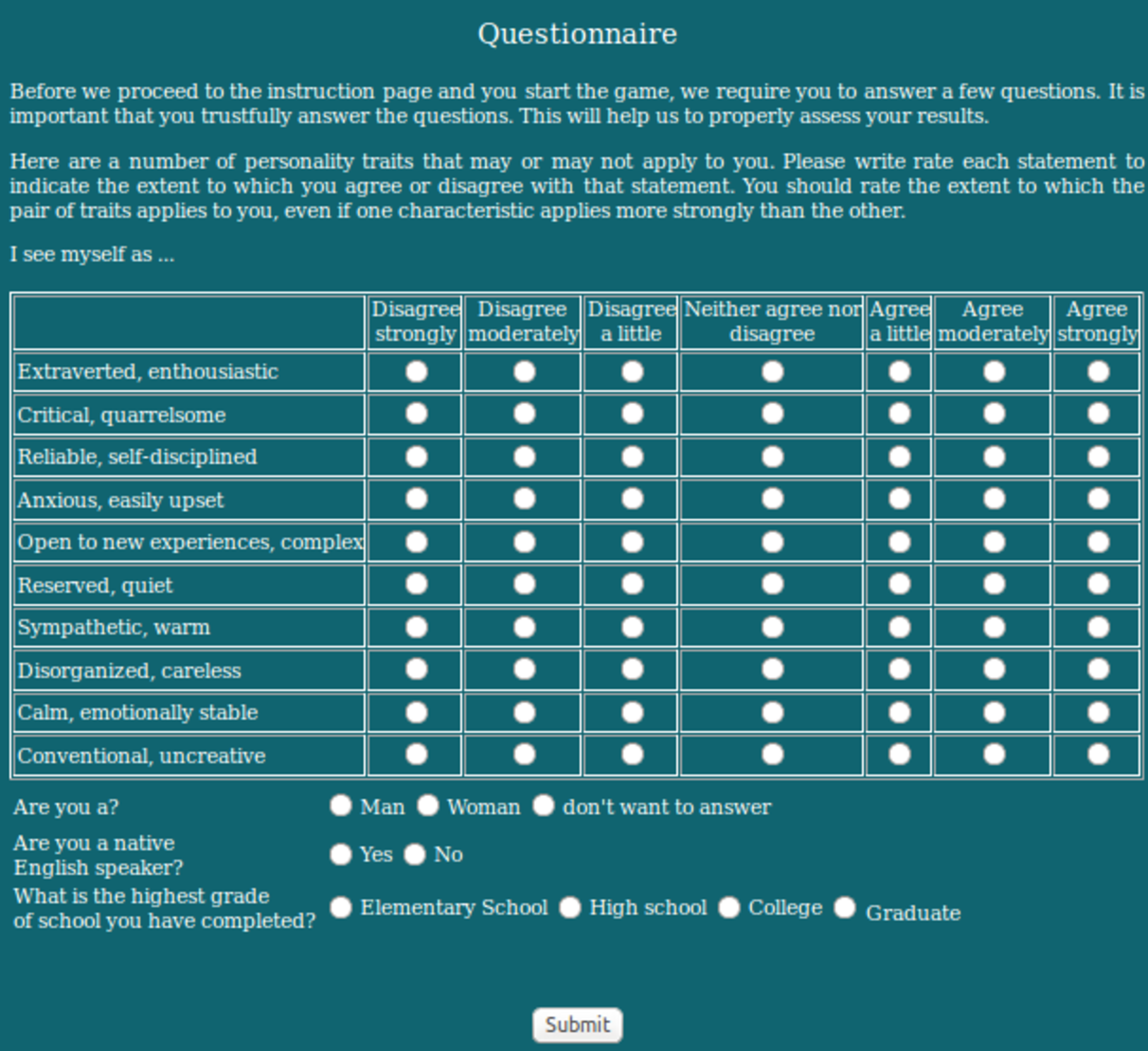}\\
\caption{Questionnaire form.}\label{fig:questionnaire}
\end{figure}

\begin{figure}[!ht]
\centering
\includegraphics[clip=true,trim=0cm 0cm 0cm 0cm,scale=0.7]{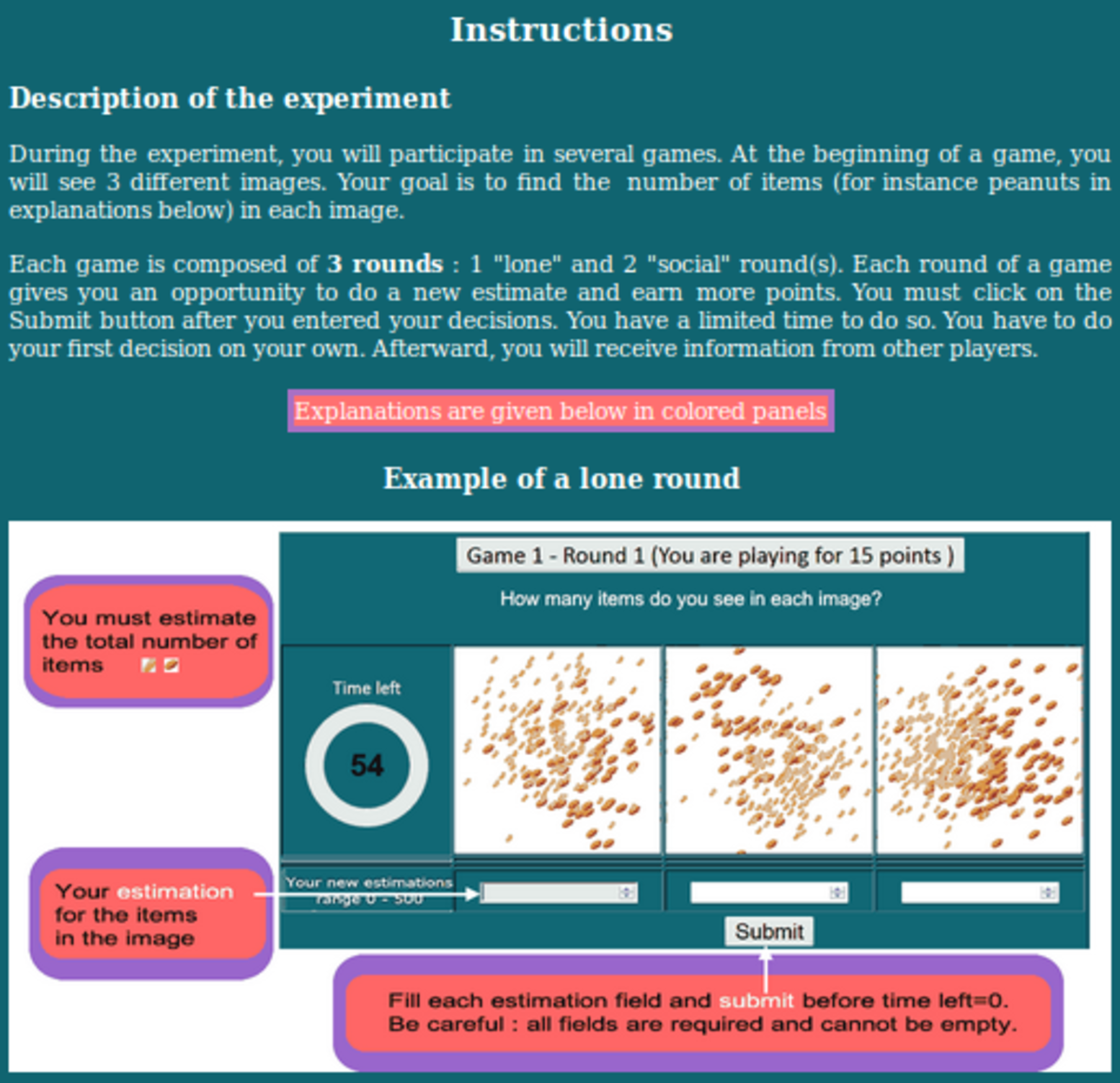}\\
\includegraphics[clip=true,trim=0cm 0cm 0cm 0cm,scale=0.7]{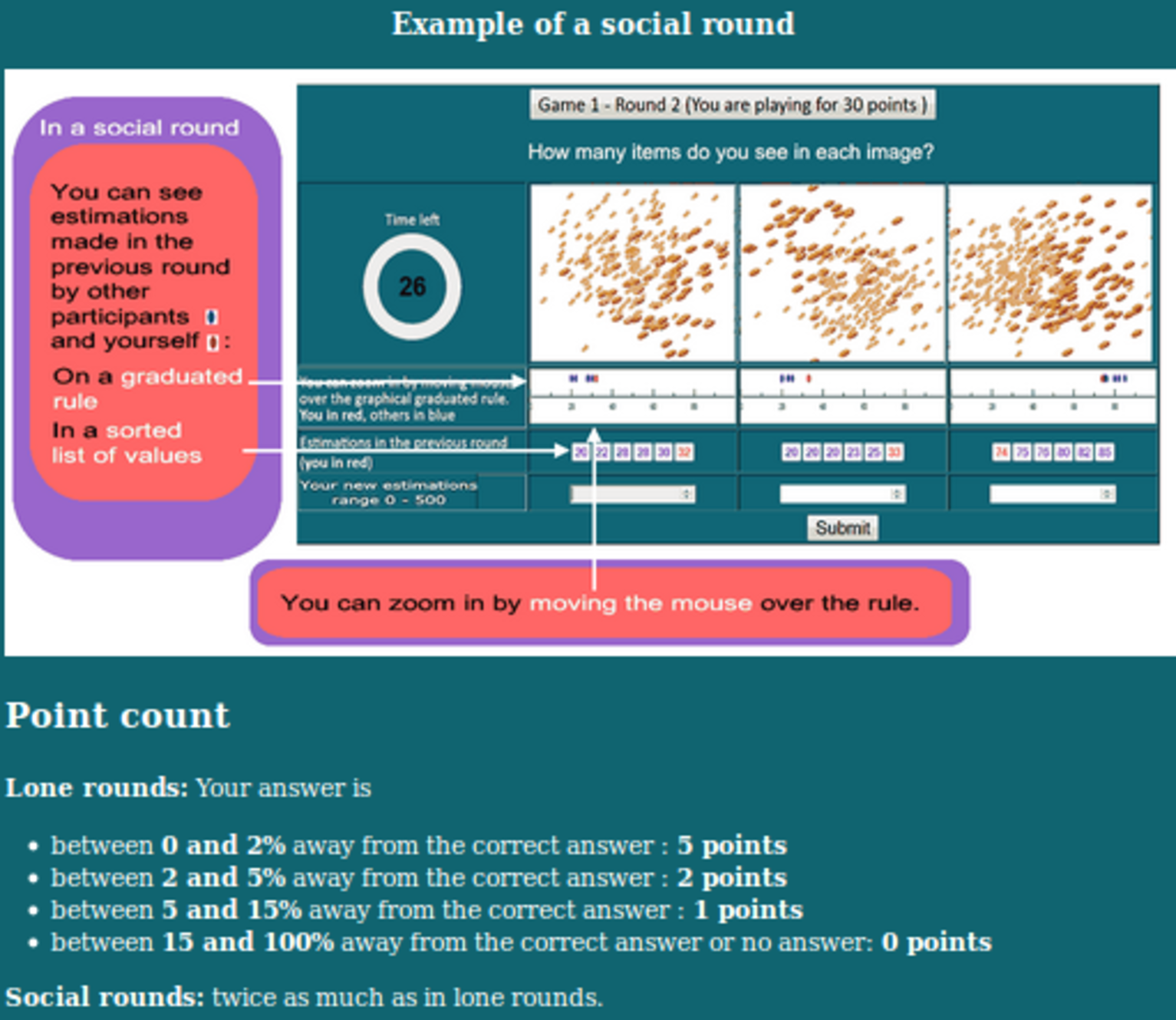}\\
\caption{Instruction page.}\label{fig:instructions}
\end{figure}


\begin{figure}[!ht]
\begin{center}
(A)\\
\vspace{0.3cm}
\includegraphics[clip=true,trim=0cm 0cm 0cm 0cm,scale=0.7]{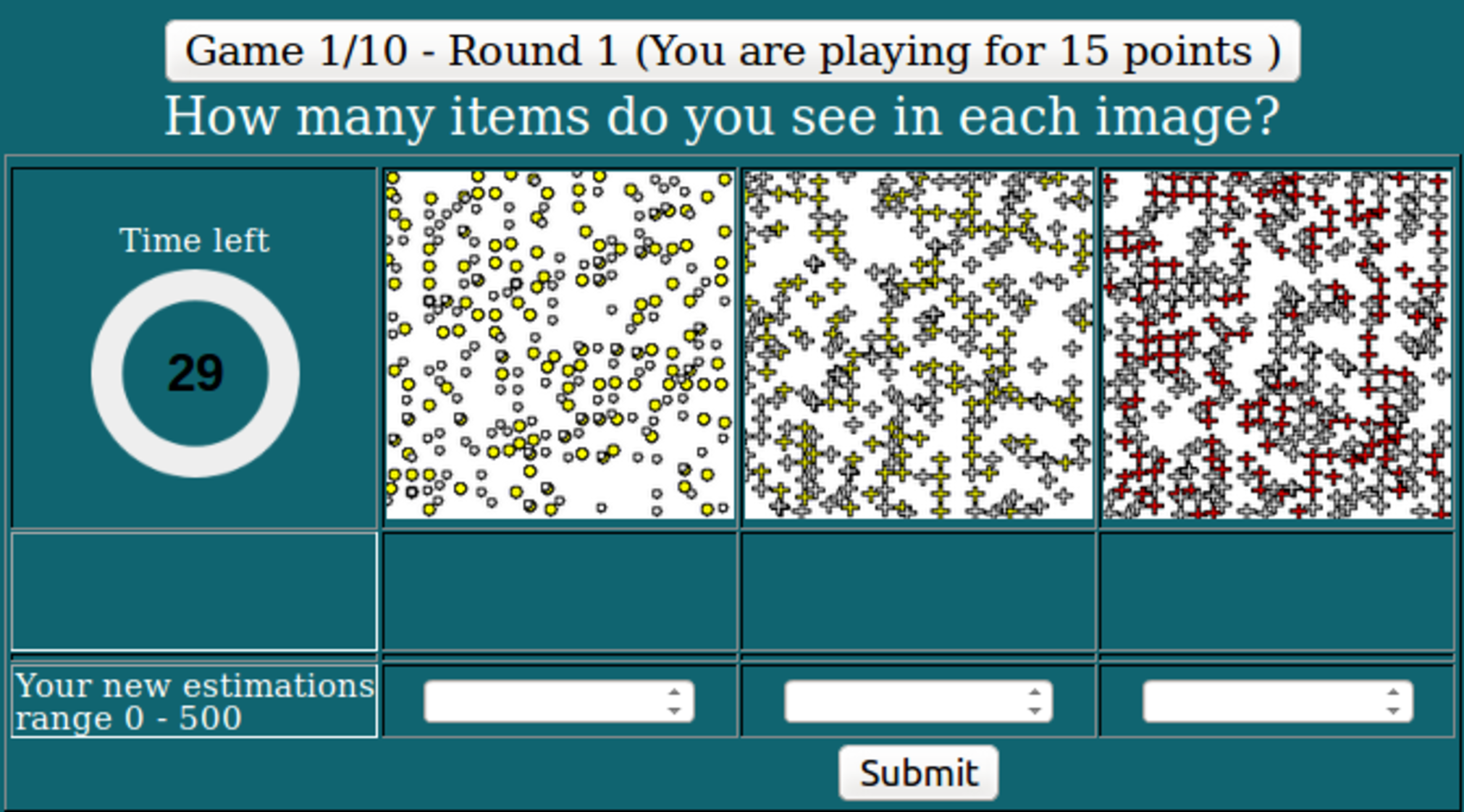}\\
\vspace{0.5cm}
(B)\\
\vspace{0.3cm}
\includegraphics[clip=true,trim=0cm 0cm 0cm 0cm,scale=0.6]{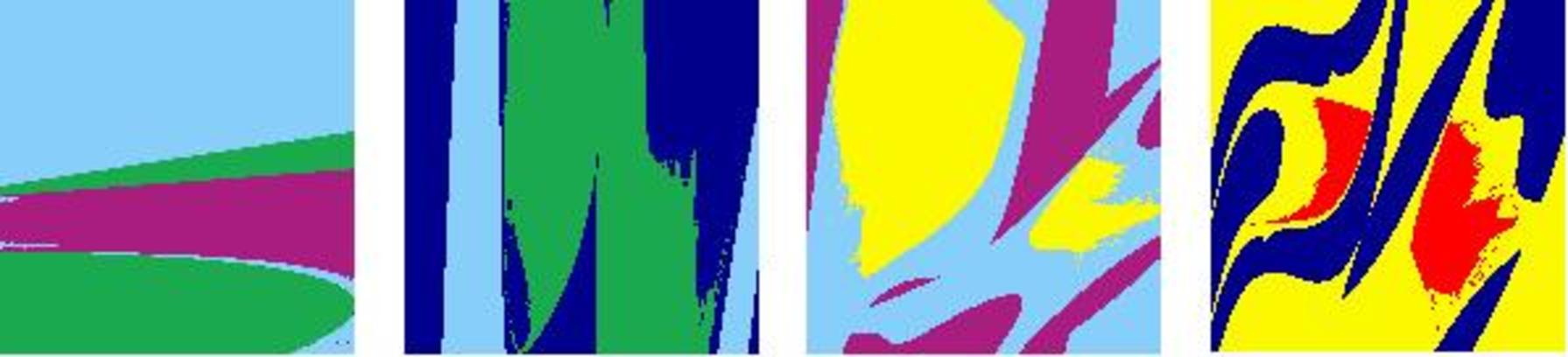}\\
\caption{(A) Interface of the first round for the \countinggame, played alone. (B) Instances of pictures for the \gauginggame.}\label{fig:lone-round}
\end{center}
\end{figure}


\begin{figure}[!ht]
\centering
\includegraphics[clip=true,trim=0cm 0cm 0cm 0cm,scale=0.6]{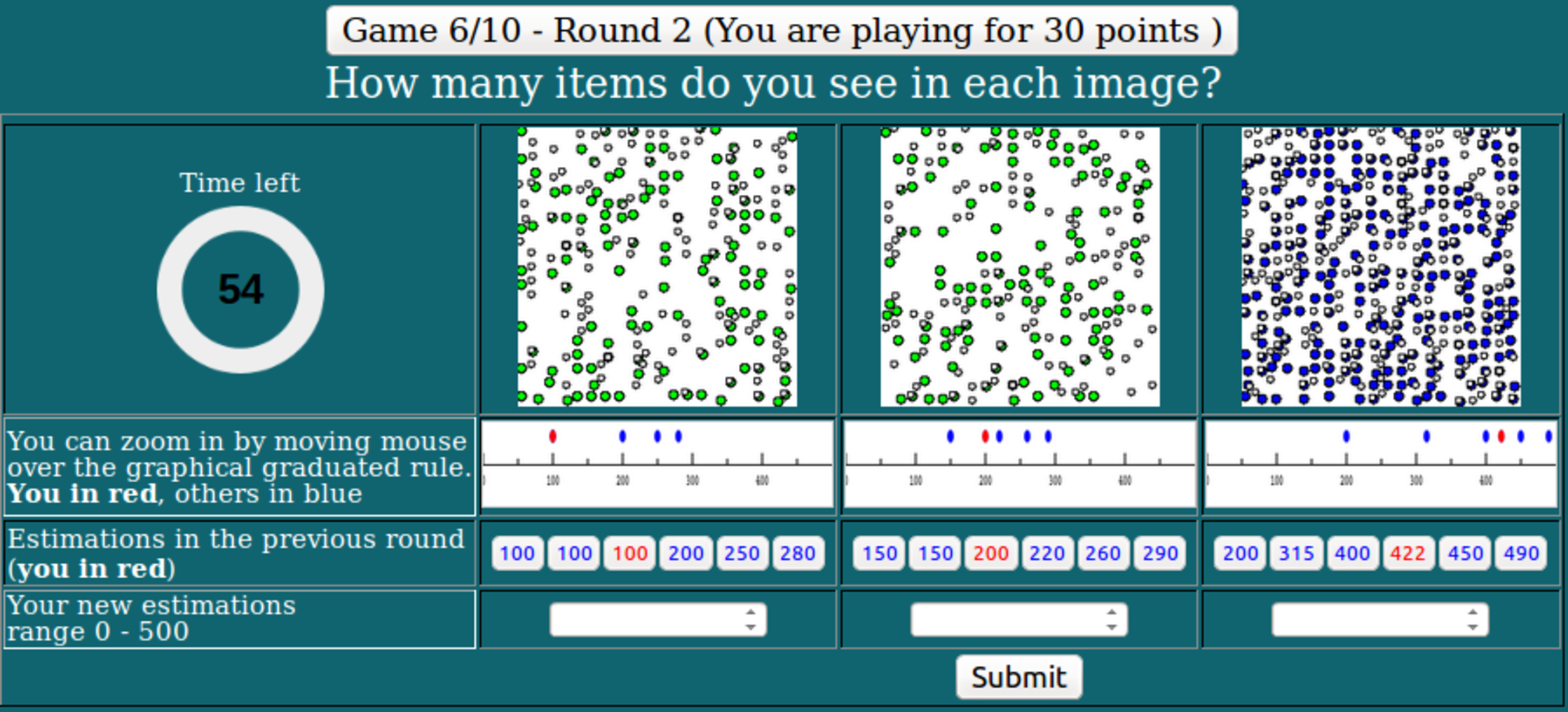}\\
\vspace{0.2cm}
\includegraphics[clip=true,trim=0cm 0cm 0cm 0cm,scale=0.6]{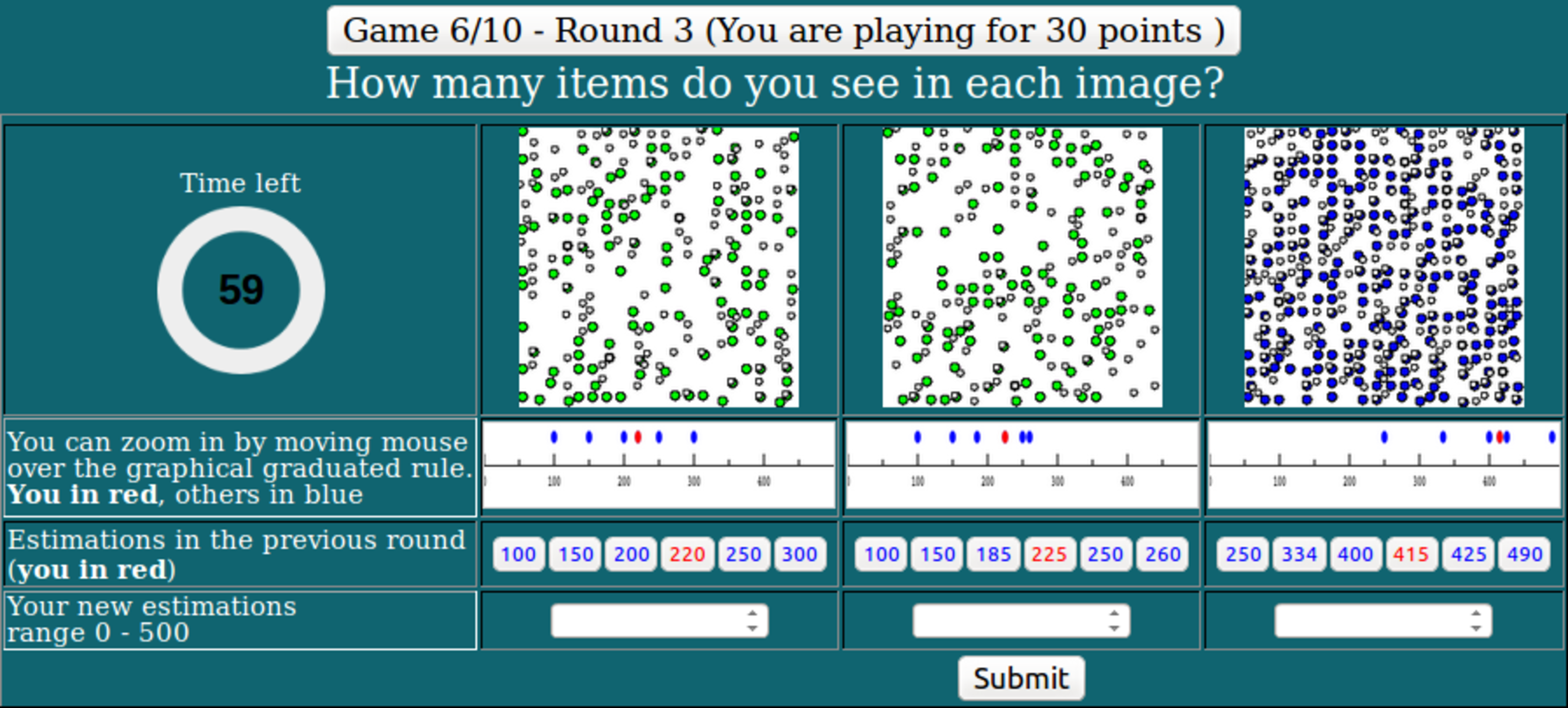}\\
\caption{Interface of the second and third rounds for the \countinggame, social rounds}\label{fig:social-round}
\end{figure}

\begin{figure}[!ht]
\centering
\includegraphics[clip=true,trim=0cm 0cm 0cm 0cm,scale=0.8]{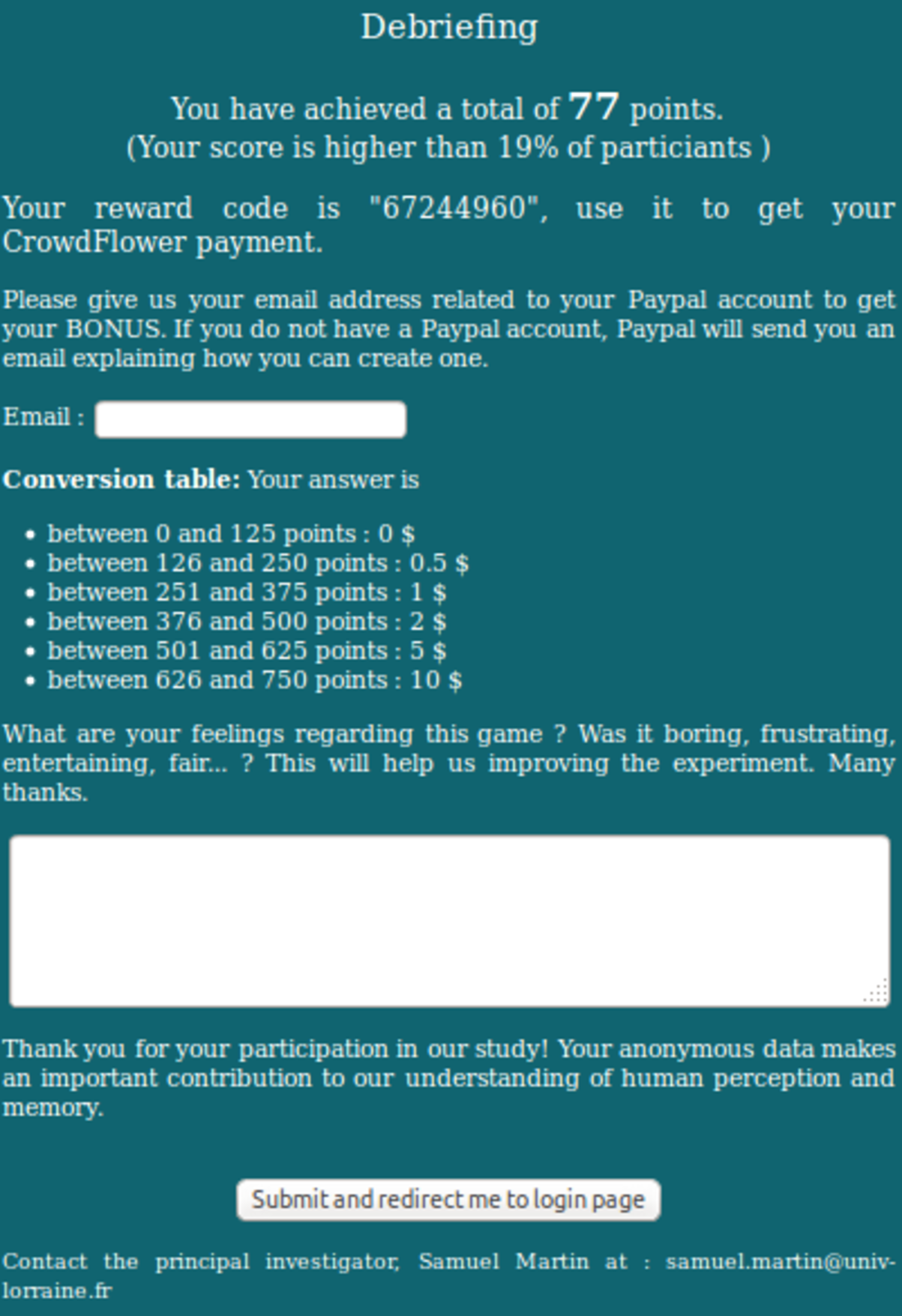}\\
\caption{Debrief page.}\label{fig:debrief}
\end{figure}



\end{document}

%% file: new_commands_and_packages.tex
\usepackage[T1]{fontenc}
\usepackage{color}
\usepackage{soul}
\usepackage{array}
\usepackage{graphicx} 
\graphicspath{{metric_paper/figures/}{}}
\usepackage{amsmath}
\usepackage{amsthm} 
\usepackage{comment}
\usepackage{amsfonts}
\usepackage{bbm}
\usepackage{amssymb}
\usepackage{dsfont}
\usepackage{mathtools}

\usepackage[autostyle]{csquotes}

\usepackage{placeins} 

\newcommand{\E}{\mathbb{E}}

\newcommand{\NN}{{\mathcal{N}}}

\newcommand{\ie}{{\it i.e.,~}}

\newcommand{\eg}{{\it e.g.,~}}

\newcommand{\corentin}[1]{\textcolor{black}{#1}}

\newcommand{\cocoreview}[1]{\textcolor{black}{#1}}
\newcommand{\samreview}[1]{\textcolor{black}{#1}}
\newcommand{\samrevieww}[1]{\textcolor{black}{#1}}

\newcommand{\samgreen}[1]{{\textcolor{black}{#1}}}
\newcommand{\corentinb}[1]{\textcolor{black}{#1}}
\newcommand{\corentinc}[1]{\textcolor{black}{#1}}
\newcommand{\corentind}[1]{\textcolor{black}{#1}}
\newcommand{\samlast}[1]{{\textcolor{black}{#1}}}
\newcommand{\samsc}[1]{{\textcolor{black}{#1}}}

\newcommand{\numhi}[1]{\textbf{\textcolor{black}{#1}}}

\usepackage{ulem}
\usepackage{soulutf8}
\usepackage{cancel}
\usepackage{lmodern} 

\newcommand{\cor}{\ifmmode coeur \else c\oe{}ur\fi\xspace}

\newcommand{\ssum}{\displaystyle\sum}

\newcommand{\influenceability}{influenceability~}

\newcommand{\gauginggame}{\textit{gauging game}}
\newcommand{\countinggame}{\textit{counting game}}
\newcommand{\gauginggames}{\textit{gauging games}}
\newcommand{\countinggames}{\textit{counting games}}

\newcommand{\DIbf}{\textbf{individual influenceability}}
\newcommand{\DIcbf}{\textbf{population influenceability}}
\newcommand{\DI}{individual influenceability}
\newcommand{\DIc}{population influenceability}


%% file: abstract.tex
{\small \textbf{
Opinion evolution and judgment revision are mediated through social influence. Based on a \samrevieww{large} crowdsourced in vitro experiment \samrevieww{($n=861$)}, it is shown how a consensus model can be used to predict opinion evolution in online collective behaviour. It is the first time the predictive power of a quantitative model of opinion dynamics is tested against a real dataset. \samreview{Unlike previous research on the topic, the model was validated on data which did not serve to calibrate it. This avoids to favor more complex models over more simple ones and prevents overfitting.} The model is parametrized by the \textit{influenceability} of \cocoreview{each individual}, a factor representing to what extent individuals incorporate external judgments.
\samrevieww{The prediction accuracy depends on prior knowledge on the participants' past behaviour. Several situations reflecting data availability are compared. When the data is scarce, the data from previous participants is used to predict how a new participant will behave.}
Judgment revision includes unpredictable variations which limit the potential for prediction. \samrevieww{A first measure of unpredictability is proposed. The measure is based on a specific control experiment.} More than two thirds of the prediction errors are found to occur due to unpredictability of the human judgment revision process rather than to model imperfection.}